# Fourteen years of software engineering at ETH Zurich

Bertrand Meyer

Politecnico di Milano and Innopolis University

Bertrand.Meyer@inf.ethz.ch

## Abstract

A Chair of Software Engineering existed at ETH Zurich, the Swiss Federal Institute of Technology, from 1 October 2001 to 31 January 2016, under my leadership. Our work, summarized here, covered a wide range of theoretical and practical topics, with object technology in the Eiffel method as the unifying thread[1].

## Overview

Computer science ("Informatik") has a brilliant past at ETH Zurich, most notoriously illustrated by Niklaus Wirth; but there has not traditionally been, nor is there now, a group or professorship with the name "software engineering". The Chair of Software Engineering existed from 1 October 2001, when I was appointed on Wirth's former position with the title of Professor of Software Engineering, until 31 January 2016. Our slogan, covering both the products and processes of software, was "**help the world produce better software and produce software better**". This chapter is an account of what the Chair did towards this goal. It is divided into the following sections:

1. Basic data.
2. Conferences, journals, summer school and other outreach.
3. Courses taught.
4. Principle and experience of teaching programming.
5. Pedagogical tools and MOOCs.
6. Methods and tools for distributed software development.
7. Language development and standardization.
8. Software process and methodology, requirements analysis, agile methods.
9. Object persistence.
10. Verification: proofs, tests, automatic bug correction.
11. Concurrent programming.
12. Software for robotics.
13. Assessment.

While this order of sections facilitates the presentation it requires some patience from the reader, since the topics that accounted for most of the chair's attention, resources and scientific results appear from section 10 on.

---

[1] This article was prepared, but rejected, as a chapter for the proceedings [288] of the PAUSE symposium, which took place on the occasion of the Chair's closing. The formatting retains some of the original Springer typesetting rules.

## 1 Basic data

**Faculty**:
- Me.
- Peter Müller, assistant professor 2003-2008 (prior to establishing his own chair).
- Jean-Raymond Abrial, guest professor 2004-2006 (to start the Rodin project which developed the Event-B rigorous program construction system). The rest of this article does not cover Müller's and Abrial's specific research.

**Completed PhD theses** in the Chair: 23 from 2004 to 2017 (see list below).

Over 60 **master's theses** (including "diploma theses", the name in the first few years before ETH transitioned to the "Bologna system" of bachelor's and master's degrees). Several hundred bachelor's theses and student project reports.

**Start-up companies** coming out in full or in part from the Chair: Comerge (software development); Monoidics (software verification, acquired by Facebook); mtSystems (language translation); Propulsion Academy (mass programmer education, propulsionacademy.com).

**Funding**: over 8 million Swiss Francs (or dollars, or euros) of funding from sources including:
- ERC "Advanced Investigator Grant" from the European Research Council on "Concurrency Made Easy", 2.5 million euros (see section 11).
- About 10 grants from the Swiss National Science Foundation.
- About 10 grants from the internal ETH funding agency.
- Foundations[2]: four grants from the Hasler foundation (including for the robotics work presented in section 12) and one from the Gebert foundation.
- Two grants from Microsoft Research, for curriculum development and concurrency.
- Two small grants from the ETH rectorate (studies administration) for curriculum innovation, including MOOCs.
- "Pioneer award" to Marco Trudel for commercial development of his language translation tools.

**Senior researchers**, defined here to encompass, regardless of actual administrative title, senior members who stayed for at least two years and exerted a significant influence on the development of our work, usually co-teaching courses and helping supervise PhD students[3]:

---

[2] One of the distinctive features of Switzerland's research scene is the presence of private research-funding foundations, usually resulting from a company founder's large bequest. They target specific areas (for example, in Hasler's case, IT and communications) and often follow a simplified process based on objectives rather than bureaucracy.

[3] The senior researchers typically continued to work on some of their earlier research themes, often with their former collaborators or supervisors. The corresponding publications appear in the bibliography but are not discussed in the text when they address topics outside of the group's research scope.

- Manuel Oriol, who helped in shaping our research culture and particularly in establishing a professional publication process, as well as participating directly in several projects.
- Sebastian Nanz, who was instrumental in the successful preparation of the Concurrency Made Easy ERC Advanced Investigator Grant project.
- Carlo Furia, who played a key role in our verification work.
- Martin Nordio, the heart and soul of the distributed programming teaching and research effort.
- Marco Piccioni, who alerted me to the rising role of MOOCs and ensured, through grueling work, the success of our three MOOCs.
- Christian Estler, who developed the CodeBoard system for on-the-cloud programming, key to that success and also to non-MOOC endeavors.
- Chris Poskitt, who introduced us to the culture of model checking.
- Jiwon Shin, coming from Mechanical Engineering, who taught us serious robotics.

**Postdocs** included (in addition to several of the PhD students listed next, staying on for a few months after the completion of their degree): Arnaud Bailly (concurrency), Đurica Nikolić (concurrency/verification), Georgiana Caltais (concurrency/verification), and Patrick Eugster (verification).

Peter Kolb (originally from ABB, then with his own company Red Expel) participated from the start in our distributed software engineering course (section 3).

The 23 **graduated PhDs**, in chronological order of defenses, are[4]:

- Karine Arnout (later Bezault): patterns and components [34], work started on 1 April 2002 and defended on 31 March 2004 (must be some kind of record).
- Markus Brändle: tools for teaching programming [46], a transfer to our group for the final part of a thesis under Prof. Jürg Nievergelt.
- Piotr Nienaltowski: concurrency [76], the first thesis on SCOOP and the inspiration for remainder of our concurrency work.
- Till Bay [95]: the Origo system for code and project management, a precursor to GitHub.
- Bernd Schoeller [96]: first dissertation on proofs, introduced the use of Boogie and model queries.
- Ilinca Ciupa [97]: automatic testing, including ARTOO (adaptive random testing for OO software), one of the first two AutoTest dissertations.
- Andreas Leitner [98]: automatic testing, including test extraction from failures and contract-driven development, the other initial AutoTest dissertation, an excellent example of two synchronized and complementary theses, both excellent.
- Martin Nordio [119]: proof-transforming compilation.

---

- Michela Pedroni [120]: principles and tools for teaching programming.
- Yi (Jason) Wei [186]: automatic testing and automatic bug correction.
- Marco Piccioni [187]: object persistence.
- Marco Trudel [213]: programming language translation.
- Stephan van Staden [214]: separation logic.
- Benjamin Morandi [235]: concurrency and its operational semantics.
- Nadia Polikarpova [236]: concepts and tools for verifying object-oriented programs, with application to the EiffelBase 2 library.
- Yu (Max) Pei [237]: automatic bug correction.
- Hans-Christian Estler [238]: tools for collaborative development.
- Julian Tschannen [239]: concepts and tools for verifying object-oriented programs.
- Scott West [240]: concurrency, with particular emphasis on performance.
- Mischael Schill [267]: concurrency with emphasis on verification and performance.
- Andrey Rusakov [268]: concurrency and robotics software.
- Alexey Kolesnichenko [269]: concurrency and GPU programming.
- Alexander Kogtenkov [272]: software verification and language design focusing on void safety (removing null pointer dereferencing).

**Other team members** included Volkan Arslan (concurrent and event-driven programming), Yann Müller (concurrency), Andre Macejko (support for MOOCs and other teaching); Claudio Corrodi (model checking), Michael Ameri (verification), Ganesh Ramanathan, Ivo Steinmann and David Itten (robotics), Lukas Angerer and Paolo Antonucci (tool support for teaching), Michael Steindorfer and Alexander Horton (automatic bug fixing), Daniel Moser (concurrency), Lukas Angerer [5]. Roman Schmocker, a research engineer in the CME project, played an essential role in the development of SCOOP, both the model and the implementation). We hosted student interns including several from the University of Cluj-Napoca, Raduca Borca-Muresan, Gabriel Petrovay and Ilinca Ciupa prior to her PhD, and from rom Nijny-Novgorod: Andrey Nikonov and Andrey Rusakov, the latter also as a prelude to his PhD. Peter Kiowski helped at the start of Informatics Europe. Three PhD students, Susanne Cech, Stephanie Balzer and Joseph Ruskiewicz, started in the Chair then moved to other groups. Patrick Schönbach, working from Germany, helped with the initial version of the Traffic library (section 4).

The Chair had two successive **secretaries**, Ruth Bürkli and Claudia Günthart, whose role in the effectiveness of our work cannot be overstated, and who provided an indispensable measure of Swissness for a "mini-United-Nations" group where Swiss citizens were always a minority. They were essential in making sure we functioned smoothly, overcame the inevitable bumps of daily life, followed Swiss practices and ETH regulations, and met our obligations to the university.

**Next employment**, partial list: associate/full professor (Eugster at Purdue, Furia at Chalmers, Oriol at York), assistant professor (Polikarpova at UCSD, Yu at Hong

---

[5] The names cited include those of master's students who co-authored some of the publications in the bibliography. They are a small subset of the many master's students we hosted.

Kong), research fellow (Poskitt at Singapore), Google Zurich (Nanz, Leitner, West, Tschannen, van Staden), Microsoft (Kolesnichenko), own startup (Bay at Comerge), Facebook (Calcagno, initially Monoidics startup), ABB Zurich (Oriol, Brändle), Roche (Nordio), Axa Rosenberg (Arnout), Praxis (Nienaltowski), Credit Suisse (Ciupa), Airbus (Brändle).

**Visitors** included Professors Egon Börger (Pisa), Hassan Gomaa (George Mason), Manuel Mazzara and Michele Mazzucco (both of Innopolis University, as part of a special agreement in which ETH helped train Innopolis's future faculty), Emil Sekerinski (McMaster's), as well as Cristiano Calcagno (Imperial College), Marie-Hélène Ng Cheong Vee (Marie-Hélène Nienaltowski) (Birkbeck), Federica Panella (Politecnico di Milano) and Per Madsen (Aalborg).

These visitors were part of our scientific relations with **outside groups** with which we had regular exchanges, including:

- Eiffel Software, the Santa-Barbara-based company which I co-founded in 1985 and where I worked prior to joining ETH: Emmanuel Stapf (chief engineer, and authors of several papers); Alexander Kogtenkov (who did a PhD as external student at ETH); Jocelyn Fiat (who provided many Eiffel-related student projects and helped guide them); Xavier Rousselot (who came up with the original idea behind AutoTest). Members of the Chair who spent internships at Eiffel Software include Andreas Leitner, Julian Tschannen and Roman Schmocker.
- University of York (UK): Jim Woodcock, Richard Paige, Manuel Oriol (after he left us) and others.
- York University (Another York, in Canada): Jonathan Ostroff.
- University of Nantes: Jean Bézivin (co-founder of the TOOLS conference series), Frédéric Benhamou.
- University of the Sarre, Saarbrücken: Andreas Zeller (who was instrumental in supporting our forays into empirical software engineering and automatic program repair).
- University of Karlsruhe: Walter Tichy.
- Technical University of Munich: Manfred Broy, Martin Wirsing.
- University of Limerick (LERO), Ireland: David Parnas.
- Politecnico di Milano: Carlo Ghezzi and Dino Mandrioli (since the late seventies!), Elisabetta di Nitto, Raffaela Mirandola, Giordano Tamburrelli (particularly in the context of the DOSE projects described below), Letizia Tanca.
- PUCRS (Brazil): Rafael Prikladnicki (also part of DOSE).
- University of Pisa: Egon Börger.
- Closer to home, the computer science department of the University of Zurich: Martin Glinz (requirements analysis), Harald Gall (empirical software engineering), Abraham Bernstein.
- On the other side of the Gotthard: at the University of Lugano, Mehdi Jazayeri, Mauro Pezzè.
- University of Bern: Oscar Nierstrasz.
- University of Lucerne (formerly Hochschule Luzern, partner in the Smart-Walker effort, section 12): Alexander Klapproth, Dieter von Arx.
- Czech Technical University, Prague: Pavel Tvrdik.

- Poznan University of Technology (Poland): Jerzy Nawrocki and Bartosz Walter (who participated in one of our research retreats).
- University of Cluj-Napoca (Romania): Iosif Ignat (who, in particular dedication to his students, sent us a succession of excellent students for internships).
- Imperial College: Sophia Drossopoulou, Jeff Magee, Jeff Kramer, Alexander Wolf.
- University of Nice: Philippe Lahire (object-oriented techniques, Denis Caromel (concurrency), Isabelle Attali.
- IRISA (Brittany): Jean-Marc Jézéquel, Noël Plouzeau, Yves Le Traon, Benoît Baudry (developments of Design by Contract).
- University of Luxembourg: Lionel Briand, Nicolas Guelfi, Yves Le Traon.
- JAIST (Kanazawa, Japan): Hiroshi Futatsugi (formal methods).
- Rio Cuarto (Argentina): Nazareno Aguirre (distributed development).
- Jiao Tong (Shanghai): Jianjun Zhao (verification).
- University of Hong Kong: T.H. Tse (testing).
- Polytechnic University of Madrid: Natalia Juristo (empirical software engineering).
- IMDEA Software Institute, Madrid: Manuel Hermenegildo, Gilles Barthe.
- University of Washington: David Notkin, Michael Ernst.
- MIT: Daniel Jackson.
- Berkeley: Armando Fox (author of a highly successful programming MOOC).
- Stanford: Gene Golub, Donald Knuth.
- Purdue: Patrick Eugster, Jan Vitek.
- University of British Columbia: Gail Murphy, Philippe Kruchten.
- Polytechnic University of Catalonia: Pere Botella.
- State University of Saint Petersburg: Andrey Terekhov.
- ITMO University (also in Saint Petersburg, where I had a part-time position as head of the Software Engineering Laboratory from 2011 to 2014): Anatoly Shalyto (who recommended to me Nadia Polikarpova, first as a master's student), Vladimir Parfionov, Vladimir Vassiliev (rector).
- Polytechnic University of Saint Petersburg: Mikhail Itsykson, Irina Shoshmina, Yuri Karpov.
- Other Russian institutions including the universities of Nijny Novgorod (Vladimir Gergel, Iosif Meyerov) and Tver (Vladimir Billig, skilled translator of no fewer than three of my books into Russian). Ever since I got to know the great Soviet computer scientist Andrey Ershov in 1975 and spent time in his Novosibirsk institute in 1977, I kept ties with the Russian CS community, including Novosibirsk (Alexander Marchuk, Nikolai Shilov).
- In Israel: Avi Mendelson (Intel Haifa then Technion), Amiram Yehudai (Tel-Aviv), Joseph Gil (Technion, like Yehudai a long-time participant in TOOLS), Mordechai Ben Ari (Technion), Dan Berry (Technion, later Waterloo).
- Monash University (Melbourne, Australia, where I had an adjunct position until 2003): Christine Mingins, David Abramson, John Rosenberg.

- University of New South Wales: John Potter (object-oriented programming and formal methods), Carroll Morgan (formal methods).
- Victoria University (New Zealand): James Noble (object-oriented methods).
- IFIP working group WG2.3 on programming methodology, of which I became a member in 2003: everyone but particularly Pamela Zave, Michael and Daniel Jackson, Rustan Leino, C.A.R. Hoare, Ralf-Johann Back, Gary Leavens, Jay Misra, Patrick Cousot, Michel Sintzoff, Natarajan Shankar, Carroll Morgan, Michel Sintzoff (until his untimely death in 2010), Andreas Podelski…
- IFIP TC2 (the oldest technical committee of IFIP, devoted to programming, where I was Swiss representative then chair): Robert Meersman (previous chair), Judith Bishop (previous secretary), Michael Goedicke ("my" secretary and successor as chair), Michel Sintzoff, Bashar Nuseibeh.
- Ecma, the standards organization we discovered on the occasion of .NET standardization and chose for Eiffel standardization: Jan van den Beld (who as general secretary helped us start the work), Patrick Charollais, Istvan Sebestyen.
- Specifically, the Ecma Eiffel committee (TC49-TG4), which produced the ISO Eiffel standard as discussed in section 7. The most active members of this committee were Emmanuel Stapf and Alexander Kogtenkov from Eiffel Software, Mark Howard initially from Axa Rosenberg (an equity management firm in California), Eric and Karine Bezault also of Axa Rosenberg, Dominique Colnet and some of his colleagues from the University of Nancy (authors of the SmartEiffel compiler) and (more episodically) Christine Mingins from Monash University, Kim Waldén from Enea Data in Sweden, Paul-Georges Crismer from Groupe S in Belgium and Roger Osmond from EMC. In addition, members of the Chair who often participated in TG4 language discussions included Volkan Arslan, Bernd Schoeller and Julian Tschannen.
- Microsoft, particularly Microsoft Research (MSR) but also product groups. First the .NET effort (Project Seven), particularly James Plamondon and Jim Miller. Then verification researchers: Wolfram Schulte (thanks to whom I spent two fruitful short stays in Redmond), Rustan Leino, Michał Moskal (the last two supervised Nadia Polikarpova in an internship at MSR), Yuri Gurevich, C.A.R. Hoare, Mike Barnett, Tom Ball, Erik Meijer, Clemens Szyperski, Nikolaj Bjorner and Pelli de Halleux from the PEX project, Nikolai Tillmann… A special mention is due to Judith Bishop, indefatigable convener of thrilling summer schools and other events, and indefatigable sponsor of our work.
- Informatics Europe, particularly the founder group including Jan van Leeuwen (Utrecht), Christine Choppy (Orsay), Willy Zwaenepoel (EPFL), Hans-Uli Heiss (Berlin), Antoine Petit (INRIA), Manfred Nagl and Gregor Engels (Paderborn), Enrico Nardelli (Rome), Letizia Tanca (Milan), and later Carlo Ghezzi (Milan) and Lynda Hardman (Amsterdam), Letizia Tanca (Milan), Victor Gergel (Nijny-Novgorod), Mark Harris (thanks to whom Intel generously sponsored the Informatics Europe Best Practices in education award), as well as the general secretary of the organization, Cristina Pereira, thanks to whom the organization was able to produce its yearly milestone reports on the state of things in European computer science education [203] [227].

- ACM Europe (I was a member of the founding Council), particularly its president Fabrizio Gagliardi (initially with Microsoft Research), Matthias Kaiserswerth (then director of the Zurich IBM lab, later head of the Hasler Foundation), Wendy Hall, John White (CEO of ACM), Vinton Cerf (President of ACM), Paul Spirakis, Michel Beaudouin-Lafon, Alain Chenais (later president of ACM), Alexander Wolf (also an ACM president during our tenure), Mateo Valero, Gabriele Kotsis, Marc Shapiro (who in another context had exerted a significant influence on the design of Eiffel's exception mechanism), Paola Inverardi (one of the first computer scientists to become Rector of a major university), Avi Mendelson, Ricardo Baeza-Yates, Serdar Tasiran, Mark Harris and Carlo Ghezzi again, and in fact many of the other computer science luminaries cited elsewhere in this article.
- Semat: Ivar Jacobson, Michael Goedicke; OMG: Richard Soley.
- EasyChair: Andrey Voronkov.
- Springer: Alfred Hofmann (publisher of the LNCS series), Hermann Engesser (initial editor of *Touch of Class* [110]), Ralf Gerstner (his successor and patient editor of *Agile!* [228] and several other works).
- The Hasler Foundation, one of the important sources of funding for research in computing in Switzerland, not only funded us but provided frequent opportunities for interactions with other research groups. I co-edited the volume that resulted from one of the Hasler workshops [57].
- Within ETH, my teaching benefitted from expertise on algorithms by Peter Widmayer and on numerical applications by Walter Gander. (Both of them were my predecessors as department heads and also provided advice in that capacity.) To get started at ETH, I received help from Hans Hinterberger and Jürg Nievergelt. Niklaus Wirth was always a font of experience and support.

**Awards** that I received during the existence of the Chair include ACM Software System Award, IEEE Harlan Mills prize, ACM Fellow, Ershov Lecture [191], and two honorary doctorates (University of York and ITMO University).

A word is in order regarding the **atmosphere** in the group. After a year or two of setting up (and, for me, learning the job), the group quickly reached a smooth mode of functioning, with no visible tension − not a given for a team of sometimes as many as 20 members, coming from widely different cultural backgrounds and subject to many external pressures. Members socialized a lot with each other. We had occasional research retreats in the neighboring Alps − I can recommend "PowerPoint Karaoke" for the evenings of such meetings − and (prompted by Manuel Oriol) yearly publication-planning workshops. The weekly research meeting, Tuesdays 14-16, involved organizational discussions, presentations by members of the group on their current work, students' final reports on master's and other projects (only for the most interesting reports, since we supervised too many students to fit them all[6]), and visitors' talks.

---

[6] When we had many student projects coming to fruition at the end of a semester, we organized special meetings devoted entirely to their presentations. The idea came in part from

With presenters ranging from novice (a 2nd-year student) to world-class guests, the technical interest of these weekly sessions and their relevance to each group member naturally had their ups and downs, but holding a regular meeting promoted communication and collaboration. Otherwise the risk would have been, given the breadth of topics addressed in the group, of splitting into micro-groups of one to three or four people. Instead, cross-fertilization regularly occurred. As suits a software engineering group in academia, our work was partly theoretical and partly applied, but there never were a theory group and an applied group. Two examples of this intellectual mobility:

- After a thesis on object persistence and databases, Marco Piccioni turned his interest to educational issues and MOOCs, triggering the interest of Christian Estler who had been working on tools for distributed programming, and Martin Nordio whose own thesis work was on formal modeling of programming language concepts but who was also managing the DOSE distributed programming project with students from ETH and other universities. The resulting collaboration, benefiting from all three contributors' expertise, led to the Eiffel4Mooc, comcom and Codeboard platforms for cloud-based teaching of programming.

- These ideas also attracted the interest of the formal methods and verification team, notably Julian Tschannen, Nadia Polikarpova and Carlo Furia, leading to the cloud-based version of the AutoProof program verification system.

An example of the benefits of cross-fertilization was the spread of formal techniques. Only a minority of the group members had formal (mathematical) specification and verification techniques at the center of their research topic. But they spread the formal-methods mindset throughout the group. Soon many others who came from a background of more applied software engineering, with little or no prior knowledge of mathematical techniques started incorporating formal aspects to their work. The other way around, Martin Nordio is an example of a group member who came to do formal work and continuously produced results in that area even after his PhD while getting into more and more applied topics.

These outcomes are just specific instances of a practice of **collaboration** without which the group could not have functioned properly. Teaching introductory programming required particular care because of the scale of the effort (up to 25 laboratory groups, each under the responsibility of an assistant) and the attention focused on first-year courses, whose final exam determines whether students can go on (about 40% do not). Every detail must be addressed to ensure that the administration does not get a dreaded "*Rekurs*" (a formal complaint by a student who failed).

For many of the PhD students this stint as assistant was their first teaching experience, and they needed all the help they could get from their seniors. The **culture** developed over the years made it possible to train and integrate new members

---

watching such sessions at Saint Petersburg State University (when visiting Terekhov). Another idea that I took from that Russian model was to insert after every final project presentation a short assessment by the student's direct supervisor (PhD student or postdoc), placing the work in context and discussing its contributions. Maybe this practice is common but I had not seen it before; I found that it increased the value of the sessions.

quickly. It involved both actual artifacts (an internal Wiki with Howtos, heaps of training materials from previous years, example exam texts in English and German written to minimize the possibility of student misunderstanding and complaints) and more immaterial elements in the form of advice and recommended procedures.

Having a solid group culture was just as essential for **research** as for teaching. The typical beginning PhD student has great ideas − the reason for being hired as a PhD student in the first place − but no concept of how to write a scientific paper. (Occasionally, you find a born writer, able from the beginning both to conceive ideas and express them. But that is the exception.) This skill has to be learned; and, beyond general writing, the art of writing for this or that conference with its peculiar tradition and requirements.

PhD students also need to understand the **practical nature of research** and their own personal role. A kind of rite of passage for the budding researcher happens the day you understand you are probably not going to discover relativity or vaccines or quanta or natural selection or NP-completeness. The consolation is that if you work hard you can add an increment or two to the existing body of knowledge. The group can help junior members survive that phase.

**Mentoring** is an important part of such a culture. I quickly corrected my initial mistake of hiring (on the five positions, in addition to administrative assistant, that ETH gave me on arrival, and others soon obtained through grants) PhD students only. For their supervision, it was essential to have postdocs and senior researchers, equipped with more patience, plus fresh knowledge of the latest literature in their fields of expertise.

Our group culture did not exist in a vacuum but benefited from our insertion in the department of computer science and ETH as a whole. While we may occasionally have balked at some aspects of that institutional culture, being part of a university focused on excellence in teaching and research, with resources commensurate with those ambitions, was a tremendous inspiration and daily boost. It was always interesting to see how other groups in the department were conducting their own business, and learn from them. Aside from the official communication channels, much peer-to-peer interaction occurred between groups at the level of PhD students; in particular, when Peter Müller left us to form his own group, his students remained close to ours and often served as teaching assistants in our courses.

Within the group, much of the culture was the result of **self-organization** at the level of PhD students, senior researchers and the administrative assistant, so that many day-to-day problems would simply be resolved without even reaching me.

The group culture as just described did not arise from a conscious design, it just grew **organically** as a result of the group members' intelligence and dedication. I do not recall any discussion of "culture" as such. It just underpinned our mode of working. Only after leaving such an environment and attempting in vain to recreate similar conditions in a different setting does one understand how it was: how elaborate, how critical and how elusive.

## 2 Conferences, journals, summer school and other outreach

The following is a partial list of "community" efforts (as they are sometimes called) in which we were involved: Informatics Europe; conferences organized; summer school; guests; journals; publications.

### 2.1 Informatics Europe

As department head in 2005 I was approached by Willy Zwaenepoel, my counterpart at EPFL, to explore ways of making the voice of academic computer science (informatics) better heard in Switzerland. I suggested that the proper arena was bigger: Europe. Zwaenepoel brought up his experience with CRA, the successful Computing Research Association in the US. We decided to hold a "European Computer Science Summit" (ECSS), gathering department heads and senior faculty from all over the continent. Finding them was very manual work, mostly from browsing the Web. We held the first ECSS at ETH Zurich in October 2005; the event was a resounding success, the first time people in charge of academic policy, confronted with similar issues in education and research, could talk with their peers across Europe. The decision to found a permanent association, Informatics Europe[7], happened then; we reported on our effort in the *Communications of the ACM* [51]. For the first years of its existence, Informatics Europe was hosted by our group; members of the Chair, notably Marco Piccioni, helped with the organization, and we hosted three of the first ECSS in Zurich. (The conference continues to take place every year, moving between European cities.) When the association was able to hire a general secretary, Cristina Pereira, she also worked in our offices for several years. Nowadays, of course, Informatics Europe has shed its ties to ETH and has its own office (still in Zurich).

I was the first president, from 2005 to 2010, working closely with two vice presidents, Christine Choppy from Paris (Orsay) and Jan Van Leeuwen from Utrecht. We published (with Jørgen Staunstrup from ITU in Denmark), again in the *Communications of the ACM*, a paper on researcher evaluation [101] which attracted significant attention. The subsequent presidents have been Carlo Ghezzi from Milan and Lynda Hardman from Amsterdam, to be followed in 2018 by Enrico Nardelli from Rome. Today the association is strong, with over 115 members[8] from all over Europe, with a solid financial base, and growing.

Informatics Europe took on itself the task of providing a forum and action center for computing education and research in Europe. (The CRA is focused on research, but we felt we had to cover education as well.) The ECSS conferences have been a magnet for senior people in the field, with an impressive roster of invited speakers from top university management, education policy, European research funding, the CRA (some of our early keynoters from CRA provided great guidance), industrial research policy (Microsoft Research, Google, Intel and others were members from the start), entrepreneurship (such as the founder of Skype). There is no other conference of that kind.

---

[7] We played with a few names including one that, I am ashamed to admit, had my preference: euroTICS. I retrospectively shudder at the mockery we would have endured. Jan van Leeuwen fortunately brought us to our senses and came up with the final name.

[8] Members are not people but organizations: departments of computer science/informatics/IT etc. of universities, research labs of companies etc.

From the start we engaged into the hard work of collecting basic data on the field; in particular, through the grueling and exacting work of Cristina Pereira, Informatics Europe produces (and continues to produce today) detailed reports on matters such as degrees, student numbers, representation of women, faculty salaries; see [203] and [227], and www.informatics-europe.org for the more recent reports.

The report on salaries was particularly controversial; we hesitated before making this information public, but decided it was important. For comparable positions, the ratio of professor salaries in different countries is over 4 to 1, even within Western Europe with a comparable standard of living. PhD students and postdoc salaries follow a similar pattern. My interpretation of these figures is that European countries divide themselves into two categories: those that treat university professors as top talent, similar to senior engineers or higher management in companies; and those that view them as somewhat ameliorated high-school teachers. The difference is profound, and reflected in more than salaries. But that is for another article.

Besides salaries, the reports provide unique data, growing every year, on other issues such as the (by far insufficient) number of CS graduates produced every year, and reveal the need for better branding of the discipline, which is known, even after translation into English, under some 20 different names in Europe.

Gathering the data was a Herculean effort. There is simply no standard source. Each country has its own, often several of them, sometimes a ministry, sometimes a statistical institute, sometimes a professional association, sometimes individual institutions. Pereira made a crucial design decision at the start: we would only report results in which we had full confidence. My own idea was different: try to get as good as possible a picture as we could of the entire continent, then make it better in successive editions. She disagreed, and was right. The result that in the first report we could only cover a few countries, but with ironclad data. Then over the years the set of countries has grown, as colleagues find out with a shock that their country is not covered, complain, and are asked to provide a conduit to reliable data sources. But what the reports do include is supported by extensive research, attested by dozens of footnotes in the text. Again and again we have witnessed the following incident: a colleague from country X sees the report and protests: "I work in country X and I know! Your figures (e.g. the for salary ranges) are wrong!" Each time the result is the same: it is not because you work in X that you know everything about X. The information has been checked and validated from official sources, and the footnotes explain the conventions. (For example the reports do not attempt to account for differences in the cost of living, taxation levels etc., since this is a task for economists and subject to subjective appreciation; they limit themselves to raw data and the relevant side factors.)

Informatics Europe undertook several other reports, in particular the already mentioned work on researcher evaluation for informatics [101] [102]. In the US, top institutions and government agencies have recognized since the nineties that computer science has its own modes of operation not exactly amenable to the evaluation criteria of physics or mathematics. For example, whether that is healthy or not, some conferences are more selective and prestigious than most journals. In Europe the battle still had to be fought, and our report, I believe, played a big role in explaining how to assess CS research. (The battle is ever recommenced. At the time of this writing, several countries, including major ones, are destroying any hope of ever joining the top ranks of research in CS and IT − and I do mean *destroying* − by imposing the Scopus or Web-of-Science h-index as the measuring stick. Goodbye, my friends. Serious institutions in scientifically leading countries know better.)

One of the most visible Informatics Europe services is the annual Best Practices in Education award, which recognizes innovative approaches to teaching[9].

Yet another Informatics Europe initiative took part in concert with ACM Europe, a new section (launched at ECSS 2009 in Paris) of the US-based ACM. The topic of our first joint effort was the teaching of informatics in schools. The (sometimes animated) discussions led to a report under the leadership of the committee's head, Walter Gander of ETH, and with my participation [190]. It makes a clear case for introducing informatics at the secondary and a primary school levels. Shortly before our report appeared, a British effort for the Royal Society, directed by Simon Peyton-Jones, made a similar point; the idea was in the air. One of the contributions of the ACM-Informatics Europe work was to emphasize the distinction between teaching "digital literacy" (word processing and the like) and teaching informatics as a scientific discipline. The public and political decision-makers often confuse the two. Teaching teenagers to browse the web, hardly necessary anyway, is not the same as teaching the seminal concepts of informatics, what Jeannette Wing (a keynoter at the 2007 Berlin ECSS) famously calls "computational thinking".

Besides such concrete results, Informatics Europe produced an immense amount of good will, collaboration (including between the Western and Eastern sides of the continent) and friendship. One of the most rewarding parts of the experience for me was the ability to step out at the expiration of my last term as president, let others take over a solid, stable and growing organization, and from the sidelines watch them take it to new heights.

## 2.2 Conferences organized

A partial list of conferences we organized:

- We organized the first ECSS conferences (European Computer Science Summit) of Informatics Europe, just discussed: Zurich 2005, 2006 and 2008; Berlin 2007 with Hans-Uli Heiss; Paris 2009 with Christine Choppy; Prague 2010 with Pavel Tvrdik,

- I had run the TOOLS conference series (Technology of Object-Oriented Languages and Systems) since 1989, with sessions in Europe (led by Jean Bézivin), the US (Santa Barbara), Australia (TOOLS PACIFIC, Sydney or Melbourne) plus a couple in China and Eastern Europe. TOOLS played an important role in the development of object technology, providing a friendly and bias-free environment for discussion of new ideas; many important concepts, such as design patterns, were first presented at TOOLS. The most prestigious scientists and technology leaders gave keynotes, including Adele Goldberg, Alan Kay, Philippe Kahn, Oscar Nierstrasz, Robin Milner, Tony Hoare, Eric Gamma, Dave Thomas, Ivar Jacobson, David Parnas and many others. We continued the conference series at ETH, organizing several sessions in Zurich, as well as one in Malaga, courtesy of Antonio Vallecillo, and one in Prague, courtesy of Pavel Tvrdik. The Prague conference in 2012 was number 50; I felt it was time to declare victory —the title of the conference was "*The Triumph of Objects*" — and end the series on a high note.

---

[9] I have a weakness for the project that received the 2013 award: a collection of videos available on YouTube (www.youtube.com/user/AlgoRythmics), illustrating the main sorting algorithms in the form of Central-European dances. See e.g. the Quicksort dance.

- In its last years the quality and friendly atmosphere of TOOLS attracted a growing set of satellite conferences. One of them was TAP, Tests And Proofs. Realizing that the old fight between dynamic and static verification was increasingly turning into complementarity, Yuri Gurevich and I started TAP in 2007 [68] in association with TOOLS. It has become an important annual event and continues today in another setting.

- With our growing interest in distributed development, we started, also in 2007, and also as a satellite event to TOOLS, the SEAFOOD conference series, Software Engineering Approaches For Offshore and Outsourced Development. It ran through several successful sessions, with Springer proceedings [72] [103] [107] [130]. Playing our part in limiting conference inflation, we felt that SEAFOOD's mandate could safely be handed over to the International Conference on Global Software Engineering, which had been started independently with IEEE support and in which we also participated actively (we received three successive Best Paper awards at ICGSE, 2012 to 2014, and Martin Nordio was PC chair in 2015).

- Other conferences co-located with TOOLS (but organized by others) included the International Conference on Model Transformation (ICMT) and Software Composition.

- A "Future Of Software Engineering" symposium, FOSE, was held in 2010, with speakers by invitation: David Parnas, Barry Boehm, Manfred Broy, Patrick Cousot, Erich Gamma, Yuri Gurevich, Michael A. Jackson, Rustan Leino, David Parnas, Dieter Rombach, Joseph Sifakis, Niklaus Wirth, Pamela Zave and Andreas Zeller. Sebastian Nanz edited the proceedings, published as a Springer book.

- We organized the first VSTTE conference (Verified Software: Tools, Technologies, Experiments), the embodiment of Tony Hoare's Verified Software Grand Challenge ([283], see section 10.1) in 2008, leading to the first VSTTE volume co-edited with Jim Woodcock [86]. The VSTTE series has become one of the principal conferences in software verification.

- With Nadia Polikarpova I organized the 2013 European Software Engineering Conference (ESEC-FSE, with the Foundations of Software Engineering Conference) in Saint Petersburg. It remains the only major international software engineering conference ever organized in Russia.

- In 2015, I was in charge, together with my colleague Walter Gander and with the help of Chris Poskitt and Jiwon Shin, of a symposium organized by the computer science department in honor of Niklaus Wirth's 80th birthday. The speakers, all by invitation, were Vint Cerf, Hans Eberlé, Michael Franz, me, Carroll Morgan (on the later developments of the program-refinement idea pioneered by Wirth), Martin Odersky, Clemens Szyperski and Wirth himself. A memorable event.

- We organized various Eiffel-related meetings, in particular several workshops on SCOOP and concurrency, with colleagues such as Jonathan Ostroff and Richard Paige, and a 2012 "Workshop on Advances in Verification for Eiffel" (WAVE).

- With Richard Soley, I helped start Ivar Jacobson's SEMAT initiative (Software Engineering Methods And Tools, I think I came up with the name), and with the help of Carlo Furia organized the first meeting. I stepped out of SEMAT after a while; see 8.6.

- Jerzy Nawrocki from Poznan asked me to help grow the Polish software engineering conference to which I had given a keynote in Warsaw in 2006. I suggested to broaden the scope to the whole region, leading to the CEE-SET (Central and East European Conference on Software Engineering Techniques) series. I have also been associated for many years with a conference bearing a similar name but based in Russia, CEE-SECR (Nick Puntikov, Andrey Terekhov), where I sponsor the best-paper award.

- For some years we (particularly Bernd Schoeller) organized a regular seminar, FATS (Formal Approaches To Software), with both internal speakers and guests (se.ethz.ch/old/events/fats/).

- I organized four meetings of the IFIP WG2.3 (Programming Methodology) working group: Prato, Italy in 2004 (with Karine Arnout), Santa Barbara in 2011, Saint Petersburg in 2013 (with Nadia Polikarpova), Villebrumier in 2016.

- As chair of IFIP TC2 (Technical Committee on Programming) I hosted a meeting at ETH.

## 2.3 LASER summer school

Instructed by the experience of others' summer schools[10], we decided to organize our own starting in 2004: the **LASER summer school** (Laboratory for Applied Software Engineering Research), held every year except 2016 in Elba Island, Italy, with prestigious speakers including four Turing award winners (not counting future ones). LASER has become a well-known event. The themes and speakers (in addition to me, since I have lectured in every session) have been:

- 2004 – Practical Techniques of Software Quality: Jean-Raymond Abrial, Ernie Cohen, Erich Gamma, Carroll Morgan, Pamela Zave (this kick-off session set the tone for the entire series by including speakers from both the practical and theoretical schools of software engineering).

---

[10] Particularly two in which I had lectured: the venerable Marktoberdorf summer school in Germany, created by F.L. Bauer (I was a student in 1975) and continued by Manfred Broy then our former ETH colleague Alexander Pretschner; and Alfredo Ferro's long-running Lipari summer school, organized in the year I was there by Egon Börger.

- 2005 – Software engineering for concurrent and real-time systems: Laura Dillon, Jay Misra, Amir Pnueli, Wolfgang Pree, Joseph Sifakis (the first of several sessions devoted to concurrency).
- 2006 – Practical Programming Processes: Ralph-Johan Back (refinement), Miguel de Icaza (agile), Erik Meijer (agile), Mary Poppendieck, (Lean), Andreas Zeller (debugging) (another striking combination of theoretical and practical contributions).
- 2007 – Applied Software Verification: Thomas Ball, Gérard Berry, C.A.R Hoare, Peter Müller, Natarajan Shankar (star speakers on the state of the art in program verification).
- 2008 – Concurrency and Correctness: Tryggve Fossum, Maurice Herlihy, C.A.R Hoare, Robin Milner, Peter O'Hearn, Daniel A. Reed (concurrency again, with a new set of top contributors to the field).
- 2009 – Software Testing: The Practice And The Science, Alberto Avritzer, Michel Cukier, Yuri Gurevich, Mark Harman, Tom Ostrand, Mauro Pezzè, Elaine Weyuker (a review of testing techniques from a diverse range of viewpoints).
- 2010 – Empirical Software Engineering: Victor Basili, Joshua Bloch, Barry Boehm, Natalia Juristo, Tim Menzies, Walter Tichy (empirical software engineering has grown explosively in recent decades, and the school gathered some of the people who created and developed the discipline).
- 2011 – Tools for Practical Software Verification: Ed Clarke (model checking), Patrick Cousot (abstract interpretation), Patrice Godefroid (model checking), Rustan Leino, (Boogie) César Muñoz (PVS), Christine Paulin-Mohring (Coq), Andrei Voronkov (Vampire) (a dazzling review of verification tools).
- 2012 – Innovative Languages for Software Engineering: Andrei Alexandrescu (D), Roberto Ierusalimschy (Lua), Ivar Jacobson (UML), Erik Meijer (MSR languages), Martin Odersky (Scala), Simon Peyton-Jones (Haskell), Guido van Rossum (Python) (this session with the "rock star" programming language designers was the all-time best-seller).
- 2013 – Software for the Cloud and Big Data: Roger Barga, Karin Breitman, Sebastian Burckhardt, Adrian Cockcroft, Carlo Ghezzi, Anthony Joseph, Pere Mato Vila.
- 2014 – Leading-Edge Software Engineering: Judith Bishop, Harald Gall, Daniel Jackson, Michael Jackson, Erik Meijer, Gail Murphy, Moshe Vardi,.
- 2015 – Concurrency: the Next Frontiers: Manfred Broy, Maurice Herlihy, Jeff Kramer, Jayadev Misra, David Parnas (another "rock star" session, preceded by a meeting of the scientific advisory board of the CME project (section 11), whose members were also the LASER speakers).
- 2017 (forthcoming at the time of this writing) – Software for Robotics: Davide Brugali, Rodolphe Gelin, Ashish Kapoor, Nenad Medvidovic, Issa Nesnas.

From 2017 on, the newly established nonprofit LASER Foundation has taken over the organization of the school.

### 2.4 Guests

In addition to the speakers at LASER and the other conferences listed above, we had a constant influx of invited talks, some in the computer science department's weekly seminar and others just hosted by us. One of the sources of talks, particularly in later years, was thesis defenses: the thesis committees included many prestigious international colleagues and whenever possible we took advantage of their presence in Zurich to have them give a talk. We had too many guest speakers to list them all, but a few that come to mind are Gilles Barthe, Jean Bézivin, Gérard Berry, Armin Biere, Roderick Bloem, Manfred Broy, Marsha Chechik, Patrick Cousot, Miguel De Icaza, Tom De Marco, Brian Fitzgerald, Patrice Godefroid, Michael Gordon, Adele Goldberg, Susanne Graf, Yuri Gurevich, Reiner Hähnle, Mark Harman, Jean-Marc Jézéquel, Joseph Kiniry, Rustan Leino, Carroll Morgan, Erik Meijer, Scott Meyers, Jay Misra, Hausi Müller, Jonathan Ostroff, Joel Ouaknine, Richard Paige, Matthew Parkinson, David Parnas, Mauro Pezzè, David Redmiles, Martin Robillard, Bill Roscoe, Emil Sekerinski, Natarajan Shankar, Mark Shapiro, James Whittaker, Jim Woodcock, Andreas Zeller.

For the CME Advanced Investigator Grant project of the European Research Council (section 11), we established a Scientific Advisory Board consisting of Manfred Broy (Munich), Maurice Herlihy (Brown), Jeff Magee (Imperial College), José Meseguer (Illinois), Jayadev Misra (Austin), David Parnas (Limerick), Bill Roscoe (Oxford) and Jeannette Wing (Microsoft). The Board held two meetings, one in Zurich in March of 2014, accompanied by seminars by members of the Board, and the other in Elba , on the occasion of the LASER summer school where the members were invited to speak, in September of 2015.

### 2.5 Journal of Object Technology

The Chair created the Journal of Object Technology (JOT) (http://jot.fm) and was for many years its publisher. Before JOT, the Journal of Object-Oriented Programming (JOOP), started when object-oriented programming first became widely known after the first OOPSLA conference in 1987, played a major role in the development of the field. I had participated in JOOP from the beginning, publishing numerous articles and columns and sitting on its editorial board. JOOP was part of a commercial enterprise, sold around 1999 to another company which promised to keep the journal alive but soon dumped it, at just about the time I arrived at ETH. The disappearance of JOOP left a gap which had to be filled; I was able to use the resources of the chair to start JOT. We (particularly Susanne Cech and Bernd Schoeller) put in place editing tools which enabled us to publish the journal online, with a pleasant graphic design and templates enabling the authors to do much of the job; when I advertised for an administrative assistant I specified that half of the position would be devoted to putting together JOT, and indeed both Ruth Bürkli and Claudia Günthart served as editorial assistants for JOT. I was the publisher and Richard Wiener from the University of Colorado was the editor-in-chief, as he had been for JOOP. Since 2002 JOT has served as an important resource for the programming community.

There is a codicil to the JOT story. With the bureaucratization of science that defines the evolution of the academic community in this century, authors increasingly told us that endorsement of publication venues by the Thomson-Reuters Web of Science (WoS) database was essential to their careers. I submitted JOT to WoS, a year-long process that ended with a negative answer, frustratingly since the process is opaque, and inexplicably since JOT was clearly better than many computer science publications accepted by WoS. This rejection was a significant blow to the further growth of the journal. In 2010, an enthusiastic group of colleagues led by Oscar Nierstrasz from the University of Berne took over from us, introducing a new formula and a different reviewing process. The journal continues to enjoy a great reputation, at least among the cognoscenti since it is still lacking WoS recognition. A pioneering feature of JOT is that it has always been a truly "free" and "open" journal; not in the deceptive sense of "golden open access", which is simply a different business model transferring the costs to taxpayers, but free to both authors and readers thanks to clever use of technology. Initially, as noted, the resources of the Chair were needed, but the new team was able to avoid even this small institutional support by streamlining the publishing process further.

## 2.6 Publications

A list of the Chair's **publications** appears at the end of this article. During the Chair's existence I published two single-author **books**: *Touch of Class* (Springer, 2009) [110], the introductory programming textbook resulting from my teaching the Introductory Programming class, and *Agile! The Good, the Hype and the Ugly* (Springer, 2014) [228]. Carlo Furia was one of the authors of a book on *Modeling Time in Computing* (Springer, 2012). Sebastian Nanz, as noted, edited the FOSE book. Manuel Mazzara edited the PAUSE proceedings (this volume). We (in each case one or more of Carlo Furia, Sebastian Nanz, Martin Nordio, Manuel Oriol and I, sometimes with external co-editors) edited some 20 proceedings volumes, most of them with Springer: TOOLS, SEAFOOD, TAP, LASER, VSTTE.

As to **conferences organized by others**, we published in just about every major international conference on software engineering, programming languages, object-oriented programming, robotics software, empirical software engineering and computer science education. Best paper awards include ICGSE as mentioned, ICST (International Conference on Software Engineering), Intelligent Environments, ESEM (Empirical Software Engineering and Measurement).

**Keynote invitations** include ICSE (International Conference on Software Engineering), ICSE Education track, ECOOP (European Conference on Object-Oriented Programming, 2005 and 2015), the software engineering education conferences of Spain, Hungary, Israel, Russia and France, IEEE education conference (CSEE&T, twice), ETAPS, Memocode, CBSE (Component-Based Software Engineering), PSI (Ershov conference) and others. Several summer schools (Marktoberdorf twice, Turku, Lipari etc.).

## 3 Courses taught

Our Chair was responsible for many of the software courses offered by the ETH computer science department; see the close-to-full record at <u>se.ethz.ch/courses/</u>. By far the most work and passion went to the introductory programming course (2004-2015), to which this article devotes a separate discussion (section 4).

### 3.1 Environment constraints

A word about the administrative context, particularly in comparison with US practice. In the period covered, the ETH went from a five-year "*Diplom*" degree to the so-called "Bologna" system of a three-year Bachelor's and a two-year Master's degrees. In most respects this was a theoretical change only, since we hastened to tell the students that only the master's had any value[11]. So the typical US/UK/Australia distinction between undergraduate (bachelor's) and graduate (master's, PhD) courses does not apply. Instead, there are courses, essentially obligatory, in the first two years (known as "Grundstudium" or ground studies), and elective ones in the remaining years. PhD students must fulfill some (light) course requirements but there are no specific PhD courses.

The "move to Bologna", little more than a sleight of hand for our own students, did have a significant effect: at the master's level it led to an influx of students from other institutions, attracted by the reputation of ETH, since the Bologna system encourages so-called "student mobility". As a result the makeup of a typical post-Bologna 4-th or 5-th year class at ETH is highly diverse, in contrast with the homogeneous nature of the mostly Swiss-German student body in the first two years. Diversity makes for fun but the change raised pedagogical challenges, since we could no longer assume anything about the master students' prior knowledge[12].

### 3.2 Software architecture engineering

Back in 2001-2002 there was little object-oriented content in the existing curriculum and, even more surprisingly, no actual software engineering course, the closest being the 2nd-year "*Programming in the Large*" course. I gave some guest lectures on OO programming in Moira Norrie's "*Programming Languages Paradigms*" and was soon asked to take over "*Programming in the Large*", which I turned into a course on software architecture, covering in particular:

- Abstract data types (as the theoretical basis for object-oriented programming, using material from my earlier "OOSC" book [299]).
- Design by Contract.
- Design Patterns, covering most of the classic Gamma et al. patterns, with a strong emphasis on equipping patterns with contracts.
- Object-oriented topics (after a year or two no one could complain any more about insufficient coverage of OO at ETH).
- A course project using Eiffel.

---

[11] I was in favor of a true move to separate degrees, but in practice very few students, even in a discipline characterized by the insatiable thirst of the IT industry for software engineers, stop at the bachelor's.

[12] Another technical point: I refer to "fall-semester" and "spring-semester" courses even though up to 2006 we had different semester dates with a "winter" and a "summer".

- A sprinkling of non-programming software engineering topics such as project management, metrics, testing.

When a reform of the CS curriculum in 2004 led to the replacement of "Programming in the Large" by "**Software Architecture**", the content was ready. I was reluctant about the next transformation, ca. 2009, into "*Software Engineering and Software Architecture*", since I felt the topics required two separate courses; but in practice it was not hard to broaden the scope of the existing course.

### 3.3 Software verification

We introduced several new electives. Since the very beginning I taught a **Software Verification** course, which started out as "Trusted Components" (the new title was used from 2008 on). Software Verification was a survey course, increasingly benefitting from the expertise of the co-lecturers (Till Bay, Manuel Oriol, Carlo Furia, Sebastian Nanz, Chris Poskitt, Đurica Nikolić) in areas such as abstract interpretation, other static analysis methods, model checking and real-time systems. It made it a matter of principle to cover not only static verification techniques but also testing (with the help of Ilinca Ciupa, Andreas Leitner, Yi Wei, Yu Pei and the AutoTest tool they were developing, see section 10.3). It became increasingly exciting in the last years as our own AutoProof proof system (section 10.7) matured and we could use it for ambitious course projects involving the verification, by students, of significant programs.

### 3.4 Concurrency

In the same spirit, the spring-semester **Concepts of Concurrent Computation** course, starting as a seminar in 2005, covered major approaches to concurrent programming and increasingly used our own SCOOP development for the project. I initially set up the course with Piotr Nienaltowski; then Sebastian Nanz took on a key role as co-lecturer, joined in the last years by Chris Poskitt. The predictably tentative nature of the first SCOOP implementations made the project somewhat of a challenge for the first generations of students; with the impulse of the CME project the later versions made advanced concurrent programming projects possible.

### 3.5 DOSE

As explained in section 6, we developed an increasing interest in issues of offshore, outsourced, and more generally distributed software development; "distributed" in the sense of projects whose developers are spread over several locations, often on different continents. The fall-semester course that finally became the Distributed Software Engineering Laboratory started in 2004. The succession of names reflects the broadening of its scope: "Offshoring and Software Engineering", then "Software engineering for outsourced and offshore development" from 2005, then "Distributed and Outsourced Software Engineering" (DOSE) from 2008 to 2012. The course covered generic software engineering topics − as noted, there was no standard software engineering course −  but focused on the issues of distributed development. We had to improvise much of the content since there was little literature on the topic, and no textbook. Peter Kolb, with his industry experience and particularly his practice of CMMI, was co-lecturer right from the start.

In the 2007 session, we decided to practice what we were teaching and turn the project into a distributed development effort between several student teams (in most years, three) from different universities. The number of universities involved in any given year reached 12, with some 20 different institutions over the years. Some of the mainstays were Politecnico di Milano (Elisabetta di Nitto, Raffaela Mirandola, Giordano Tamburrelli, Carlo Ghezzi), which quickly became co-leader with us; University of Rio Cuarto, Argentina (Nazareno Aguirre); Polytechnic University of Madrid (Natalia Juristo); PUCRS in Brazil (Rafael Prikladnicki); University of Nijny Novgorod (Victor Gergel); and other universities in Switzerland, Russia, China, Korea, Vietnam and India. As one may guess, the logistics was a challenge; Martin Nordio was the one who made the whole thing work, assisted in various years by Roman Mitin (then a PhD student of Jürg Gutknecht), Julian Tschannen, and Christian Estler whose Codeboard development (section 5) was closely connected with the course.

DOSE provided many moments of excitement and many of anguish, and, as a living experiment in distributed development, material for a whole set of joint papers. Three typical anecdotes:

- We carefully tracked commits by the various groups to the development servers. Seldom have curves be more flat in ordinary times, and more stratospheric in the hours, minutes and seconds before the strictly enforced deadlines.
- The teaching team took care of dividing the work between the three teams of each group: typically, user interface, business logic (e.g. game logic since the projects often involved games) and database. The first time, we encouraged them to use Design by Contract mechanisms. Encouragement does little, so some groups ended up with date modules that did not check for invalid dates such as 31 September (that's the job of the clients, of course), and client modules that assumed the dates they were getting would be meaningful (that's the job of the date module, of course). We learned our lesson: in subsequent years the use of contracts to specify and protect APIs was no longer optional.
- Those of us who thought they had seen all possible student excuses for missed deadlines were in for a bit of cultural diversity, as in "*last night my Internet connection was down because the cable was eaten by a drunken bear*".

Each session of the course went with its own drama, with the initial euphoria of students meeting new friends from the other end of the world, followed by despair and noises of dropping out at the time of the first code deliveries, and a generally happy resolution when they finally were able to deliver a result.

3Many former students have told us how useful a preparation the course was for the challenges of software development in companies.

### 3.6 EiffelStudio projects: open-sourcing 3 million lines of code

Eiffel Software's EiffelStudio environment was initially closed-source. We decided to make it open-source in 2006, and in a lecture of the 2006 Software Architecture course I clicked the "big red button" on a slide:

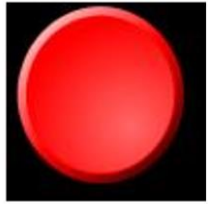

revealing the password for the Subversion repository. Much of the information about Eiffel is nowadays at eiffel.org [280]. Along with Emmanuel Stapf from Eiffel Software, Till Bay had been arguing for that change, and I took up his suggestion of letting ETH students loose into the 3 million lines of EiffelStudio code.

From that semester on, students started (initially as part of the Software Architecture course project) making interesting contributions, proving that with strong information hiding, contracts and other Eiffel properties it is possible for a newcomer to dive into a complex software system and zoom in on the relevant parts.

Every semester (not just every year) since Fall 2006, we ran the "Software Engineering Laboratory: Open-Source EiffelStudio" project course, in which students contributed extensions, often taken over for improvement by other students in later semesters. Some important components of today's EiffelStudio, such as the "Eiffel Inspector", an extensive style checker and static analyzer, arose from such projects.

### 3.7 Projects

The EiffelStudio projects were just a part of numerous student projects at all levels. All PhD students and other researchers took part; a particular mention is due to Till Bay, who started the EiffelMedia project and had several dozen students contributing pieces (sound, video, advanced user interfaces), on which several of our courses relied. Michela Pedroni was another large provider of projects in connection with TrucStudio (section 5) and other pedagogical initiatives.

In the early years the semester dates allowed us to give optional student projects over the break; many students, including first-semester students, rose to the challenge and produced impressive games using EiffelMedia. I was always envious of the impressive student presentations in Mechanical Engineering, showcasing robots and other stunning devices; now we had our opportunity to make our students' work visible too and I booked the central hall of the ETH:

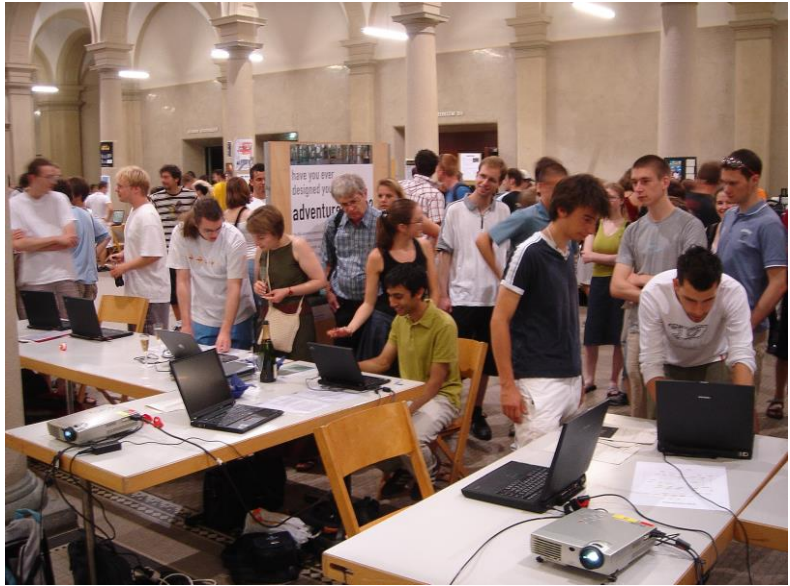

Today using game development for teaching is common in universities including ETH (the CS department has a full program of game programming), but I believe we were the first to ask students to produce games on a large scale. Some of the games they developed were particularly impressive and sophisticated.

### 3.8 Language courses

Because I used Eiffel in my programming courses, I felt I had a responsibility to provide language-focused courses on other languages. I would have liked to offer a full "languages in depth" series covering all major languages, but that was too ambitious. C++ was addressed by a course on generics taught by Eugene Zouev (in Jürg Gutknecht's group). We set up a Java course and a C# course, soon merged into "Java and C# in depth", and taught by a team that included Manuel Oriol, Carlo Furia and Marco Piccioni.

I also offered an "Eiffel in depth" course focused on the language itself, as opposed to the courses using Eiffel just as a tool to introduce general programming and design concepts.

### 3.9 The Robotics lab

A course that generated a level of challenge and excitement reminiscent of DOSE (or of Introduction to Programming as discussed next) was the Robotics Programming Laboratory (RPL). We taught it three times, in the Fall semesters 2013-2015, under the leadership of Jiwon Shin and with the extensive involvement of Andrey Rusakov as assistant, in connection with the development of our robotics software activities (section 12). Everyone in universities nowadays praises interdiscipli-narity: do away with discipline and department boundaries! RPL was an interdisciplinary course, and showed that the practice is not so easy. It was open to students from computer science, mechanical engineering and electrical engineering (with actual proportions of about 40%, 40% and 20%), and we had to contend with different regulations in each department. (In the end we gave up trying to define a single scheme, letting each department, for example, decide how many units the course was worth for its own students — for the same amount of work! — based on its own practice.) The pedagogical challenges were interesting: computer science students had to be taught how to deal with real, physical hardware, and find out, for example, that a robot that you sent to point [x, y] is not necessarily, even in the absence of bugs, *exactly* at [x, y]. In their mindset, either there is a bug or the robot must be right where it is supposed to be. To mechanical engineering students, we had to teach the essentials of programming. The course as a whole is a combination of programming techniques, software engineering and architecture (particularly design patterns), basics of robotic hardware, and robotics algorithms. We found few equivalents when looking at other institutions: there are lots of robotics courses, but not many focused on the software aspects.

Like DOSE, RPL is project-based. The project consists of programming successively harder tasks for a Thymio robot, an EPFL design intended as an educational toy. Since the limited Thymio sensors does not suffice for the project's advanced tasks such as obstacle avoidance, Rusakov concocted some hand-attached additions:

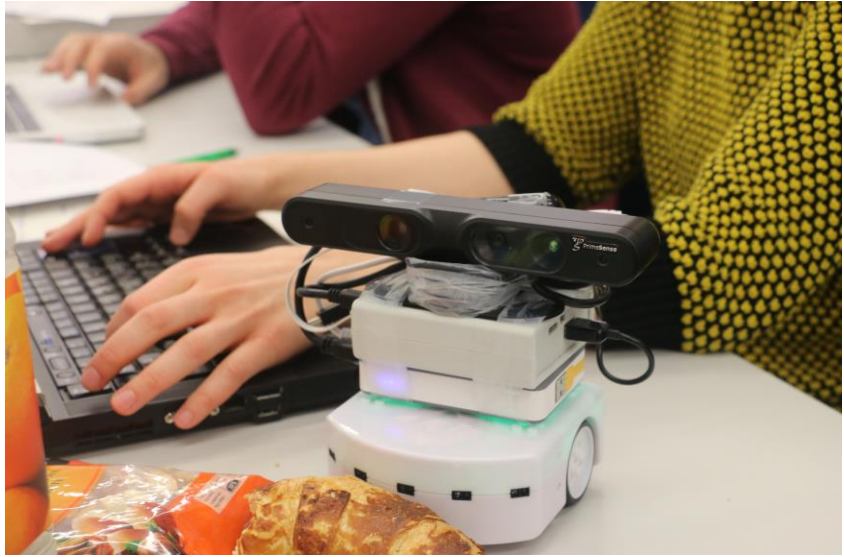

Software development in the course uses the Roboscoop framework that we developed (section 12) and a mix of other Eiffel and C++ tools.

RPL is a demanding course. Since the number of seats was limited to 16 and we had a waiting list, we tried to spot early those students who would later give up, so as to free positions for others. The first session was cumbersome for students as well as instructors because when leaving the lecture room they had to unmount and remove the whole setup. In the next two years we managed to secure a suitable room just for us; the atmosphere of the course improved considerably, with the room becoming a mini-lab, equipped with a coffee machine and such, where students could work at any time of day or night.

The students had a hard time, but increasingly succeeded in solving the advanced tasks of the project, and those who completed the course told us they learned a lot. Several went on to a robotics job. We did not just rely on such informal feedback but, as with other courses, published an article about the pedagogical experience [226].

### 3.10 Other courses

Among other courses (the list given here is not exhaustive, see the more extensive record on the course page cited), we ran multiple times, initially with my colleague Daniel Kroening (now at Oxford) and then just by ourselves, the "Software Engineering Seminar", required of all master's students in the software engineering track, in which they study and present a research paper from the recent literature.

### 3.11 Teaching and research

A general observation about our courses is that (in line with the Humboldt idea of a research university) we always thought of research and education together. We were fortunate that all the courses we taught (I cannot think of any exception!) were courses we *wanted* to teach. We constantly included elements of our research into our courses, including, as will be noted, in the very first semester; the other way around, we used the observation of students' programming in our courses to inform our research, publishing (more, I think, than other groups in the department) in CS education conferences and journals. The DOSE course alone yielded numerous research papers [78] [105] [124] [148] [157] [171] [198] [222] [225], many of them with colleagues from other institutions collaborating in the project, as described earlier.

Such use of student observation for research requires, of course, great care. There are ethical issues involved: students are not guinea pigs; the purpose of a course is not to help the professor's research but to educate students. Never should the first goal damage the second one. For example, it's OK to study how well students perform under method A or method B, but you cannot ask one half of the class to use A, the other B, and look at the results: if one is indeed better, you are putting the other group at a disadvantage! Similarly, if you are testing the students, the test should have a pedagogical value for *them*: it cannot be just for your own enlightening. But again it can be for both. An example trick we devised was to have a group learn A then B, with tests (interesting for them) in-between and at the end, and the other group do the same with B then A. If each step is short, say an hour of instruction, we reasoned that any pedagogical harm produced by following the less ideal of these two orders (assuming one is indeed better) would be minuscule.

An example of such a situation occurred when we tried to find out if our SCOOP method of concurrency (section 11) was, as claimed, "easy to learn". We were introducing some concurrency topics into the Software Architecture course, which provided an ideal testbed to assess the ease of learning of SCOOP versus Java Threads, the other mechanism we were teaching. But we had to be doubly cautious, not only because of the ethical and pedagogical considerations just mentioned but also to guard against our own bias, since we could obviously be suspected of *wanting* SCOOP to be easier to learn. Sebastian Nanz, Michela Pedroni and I thought very hard about how to devise a study that would guard against this risk. The basic idea was to put all the possible biases *against* us. For example, students were new to SCOOP but many of them already had had some Java Threads exposure, which went in the right direction. We made sure that the Java concepts were described in a clear tutorial and so on. We involved, for a control study, Faraz Torshizi from York University in Canada, where students, unlike ours, had only limited Eiffel experience. With the deck stacked against it, SCOOP still won by a small but significant margin. We were particularly pleased that the resulting paper [152], which explained in detail the methodological challenge and how we addressed it, won the best paper award at the Empirical Software Engineering and Measurement symposium (ESEM 2011) and was extended into a journal article [192].

Course-based studies such as those leading to the DOSE and SCOOP learnability papers were a major step in a significant evolution of our group's research, worth mentioning here even though the present section is about teaching (research proper starts with section 5). Some software engineering research is concept-driven; that had largely been my background. In the past two decades, more and more software engineering work has been empirical, based on the quantitative study of software artifacts and processes using methods (in particular statistical methods) from the natural sciences. Starting with Manuel Oriol, our great postdocs, plus collaborators from other groups such as Alexander Pretschner, had extensive experience in the field and drew us to include ever more empirical backing in our work. A sequential reading of the bibliography (which is in approximate chronological order) shows this growing role, over the years, of empirical elements in work that originally was mostly concept-driven.

## 4 Teaching introductory programming

Introduction to Programming" is a first-semester course, compulsory for all entering computer science students, and part of the first-year exam which determines whether students will be allowed to continue. Not long after I started at ETH, the department head asked me if I would take over the course and teach it using Eiffel. I did not expect that request, for three reasons:

- ETH rules require the use of German in first- year teaching (with progressive introduction of English in the second year and its sole use at the master's level). Since I did not have a level of German allowing me to teach, I had considered such courses beyond my realm. (Prior to joining ETH, I had participated in a conference panel with the future colleague then in charge of the course, who joked that it was safe from me for that reason.)
- Coming from industry, I had little recent experience teaching large undergraduate courses.

- I did not intend to tread into the respected ETH tradition of using Wirth-designed languages: Pascal in the seventies, then Modula-2, then Oberon.

I did not hesitate long, however, particularly after a courtesy check with the holders of that tradition confirmed I would not cause offense provided I waited a year to let the current holder teach the course one more time. The one-year respite was of great benefit. On the matter of German, more below (4.2).

The design and delivery of the course mobilized considerable energy within the Chair; most PhD students and postdocs served at some time or other as teaching assistant, and the teaching effort influenced several of our research developments. We published extensively on the corresponding pedagogical issues, particularly in ACM SIGCSE and ITiCSE and IEEE CSEE&T, the main venues for computer science education. The course resulted in a textbook, *Touch of Class* [110], which other universities have also adopted.

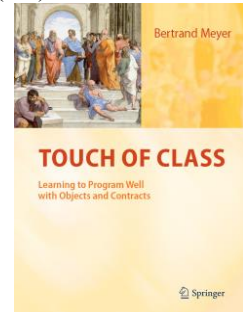

### 4.1 Background

My interest in the topic of how best to teach programming predated my ETH assignment. Years earlier, barely out of school myself, I was put in charge of a university-level introductory programming course, using Algol W and some dose of object-orientation, and published an article about our pedagogical approach [285]. More recently, I had had many discussions with academic colleagues using Eiffel in their courses, such as Jonathan Ostroff at York University in Canada and Christine Mingins at Monash University in Australia[13]. I had myself devoted considerable thought to the use of OO techniques and Design by Contract in introductory programming, developing the concept of "inverted curriculum", a name taken from a comparable approach to the teaching of hardware design, by Bernie Cohen of London's City University. My earliest papers on the topic [295] were published in 1993, and very much described the approach that would be applied at ETH, but I had written them from the comfortable position of someone in industry who can safely tell colleagues in academia what to do, without having to suffer from the results. (Usually, it is the other way around.) Now I had to apply my own precepts.

I also had the opportunity to state these precepts. Sometime during my not-quite-standard hiring process at ETH[14], someone must have panicked and asked whether with my limited professorial background I had any concept of teaching. I was asked to produce a "teaching statement". Based on the principles that (1) if you have a chore to perform you might just as well do the best job you can and (2) even if the result is intended for a small local audience, in this case the department management, you might just as well be generous and let the whole world benefit from it, I turned that pensum into an article for IEEE Computer, "*Software Engineering in the Academy*", my first publication as a member of ETH and the first entry of this article's bibliography [1].

---

[13] The *Touch of Class* textbook cited below has a fuller list of acknowledgments. See the end of the "Preface for Instructors" at touch.ethz.ch.

[14] ETH professors are selected by the president. There is a committee, but its role is advisory and the president can short-circuit it, as happened in my case. It is clear from my later experience as a member of numerous hiring committees that if I had been subjected to the standard process myself I would almost certainly have failed it. I knew (at least in some way) how to do teaching and research, the main duties of a professor, but I was not a member of the caste and did not know the codes. An academic job interview is a ritual; everyone in the business knows how to answer the ritual questions, but I did not and it is likely that I would have given a disqualifying answer at some point. It is for the reader to judge whether the non-bureaucratic, highly personal ETH hiring process for professors was, in my case, a benefit or a bane.

I taught the "Introduction to Programming" course without interruption in the fall semesters of 2003 to 2015, to some 5000 students altogether. ETH has good resources and organization, which enabled us to develop a strong infrastructure, in particular a dedicated group of assistants for teaching the lab sessions, made in part of PhD students of the Chair, who from year to year accumulated experience and training materials so that it was easy to accommodate newcomers. A weekly team meeting made it possible to check the progress of lectures and lab sessions and take the pulse of the students through the teaching assistants. More direct contact is also useful; in the later years I took to the habit of systematically visiting lab sessions myself. First-year compulsory courses are, as noted, under particular scrutiny because the stakes for the students. They require a well-oiled machinery, constant attention to mishaps that could lead to crises, and a quick reaction when they arise.

## 4.2 An expensive way to learn a foreign language15

As to the language of instruction, what German I had consisted mostly of my effort as a 12-year-old to learn enough German (and Italian) vocabulary to get the gist of Mozart's operas and Schubert's Lieder. To ask for directions in the streets of Zurich, I could use16

> *Ist das denn meine Strasse?*
> *O Bächlein, sprich, wohin?*
> *Du hast mit deinem Rauschen*
> *Mir ganz berauscht den Sinn*.

which (on top of being less effective than GPS) was not sufficient preparation for teaching programming. It was not hard to use the first minute of the first lecture of the semester to convince students, who at ETH in computer science have a good level of English[17], to let me continue in that language[18]. For seven years in a row I taught in English; no student ever complained, but in 2008 someone found out[19] and I was politely asked to cease flouting the rule. One of the techniques for learning a language is to stand in front of three hundred native speakers and teach them something you know and they don't in a language they know and you don't. It is expensive (in my case, graciously funded by Swiss taxpayers), but it works. Little by little I reached a level where the proportion of end-of-semester student evaluations consisting solely of "*let the poor guy use English!*" decreased from about 90% to something more reasonable, leaving place for comments about the actual course contents. Even after I became able to deliver the course contents in an acceptable way, the switch to German[20] was still affecting the pedagogical quality of the course: my delivery was less spontaneous, and there was less interaction[21]. In the preceding years, I had constantly increased the interactivity of the class: with the stunning amount of teaching materials we made available to our students — copies of the PowerPoint slides, video recordings of the lectures from the current year and previous ones, lab session slides, tutorial notes, sample questions from previous years' exams, homework texts and solutions, then the textbook *Touch of Class*, published in 2009 but available to students in draft while I was writing it — I could afford occasionally to depart from the official material and improvise discussion sessions. Long before we heard the buzzword "Flip the Classroom", we had the practice, a couple of times in the semester, of holding what we called a "Socratic lecture": I would make the corresponding slides available in advance and announce that next time I would not cover them in the usual way, just come and wait for questions. The trick in such sessions is not to flinch: you must have the guts to stand there and wait for someone to speak up, even if nothing happens for fifteen minutes. That was the case the first time around (this is not usual ETH teaching style, and students felt awkward), then someone raised his hand and afterwards I could hardly stop the flux of questions. It seems to have been part of a general evolution of students' attitude that there was more and more spontaneity as years went by[22]. In the German version of the course, things were more awkward. The students' nervousness may have been gone, but I had become the nervous one, fearful of misunderstanding a question or just not understanding it.

---

15 The reader may want to skip this sub-section since it pertains to my personal circumstances rather than to principles of software engineering education. I am including it (with details confined to footnotes) because of its role in my application of these principles.

16 *Is this, then, my road? O little stream, pray tell me, where to? Your rustling has enraptured me.* Text by Wilhelm Müller from Schubert's *Die Schöne Müllerin*.

---

[17] People sometimes ask why I could not resort to French, one of Switzerland's official languages. Although ETH used to be the only federal university and had French-speaking chairs, any such excuse went away with the creation of its sister institution EPFL in Lausanne in 1968. Even a suggestion to use Italian is only good for cocktail-party chatting. Contracts for new non-German-speaking ETH professors stipulate that they must learn the language within a year and can obtain financial support for that purpose.

[18] The department's studies director wrote a formal letter to the Rector requesting for me a special dispensation of the first-year German-only rule. No answer came. Surprised (ETH administration is usually punctilious), he raised the matter on meeting the Rector at some university function a few months later. The answer was that no answer would be forthcoming, since a rule is a rule and the answer could only be no, which would wreck the already scheduled course. As long as I could find an arrangement with the students and the Rector did not get any complaints, he had not received the letter.

[19] Career advice for beginning academics: *close the door*. It seems that one day I left the auditorium door open and an administrator walked by, heard me, and raised a rumpus. The atmosphere had changed by then; ETH management was experiencing more pressure from politicians to use German. It is possible that without this chance discovery I could have continued in English.

[20] The reference here is to standard ("high") German (*Hochdeutsch*), not "Swiss German" (*Schwiizertüütsch*), a separate language. Swiss-German speakers learn German at school, and it serves as the written language, Swiss German having no standardized written form.

[21] While you can learn to speak a foreign language through sheer hard work, *understanding* native speakers is a different challenge, compounded in the Swiss case by the unique variety of local variants, including those of foreign students from various parts of Germany and Austria. Understanding was always the harder part for me.

### 4.3 The right programming language

Unlike the language of delivery, the programming language was never a problem. While countless educators have used Eiffel, others have told me over the years that they would really like to but cannot because management, or students, or their families (the supposed culprit varies) would never accept it[23]. In most cases this supposed hostility of others is an excuse. Some ETH student evaluations criticized the choice of language, but they were no more numerous than what one would expect had the course used Java instead of C++ or C++ instead of Python. I always took care of explaining the pedagogical reasons behind the choice and pointing out that learning one more language, whether or not it ends up as your favorite, is a benefit, not a burden. Plenty of courses in the ETH curriculum give the students opportunities to learn Java (which many of them encounter *before* joining the program), C++, C, Javascript and other languages in wide use. In practice many students enjoyed

---

[22] In end-of-semester evaluations the question about the merit of the Socratic lectures generally received a negative assessment. We can only surmise why; maybe a feeling that somehow the professor was not doing what is expected of a professor — profess — and instead unfairly asking the students to work. But after the initial awkwardness the sessions were in fact very lively, full of good questions that might not otherwise have received an answer, and in talking to students we felt the sessions were productive, even if they refused to acknowledge it in their evaluations. One of the benefits of tenure is that you do not have to kowtow to students for fear of negative evaluations. Instructor assessment is necessary, in particular to detect truly bad teaching, but should not discourage pedagogical innovation. After all, most adults can remember a teacher whose lessons turned out to be important in the long run but whom they resented at the time. The ETH policy on assessments is commendable: assessments are made public to the department members without comment; they cause no further action, except in the case of a grade of 2 or less (on a scale of 1 to 5), which requires a letter from the department head to the rector explaining the reasons for the bad assessment and describing measures being taken, in conjunction with the lecturer, to avoid a repetition of the problem.

[23] An excuse I heard more recently is that the *Touch of Class* textbook was too heavy for students to carry around. Indeed the first printing was big (Springer soon corrected the problem), but let's be serious.

Eiffel and came to us in later years to ask for Eiffel-related topics for bachelor's and master's theses or other projects. Most of the student contributions mentioned earlier started that way.

One benefit of using Eiffel was to talk about the language as little as possible. With Java you have to start by explaining "static void main" and other esoterica. You end up teaching a Java course or, in other cases, a Python course or a C course. I did not want to teach an Eiffel course, but to deliver on the specification of the course: *Introduction to Programming*. Eiffel enabled us to focus on the concepts.

There is, by the way, a benefit of using a non-majority programming language for teaching: this solution removes an increasingly frequent complaint of instructors, "Google programming". With Java or another dominant language, a natural reaction of many students when given a programming exercise is to search for a ready-made solution, which they often find on the Web. While professional programmers also resort to this technique, it is not appropriate for students learning the basics. The amount of Eiffel code on the Web is considerable but nowhere close to the amount of Java, C++, Ruby, C# or Python code.

### 4.4 Concepts and skills

A central question in teaching programming today, already discussed at some length in the "Software Engineering in the Academy" article [1], is whether to focus on concepts or skills. Teach only computer science concepts, and you produce people who are not ready for the practical software engineering tasks that industry needs filled. Teach only skills, and you produce technicians that do not have the big picture necessary for a successful career.

The easy answer that we should teach both is not enough: education is a fixed-pie business in which the total number of coursework hours is set. One has to make choices.

I hit an example of the need to teach more than skills when giving some lectures in 2004 at a renowned Chinese university, which was proud that its software engineering students were at the top. I devised a small questionnaire to find out for myself, and have used it often since. One of the questions, made up almost as a joke (but then the joke was on me) was:

*Your boss gives you the source code of a C compiler and asks you to adapt it so that it will also find out if the program being compiled will not run forever (i.e. it will terminate its execution). Check one:*
1. *Yes, I can, it's straightforward.*
2. *It i hard, but doable.*
3. *It is not feasible for C, but is feasible for Java.*
4. *It is not feasible for C, but is feasible for Eiffel.*
5. *It cannot be done for any realistic programming language*

It is touching to see how many enthusiastic students select one of the first four options (#4 undoubtedly from an unconscious hope that the instructor must be on to something). Now one may argue that ordinary software development does not require dealing with undecidability, but it does not sound right that a competent software engineer should have no idea of the issue. Particularly one coming out of a good university. Electricians may not know Maxwell's equations; but electrical engineers should.

Skills are critical too. Students do not have to know all the technologies of the moment; in fact they cannot, since they are too many of them, and they come and go anyway (ASP.NET yesterday, Angular JS today, something else next year). But they have to know some, both for their relevance to industry and to make sure software concepts do not remain theoretical, as they would without an actual implementation using the tools of the moment.

The philosophy we adopted was *skills supporting concepts*. Teach skills so that students get a good grip on the practice (and program a lot), and use these skills to drill the fundamental concepts of programming − [1] names some 20 of them, from abstraction and information hiding to recursion and invariant-based reasoning − into their heads.

The skills part is not to be neglected. In recent years I have become involved in a new and successful effort, Propulsion Academy ([propulsionacademy.com](propulsionacademy.com)),which complements university efforts, often in collaboration with the universities themselves, to offer intensive programming education through 3-month programming bootcamps on topics such as Full-Stack Development and Data Science. Many of the students have university degrees already, including sometimes PhDs, and want to move to IT. Industry gobbles them as quickly as we graduate them.

I did, by the way, as a sanity check, gave the termination-smart-compiler quiz to second-year ETH students, having undergone "Introduction to Programming" the previous year, and everyone gave the right answer.

## 4.5 How to teach programming

Other articles including [20], [23], [50] and the "instructor's preface" of the *Touch of Class* textbook ([110], available at [touch.ethz.ch](touch.ethz.ch)) have detailed the rationale behind our Introductory Programming course and the pedagogical issues and principles. Here are the principal ideas:

- Use of object-oriented concepts right from the beginning and throughout, including classes, inheritance (single and multiple), genericity, information hiding, polymorphism, dynamic binding.
- Emphasis on system skills, not just low-level programming.
- Balance between concepts and skills as discussed above.
- More generally, emphasis on concepts and skills that support lifelong career success. A concern that was particularly strong in the mid-2000s, leading for a few years to a decrease in computer science enrollments in Switzerland, was the competition from outsourcing to low-labor-cost countries (see also section 6). While outsourcing has never threatened programming as a profession in the West − still today, there are some 1 million unfilled IT positions in Europe − its existence clearly means that an ambitious university should teach not just low-level programming techniques (for which there will always be a cheaper programmer somewhere) but engineering skills of lasting value.

- Use of Design by Contract throughout. While using a completely formal approach in an intro course goes too far (see the Lamport citation in [110]), teaching students to document the semantics of their code through preconditions, postconditions and class invariants, and use these elements for effective testing (in the future, correctness proofs, see 10.7 and 10.8).
- More specifically, constant emphasis on quality, through application of strict design principles from *Touch of Class* and the earlier *Object-Oriented Software Construction* [292] [299]: Information Hiding, Uniform Access, Open-Closed, Single Choice, Command-Query Separation and others.
- Large coverage, including of topics often considered to fall beyond the scope of an introductory course, such as event-driven (publish-subscribe) programming, higher-level functionals (agents/closures), some design patterns (undo-redo, Observer0, elementary rules of programming with floating-point numbers, an introduction to lambda calculus and an introduction to non-programming aspects of software engineering such as requirements.
- Use of a specifically designed library of components, the Traffic library (graphical simulation of traffic in a city).
- In the later years, systematic use of a MOOC (online version of the course) and of advanced cloud-based tools allowing students to perform exercises and teaching assistants to evaluate them directly in a Web browser (see the next sections).
- "Outside-in" approach, also known as Inverted Curriculum.

The idea behind "outside-in" [295] [50] is to let students work, right from the beginning, with sophisticated software elements (the Traffic library played this role). The traditional solution is to learn from the ground up, starting with small programming exercises, which in the current state of technology (where everyone, even small children, uses computers to perform advanced tasks) are boring. With the outside-in approach, the students can rely on the given software components to produce interesting applications from day one. A pedagogical benefit is that to use the components they have to go through the official programming interface (API); they learn the value of abstraction, information hiding and contracts through practice, rather than just through theoretical exhortations. The approach assumes components designed and implemented to high quality standards; the use of contracts, as supported by Eiffel, is essential.

"Outside-in" also implies that students get exposed, early on, to lots of code. Any successful path to learning programming involves imitation of existing models. The components given to the students will play this role. The Traffic library was written very carefully, with systematic use of all the quality-enhancing techniques that we teach through the course. It provides a huge repository of models of good design, good APIs, and good implementation.

"Outside-in" also addresses an issue that vexes anyone teaching introductory programming today: the wide variety of initial student experiences. From systematic studies performed on all students over many years, we found (in a study with Michela Pedroni and Manuel Oriol [116]) the distribution shown on the right, with 18% of student having done no prior programming, and at the other end of the spectrum 10% having written an object-oriented program of over 100 classes. How do you cater to such a diverse audience? You can tune the course down to the novices and bore the experienced students, or align it with the experts and lose the beginners, even though some may have other skills (particularly in mathematics) that qualify them for successful computer science studies. Working from components helps solve the problem. The novices can proceed step by step, "outside in" in the sense of using the components first strictly as consumers, through their APIs, then through a process of "progressive opening of the black boxes" discover how they are written. Hence "outside-in". The more experienced students are welcome to go into the source of the components from the beginning. They are also encouraged to extend and adapt them. There is, then, something for everyone.

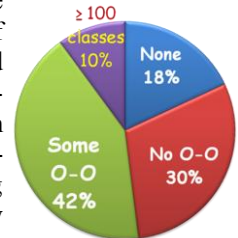

The difficulty of assessing approaches to teaching programming is that the effect is long-term: one would need an extensive study measuring the academic and industry performance of students in the years after the course. Although we performed constant evaluations of the course and the students, no such long-term study is available for our approach, or anyone else's. I believe, however, that the course was highly successful, teaching thirteen successive classes of ETH computer science students the skills necessary to understand programming in depth and become effective software engineers.

## 5 Pedagogical tools and MOOCS

Modern technology does not replace face-to-face teaching but can enhance the quality of the pedagogical experience. A combination of factors made it almost inevitable that we would invest in educational tools: we were a software engineering group with a strong focus on practical programming; we were in charge of many important courses as noted above, and the particular challenge of teaching introductory programming; MOOCS (Massive Open Online Courses) came of age during our time; I had always maintained a keen interest in issues of teaching programming; and so did several group members, particularly Michela Pedroni and Till Bay at the beginning and Marco Piccioni, Christian Estler and Martin Nordio later on (as well as Marie-Hélène Ng Cheong Vee/Nienaltowski as a visitor), who all pushed for developments in this area. The results are of three kinds, reviewed next in sequence:

- Concepts (Trucs) and tools (Trucstudio) for course design.
- MOOCs, home-made and at edX (Introduction to Programming, "Computing: Art, Magic, Science" 1 and 2, agile development).
- Tools for on-the-cloud programming (Eiffel4Mooc, Codeboard, plus CloudStudio discussed in section 6).

## 5.1 Trucs and Trucstudio

The first effort stemmed from an attempt to think about the whole domain of teaching a course with the benefit of object-oriented ideas, which provide a powerful modeling technique applicable to many areas. In January 2005 I introduced (in a paper published in 2006 by IEEE *Computer* [52]) the notion of Testable, Reusable Unit of Cognition, or *Truc*, arguing that any course or learning objective can be described as a combination of Trucs, with clear relations between them, such as usage and specialization, similar to the client and inheritance relation of object-oriented programming.

At that time, Michela Pedroni was already a member of our group; she had approached me two years earlier, while still a Diplom (master's) student, about the possibility of doing her master's thesis in the area of programming education. Her particular interest, after going through the ETH computer science curriculum, was to make the teaching of programming, which as a recipient she had found less than ideal. She went on to enroll in the PhD program and became a key contributor to the Introduction to Programming course. The notion of Truc provided a good basis for a PhD topic; she refined the concept much further and built a comprehensive environment, TrucStudio.

TrucStudio [80] [120] looks at first sight similar to a modern IDE (Interactive Development Environment) as used by programmers, such as EiffelStudio or Eclipse, but with building blocks that instead of classes, routines and other programming artifacts are courses, lectures, Trucs and other pedagogical artifacts. In the screenshot on the right, they describe the pedagogical concepts involved in teaching certain aspects of object-oriented programming (although of course there is nothing specific to programming in TrucStudio, and the example could be about the teaching of Greek or of bird migration patterns).

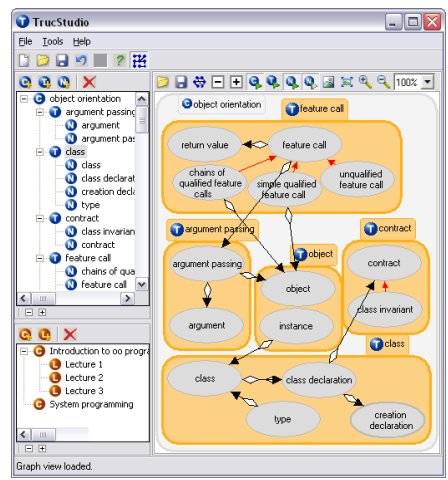

Trucstudio makes it possible in a convenient and intuitive way to create and modify courses, textbooks and other educational ventures. The tool that came out of the PhD was convincing and served to design several example applications. It attracted attention in both academic and industrial contexts internationally. The project did not continue because we failed to obtain further financing. (In the local context, educational units typically do not have the resources to fund research[24], and research-funding bodies told us that they do not fund educational projects. We were of course asking for research funding, but that did not work because the work came from a computer science department and was viewed as education research, not CS research.) Trucs and TrucStudio are, in my view, a significant advance and are there waiting for someone to pick up the concepts, the research and possibly (at Source-Forge) the tool.

---

[24] The ETH unit in charge of teaching, the Rectorate, does have an interesting scheme for funding educational project, whose name is now Innovedum (www.ethz.ch/en/die-eth-zuerich/lehre/innovedum.html), from which we did benefit for other projects. But the amounts involved only make it possible to fund some equipment, student help or other expenses at those levels, not a PhD student or postdoc over one or more years.

## 5.2 MOOCs

The next major teaching-related move was about MOOCs. Marco Piccioni in particular kept alerting me, from 2001 on, to the MOOC phenomenon and its rapid rise worldwide. If we really thought we had some of the best software courses in the world, we should make them known to the world. He was persuasive, and I came to agree. We were not bright-eyed believers in MOOCs as the final recipe for all education; we simply understood that they were a major development in pedagogy, offering extraordinary possibilities, and that we should get into it.

The institutional context was not encouraging. ETH does teaching and research. In research the reach is global, having enabled the university to reach the very top international stature[25], an extraordinary feat for a university in such a small country[26]. In teaching, the focus for the *Grundstudium* (early years of the curriculum) is local: train Switzerland's technical elites. MOOCs are by nature global, reaching out to anyone anywhere who wants to learn. For a typical US university, which wants to recruit the smartest (and fee-paying) students from all over, they are an excellent branding opportunity, showcasing their best teachers. For a government-funded institution with negligible tuition fees, getting thousands new undergraduate applications from abroad is not part of policy goals[27]. These considerations did not prevent the sister and rival institution, EPFL in Lausanne, by nature more adventurous, from jumping right into MOOCs and producing dozens of them in French and English, in particular with the Coursera MOOC company (a venture created by Stanford professors). ETH reacted differently, promoting courses open to its students only. The acronym, which did not trigger our enthusiasm, was TORQUE ("Tiny, Open-with-Restrictions courses focused on QUality and Effectiveness" [28]); we thought that Tiny as the first word and Restrictions as the fourth did not exactly reflect the institution's relentlessly self-advertised passion for excellence, and the natural echo of "Torquemada", with its hint of torturing students, did not exactly reflect our pedagogical ideals. (For such internally-oriented online courses, the education world at large uses a less gaudy acronym, SPOC, for Small Private Online Course, which the paper [224] reporting on our first experience uses in its title.)

---

[25] University rankings became an international obsession around 2003 with the first "Shanghai" ranking. They are controversial, but no one would deny that they have at least some value; they have become an inescapable factor in academic policy. ETH, both across disciplines and specifically for computer science, usually comes around number 10 worldwide, behind top US institutions such as Stanford, Berkeley, MIT and Carnegie-Mellon, and just behind Oxford and Cambridge. It is systematically number 1 for continental Europe.

[26] A superficial analysis would suggest that the budget for ETH (currently $1.7 billion, about 2.5% of Switzerland's federal budget) is outrageously high. In reality, ETH has been since its creation in 1855 one of the key engines for the transformation of Switzerland from one of the poorest countries in Europe to the richest non-natural-resource-based economy in the world.

[27] Studying at ETH is such an outstanding proposition, especially from a financial perspective as compared to top US, UK and Australian universities, that even with high entry standards for foreigners the university would be flooded with applications were it not for German-language requirements.

As late as the very last days of 2014 (after we had already produced our first homemade MOOC, presented next) many people at ETH were taken aback when reading a major interview of our brand new university president, promoted from the position of Rector where he was in charge of teaching policy. The journalist from the NZZ (the main German-language Swiss newspaper) grilled him on MOOCs, saying things like "Harvard, Yale, Stanford and others have been active in MOOCs in the past few years, hasn't ETH been sleeping during all that time?". He only got answers such as "*The idea that one can read a text or look at a Web page and understand is an illusion*". All shook their heads in disbelief.

From the beginning we thought that the MOOC (or rather no-MOOC) policy was unsustainable and that management would come around. In the meantime we were not going to wait patiently. In early 2013 Piccioni and I decided to produce a MOOC from the Introduction to Programming course, as a skunkworks project. (In most institutions I know, a negative decision from management would have been the end of the story. It is testament to the extraordinary ETH environment that I could simply shrug off the official policy and use the resources of the Chair, cushioned by the grants we had accumulated over the years, to do our own thing quietly − as we had when we built Informatics Europe, the Journal of Object Technology, the LASER summer school and other achievements − without asking for anyone's permission. Today the evolution of almost all European universities goes in the exact reverse direction: ever more regulations, ever more constraints, ever more fear, ever more control, ever more forms, ever more interdictions, ever more caveats, ever more provisions, ever more codicils, ever more bureaucrats, ever more naysayers, ever more signatures, ever more veto opportunities for those who wake up at night dreading the thought that someone, sometime, somewhere, might succeed. That, and not just the money, is why they are not ETH[29].)

---

[28] www.ethz.ch/en/the-eth-zurich/education/innovation/torques.html.

[29] I have never met, read or heard a single non-Swiss European politician who has the slightest clue on this matter. The idea of ETH is that you hire good professors, give them good resources, decision power and a reasonable regulatory framework, and mostly leave them alone, exerting *a posteriori* controls. Another concept that other Europeans are constitutionally incapable of understanding is that it is OK, in fact inevitable, to use a system that is inherently (in an Isaiah-Berlin kind of way) imperfect. Once in a while you will hire someone who is not as "good" as you thought, or even some of the "good" people will squander some money. Only rarely is an extra regulation the solution. Most of the time the right response is to accept that occasional mishaps are just the price to pay for the flexibility of the system and the creativity it fosters. One success justifies a hundred failures.

We needed a platform and contacted two of the main MOOC ventures, Coursera and edX (a consortium started by MIT and Harvard), who courteously turned us down because they work with educational institutions as a whole, not individual groups. We were running out of options when Chris Poskitt brought to our attention the existence of the open-source Moodle framework, which we did not know even though we later found out that some units of ETH were using it. Moodle turned out to do what we wanted. Over the next few months we recorded our MOOC lectures, covering the entire Introduction to Programming course.

The recording was somewhat of an obstacle course. ETH buildings are not far from the University of Zurich's hospital, the main one in the city. We had all grown accustomed to the sound of ambulances rushing to the emergency room, not a major nuisance since our building was on a quiet street a few hundred meters away. Never had we realized how loud and frequent these siren sounds actually were. We started hating the poor victims of heart attacks for spoiling our recording sessions. Even though Marco Piccioni, the recorder, was infinitely patient with my many false starts, spending many hours piecing together the takes, I again and again had to restart a lecture, usually interrupted just before the end right when I thought I had finally had done a good job on that particular topic without stuttering or panicking.

We had no guidance and were devising techniques by ourselves. In retrospect, while professionals may criticize the quality of the audio or other technical aspects, the main deficiency of this first MOOC attempt was probably the length of many of the sessions, often going beyond the eight minutes or so that are (as we learned later) standard in the field.

One of the main results of this effort was to provide an impetus for the work on on-the-cloud programming covered in the last section. The first version, Eiffel4Mooc, enabled students to perform Eiffel programming assignments from the browser, compiling and running them on the cloud. It evolved later on into Codeboard, applicable to other programming languages.

The MOOC was released in time for the 2013 session of *Introduction to Programming*; it is still available, hosted, after we gained some institutional recognition, at a page managed centrally by ETH [202]. It was first made available to our students, then released generally. The general release reached a non-ETH audience of only a few hundred people, far from the tens of thousands that one expects for a MOOC on a core topic. We would not have been able to handle support and exercises for such a large crowd anyway, but the experience taught us that to have real global impact we needed, regardless of the quality of the offering, the benefit of a global MOOC brand such as Coursera or edX.

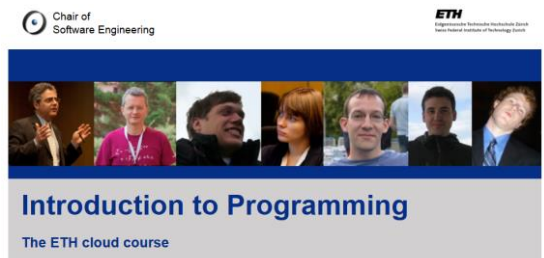

The most interesting impact at that stage was internal. The MOOC was a big hit with ETH students. One of our concerns had been that it would provide another incentive, along with all the other materials we provided (Touch of Class textbook, slides of lectures and lab sessions, videos of previous years' lectures and 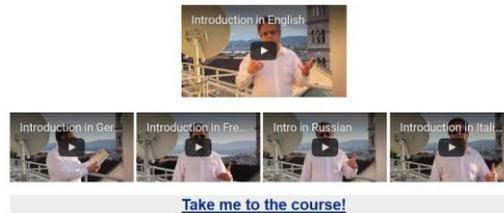 other documents), to skip class. In fact the attendance at lectures increased because, as students told us, the presence of MOOC made the learning of programming hip again. The Eiffel4Mooc tool provided a marked improvement of the interaction between students and teaching assistants: we could now see directly the students' programs and as a routine matter try to compile and run them without them or us having to set up any particular environment. We could immediately see errors and misunderstandings and provide instant feedback. In students' evaluations, the MOOC came out as one of the top benefits of the course. [224] provides an assessment of that first experience.

In the meantime the institutional attitude towards MOOCs had improved. We were able to present our skunkworks effort to management and plead that "tiny" and "with restrictions" were not in the ETH spirit. ETH had already entered into an agreement with edX. It was devoting some resources and providing a supporting structure in the form of the Laboratory for Educational Technology with which we had already been collaborating for several years; its members, under the leadership of Olaf Schulte, were video-recording the Introduction to Programming lectures. Not everything was technically perfect yet; we no longer had the ambulance sirens, but the studio was under a corridor at the Rectorate and now it was passersby with high-heels who (more rarely) still forced us to redo takes. It would take another year or two for a fully state-of-the-art recording studio to be available. But we could move on to new MOOCs at a much higher level of professionalism, with the help of experts in recording technology such as Schulte and Artan Hajrullahu.

For edX we started *Introduction to Programming* again, but generalized to an overall introduction to computing, with supplementary material such as a roundtable between members of the group and a guest interview of my colleague Peter Widmayer on the goals and role of theoretical computer science. Because of the amount of material, we produced two courses [219][245], with more advanced material moved to the second one. Developing professional-quality movies is a huge amount of work, but also fun, as when we recorded the introductory sequence on the terrace at top of the ETH building, with post-processing that made me appear by magic at one point and then 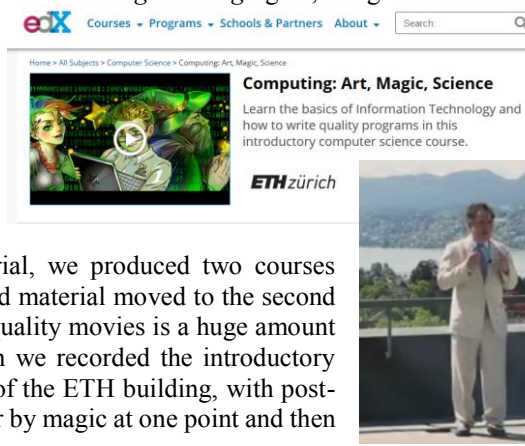

suddenly another, with Zurich in the background. The courses aired several times and in this case we did reach audiences in the tens of thousands, with a large following in several countries such as India.

Emboldened by this success, Piccioni and I produced a third edX MOOC, on agile development, based on my book on the topic [228]. We tried to do the introductory sequence ourselves, but it did not pass muster with the experts, so we had to record it again in Zurich, this time on the newly redone plaza in front of the Zurich opera, suggested by our colleagues because of the ballerina illustration on the cover of the book. The agile MOOC [270] also enjoyed a substantial following.

A MOOC, in the proper definition of the concept, is not just a video archive of a set of lectures, but an active event that runs regularly like a classical course, with exercises, laboratory sessions and the support of teaching assistants. While the three edX MOOC are available on the archive as given in the references, I have so far failed to obtain the small support that would be needed to run them again, even though this means benefiting from an investment already made, and even though at the time of writing they remain the only computer science MOOCs ever produced by ETH[30] and a substantial part of the total record[31].

For the Chair and its members, the MOOC effort has been extremely rewarding. As noted at the beginning of this discussion, we are not naïve endorsers of the myth that MOOCs will replace other forms of teaching. Training programmers, for example, requires intensive human contact (as we are providing in Propulsion Academy, a programming bootcamp-style school that colleagues and I created after my departure from ETH, see propulsionacademy.com). MOOCs are simply a remarkable new instrument in the panoply of pedagogical tools. Many European universities have neglected that instrument, or joined the movement more belatedly and timidly. All the major initiatives and companies are US-driven. This lack of action ("being asleep", as the NZZ journalist put it) is a historic mistake.

For course authors, the effort of putting together a MOOC is exhilarating but huge. Less formidable today if you have good resources (such as EPFL's "MOOC factory"), and decreasing as you gain more experience, but still huge. This is the reason while producing SPOCs (MOOCs without the "Open" part, intended just for an institution's own students) is in general not attractive: if you are already doing your job seriously for your courses, by delivering the lectures, possibly having them recorded and providing supporting material, there is no incentive for the investment of an online course. The incentive comes when you can hope to gain impact and recognition worldwide. This is the same kind of motivation that leads professors to write not just course notes but textbooks.

---

[30] Roland Siegwart from mechanical engineering and his colleagues produced a successful MOOC on Autonomous Mobile Robots.

[31] It is still the case in July 2017 that all ETH MOOCs (www.edx.org/school/ethx) come from three professors including, besides Siegwart and me, Gerhart Schmitt from the Department of Architecture with courses on cities of the future. The EPFL page, moocs.epfl.ch/, has several hundred offerings.

MOOCs have become an indispensable part of the modern battery of pedagogical tools. Our MOOC experience was immensely valuable, and many of the pedagogical lessons it taught us are applicable in other forms of education. Our ETH students told us again and again how much they benefited from the tools; just as gratifying was the ability to reach out to tens of thousands of external participants, learn from their questions and reactions on the course forums, and receive so many messages of thanks.

## 5.3 Tool support for programming on the Web

Any path to learning programming requires doing programming exercises, lots of them[32]. Prompted by our experience with Introduction to Programming, our effort to make the MOOCs effective, our distributed programming work (section 6) and the development of verification tools (section 10), we started developing tools enabling students and more generally programmers to use the browser as a development environment. The current version, preceded by Eiffel4Mooc and comcom, is Codeboard, available at [codeboard.io](codeboard.io). The earlier versions were tailored to our needs and specifically to Eiffel, but Codeboard supports a number of other programming languages.

A user of these tools can:

- Enter code through a Web browser.
- Compile the code by pressing the Compile button.
- See any compilation error message, correct the error and recompile.
- Run the code.
- Specifically, run tests. The instructor can prepare public tests, with data and expected results available to the user, and secret ones, which can be run to help ascertain whether the code is correct.

The tools can be embedded in a Web page; in the specific case of MOOCs, Codeboard offers specific interfaces to edX, Coursera and MOODLE. As an example, the screenshot below, coming from the first "Computing: Art, Magic, Science?" edX course ([courses.edx.org/courses/course-v1:ETHx+CAMS.2x+3T2015/info](courses.edx.org/courses/course-v1:ETHx+CAMS.2x+3T2015/info)), shows an exercise on control structures. The "*Your code here*" parts (as highlighted below) are the places where students must insert their own code. The rest of the structure has been pre-filled. Instructors preparing the course can decide how much pref-filled context to provide, and how much to leave to the student.

---

[32] Malcolm Gladwell's theory that mastering any topic requires "ten thousand hours" of practice is controversial, but it certainly matches my observation of good programmers (Bill Joy was in fact one of his original examples).

The student can make the result of a compilation, execution or test available to others, most importantly a teaching assistant.

The tools are not strictly limited to teaching or, for that matter, to programming. One of their applications is the online version of the AutoProof program verification system (10.7 and 10.8) at autoproof.org, serving as a tutorial but also usable simply to perform program verification online.

An interesting master's project by Paolo Antonucci [249] added further parameterization possibilities for the course preparer: a hint system enabling the student who gets lost to click a "Hint" button and get increasingly detailed hints. Such mechanisms are in line with the earlier work on TrucStudio (section 5.1); the dream was to integrate all these ideas and others into a sophisticated environment for preparing effective interactive courses.

Even without such developments, the pedagogical effect of these tools for cloud-based programming was nothing less than spectacular. They profoundly changed our interaction with the students and the effectiveness of the interaction with the teaching staff. In particular:

- A constant source of trouble, sometimes despair, for novices is the difficulty of setting up a proper environment for the development of their programs. Professional programmers do not mind installing this or that tool, downloading a script, adapting a control file, trying various fixes when something does not work the first time around; they have colleagues (and the Web) to ask for help if they get stuck. For novices (in particular the 18% of our students, noted above, who have not done any programming before), any one of these obstacles can be terrifying. In an introductory programming course we want to assess whether they can write a program, not whether they can fight with the details of downloading and installing (a skill they will need, but can acquire later). With Codeboard and our other tools, we set up the environment for them. They have nothing to install since all they need is a Web browser. Then can focus on the challenges of programming, such as data structures, algorithms and correctness, not on details of logistics. This approach frees and empowers a full new class of students who may have all the skills needed to become excellent programmers but would otherwise be blocked by side issues.

- The exercise preparer can decide to include supporting code at the exact level desired. The example above involves two classes (shown on the left in the figure), BOOK and LIBRARY; for both of them some of the code is pre-written, and the student only has to fill the parts on which the exercise preparer has decided to put the focus. The pedagogical benefits are immense: the instructor can decide on a precise order of skills to teach and test.

- Since some of the code is pre-written, it provides guidance to the students who later on have to write similar code themselves, in line with the earlier observation that much of the learning of programming occurs through imitation.

One of the fundamental changes affected how teaching assistants could help students encountering difficulties in solving programming exercises. The typical situation is "why do I get this compilation error?" or "it crashes, I can't understand why!" or "I have no idea why that test does not pass". In a standard setup the assistant who wants to help has to ask the student to prepare and send a Zip of the project, unzip it, and try to reproduce the problem. Often this is only a first step in the process: maybe the student is using Windows and the assistant Linux; maybe the assistant cannot assemble the system because the student forgot to include some library; maybe what does not work for the student works for the assistant, or the other way around, because the setups are different. With Codeboard, the student just sends a link to the project, and the assistant immediately has access to the original code in the original setup. All the troublesome losses of efficiency disappear. Again and again we witnessed this tremendous improvement in our ability to help students.

It is interesting to note the objections that we kept hearing against such tools: concerns about student privacy. One needs of course to be extremely careful, as we always were (out of obvious ethical concerns, in addition to ETH rules). But an undue concern for privacy was part of the resistance to MOOCs and delayed the introduction of mechanisms that demonstrably help students. At ETH these hesitations were particularly irrelevant, since in first-year courses the students' performance on exercises and homework has zero effect on their grades: the results, including the possibility to go on with their studies, are entirely determined by end-of-year exams.

Part of the reason for these concerns was that the MOOC organizations such as edX and Coursera are US-based, and so are their servers. If that is the problem, rejecting technology is not the solution. The best solution would have been to develop European-based MOOC technology, but since this did not happen students should enjoy the benefit of technology regardless of its origin[33].

Codeboard usage has grown far beyond ETH, involving courses in numerous universities and reaching tens of thousands of students. Like its predecessor tools, it provides a powerful example of how technology can help modern teaching.

## 6 Methods and tools for distributed software development

An early foray into tools for supporting distributed software projects was Till Bay's Origo, developed for his PhD on a topic that he had himself defined [95]. He had recognized the emergence of complex software projects being developed over different locations, and the inadequacy of the solutions available then. Origo benefitted from numerous student contributions and integrated many novel ideas. For several years it hosted all of our projects, and attracted a steadily growing number of diverse projects from all over the world. Attempts to turn it into a commercial success did not interest venture capitalists in Switzerland, and Comerge, the company that Bay had founded, initially with Bernd Schoeller, went into other directions where it is thriving today. The success of Github in recent years proved that Origo was an idea with great potential, even if others realized that potential. Our own CloudStudio, discussed below, also continues the Origo work.

A big impetus for our interest in distributed development was the extraordinary growth of outsourcing, one of the most significant developments in software engineering during the last decades. The Indian software industry, the most visible ben-eficiary of this phenomenon, grew from less than one billion dollars in the late eighties to over 120 billion dollars in 2014. As mentioned earlier, this phenomenon had a strong effect on the mood in Switzerland in the early 2000s: computer science was no longer a popular topic for entering university students; enrollment had actually dropped. There were other factors, notably the burst of the first Internet bubble, but the fear of outsourcing was seemingly the dominant cause. In the view of many parents, there were no longer any good prospects in programming; "*all the good jobs have gone to Bangalore*".

---

[33] The Wirth 80th birthday symposium, which we organized in 2014, came at the height of the Snowden affair. In one of the breaks, I asked Vint Cerf, one of the invited speakers, whether he would be comfortable using a Cisco router if he were a sysadmin for (say) the German ministry of defense. He did not directly answer but said he could not understand why no one was developing a major Europe-based cloud solution along the lines of Microsoft's Azure or the Amazon Web Services. Unfortunately, I knew the answer to that question, as does anyone who has tried to obtain venture capital for ambitious non-standard tech projects in Switzerland or Europe in general. As a typical and sad example, a request to the ETH venture kick-up mechanism to fund (for about 150,000 dollars) a preliminary effort towards commercial development of Codeboard was dismissed summarily.

In May of 2004 I gave an invited talk at a software engineering conference hosted by Siemens in its main site in Munich. Like everyone entering the huge Siemens campus that day, I got a flyer from union representatives, complaining about the outsourcing of jobs and the danger to the employment of local workers[34]:

---

[34] With the typical moderate tone of German unions, e.g. (second paragraph) "*we know that not every job can be kept and that German society benefits from the international division of labor. However…*"

Shortly thereafter, I saw the Call for Papers for the forthcoming International Conference on Software Engineering, the main annual event in our field, which was going to be held in Shanghai (and which I had the pleasure of attending). Given that China was already hosting a significant outsourcing industry (focused for a large part on Japanese customers), and more generally given the extraordinary growth of outsourcing in software, one might have expected the conference to address that matter,; but in the long list of topics for the papers being solicited, it did not even appear!

The creation of the DOSE course was the first step towards recognizing the importance of the phenomenon and integrating its study into software engineering. In our research as well as in the course, the focus soon moved away from outsourcing to encompass the more general matter of distributed development. What matters more than where parts of the team are, for example India, is that they are not all together. We explored the consequences of distribution in many research efforts relying on techniques of empirical software engineering.

Too many to list here, in fact (the list of articles relying on studies from DOSE appeared in section 3). They address such questions as the influence of time zones and distribution [157], collaborative debugging[198] and pedagogical aspects [148]. The distributed development experience, both at ETH and in my continued role at Eiffel Software, led to a new tack on the venerable software engineering technique of code reviews. A SEAFOOD contribution [87] and the resulting *Communications of the ACM* paper [91] proposed a process centered not on a meeting but on a document. Most of the work is done in advance, in writing, on the document; the meeting still takes place but is devoted to the most interesting part, points of disagreement. The document follows a standard structure going from high-level concerns of choice of abstractions and inheritance structure (these are *design and code* reviews) down to implementation, comments and style. Originating from practice developed at Eiffel Software, where the development is distributed, this review methodology and the supporting document structure were quickly and organically adopted in our group. We performed many spirited and productive reviews.

A more challenging research topic is configuration management. While it is one of the undeniable advances in modern software development that any reasonable team keeps all code under a configuration management system, and while we always taught the principles in our courses (starting with the essentials in Introduction to Programming), I have always resented the complication and constraints of configuration management systems such as Subversion and Git. The whole idea is very much 20-th century: you manually track versions, and manually reconcile changes with those of other developers, an often painful and always tedious process. I do not want to bothered by any such nonsense. I want to develop my code in my IDE (such as EiffelStudio), and trust that it will keep track of what I do and warn me when one of my actions might conflict with a change made by another developer. Prevention rather than cure; and when cure is necessary (if incompatible changes were made despite the warnings), as much help as possible to reconcile versions. Never again to I want to go line by line through a "diff" to decide what to retain and what to reject:

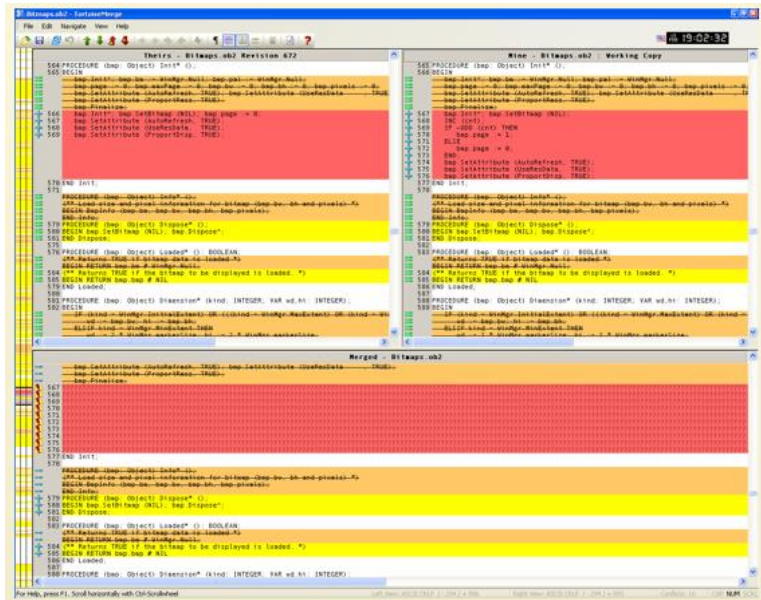

To address that challenge I proposed the concept of view: each developer works on his own view of the code, but can also see other developers' views to spot divergences early. The IDE supports these concepts directly:

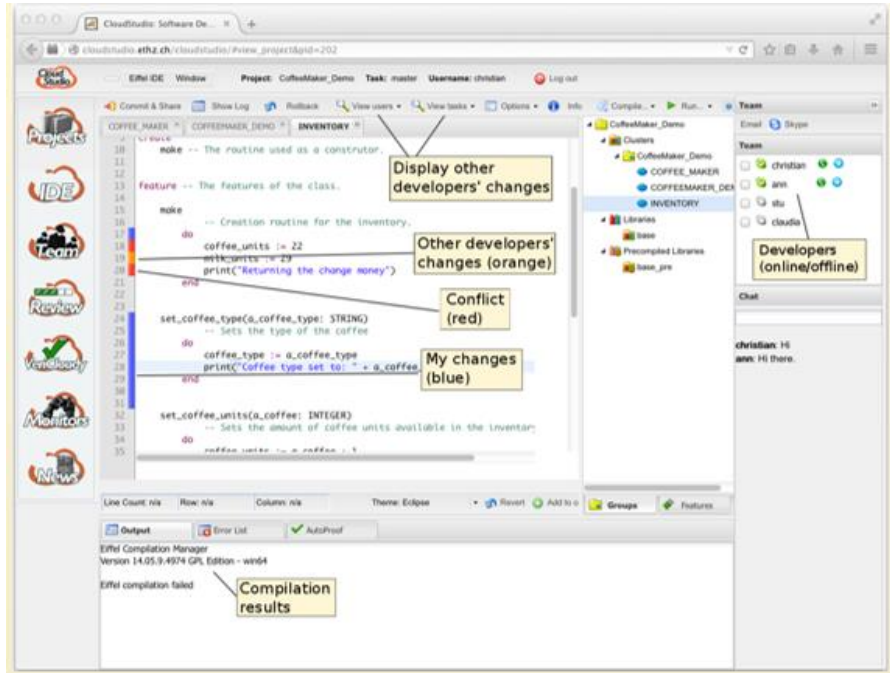

This screenshot is taken from CloudStudio, developed by Christian Estler on top of EiffelStudio for this thesis. CloudStudio provides a flexible environment for collaborative development, essentially avoiding merge conflicts.

We found these ideas and the application exciting and started going out to present them to companies. The reaction was sobering. In short, no one was interested. We were told that configuration management tools are good enough and that merge conflicts do not arise in practice. I do not believe it, but as Jules Romains's Knock knew, in the land where people think they are healthy doctors have no business.

We soon noticed, however, that the presentations were not only falling flat: as soon as we talked about applications to teaching rather than industrial development, people got excited. We decided that for the moment this is where the applications were, and the CloudStudio effort took on a new life in the form of the cloud-based programming education tools discussed previously (5.3). So in spite of industry's rejection this work blossomed.

I remain convinced that the multiview approach addresses a critical problem and solves it effectively. We have to wait until the patient realizes he is sick.

## 7 Language development and standardization

Much of programming (do not believe Leslie Lamport) is about languages. The core language of our chair, although by no means exclusive of others (C, Java, C#, Javascript…) was Eiffel, used in particular as the vehicle for the basic teaching of programming and much of our research developments.

Over the Chair's life Eiffel underwent significant changes (7.1) and a process of international standardization (7.2). We will also look at efforts at language specification (7.3) and translation (7.4, only partly Eiffel-related). Finally (7.5) I will mention recent work towards a general theory of programming and programming languages.

## 7.1 Modern Eiffel[35]

This is not the place for a detailed description of Eiffel and its evolution; for up-to-date information see the eiffel.org site [280]. Because many people still keep their impression of the language from early publications [292] [293] [299] it is important to mention that the language has considerably advanced. Some of this evolution has been in new expressive features, particularly agents, which represent an object-oriented version of "closures" in functional languages (see also [99], discussed at the end of section 8). C# similarly has a concept of "delegate". What we found, in response to early concerns, was that agents do not compete with standard OO mechanisms but complement them.

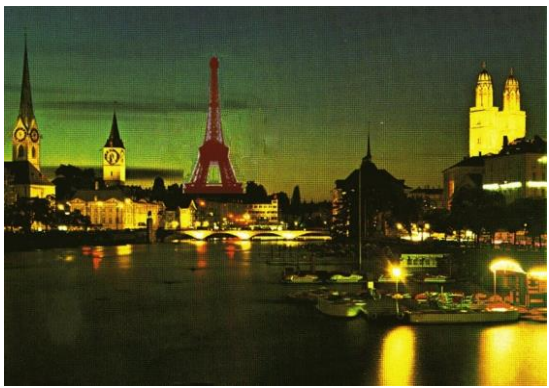

The development of Eiffel, particularly in the standardization processed described next, follows explicit principles:

- Simplicity.
- Consistency.
- Smooth evolution, respecting compatibility but not precluding progress.
- Precision of the language specification.
- Helping programmers develop software of high quality, particularly correctness.

. The view of simplicity is not the same as the well-known Wirth language design philosophy of devising *small* languages (Pascal, Modula 2, Oberon). Size is not the goal per se; ambitious languages cannot be miniatures[36]. Our goal with Eiffel is to maximize expressive power, using the motto that the language must "*provide one good way to do anything*". The other key criteria are that the language should help programmers avoid mistakes, and avoid surprises (natural modes of expression should provide the naturally expected results).

---

[35] The adapted photograph of the Zurich skyline (showing the limits of my Photoshop skills at the time) is from my 2002 inaugural lecture [11], which yielded an interview [12] in the Neue Zürcher Zeitung (NZZ), the main Zurich newspaper.

[36] At the time of Ada's release, some people were criticizing the language, in a Wirthian mindset, for the size of its definition. Asked by an interviewer, Jean Ichbiah, the lead designer of Ada, responded that small languages solve small problems.

C++ provides a typical counter-example of the "*one good way*" principle. For the typical case of an operation that must be adapted to the run-time type of its target, you have two ways to proceed in C++: the traditional solution of indexing into an array of function pointers; and the OO mechanism of dynamic binding. How is a programmer, perhaps a novice, expected to choose? While such conflicts may matter less, in the C++ worldview, than the language's goal of compatibility with C, Eiffel's design tries to avoid them. There are no function pointers to begin with. As another example of the one-good-way principle, Eiffel has only one loop construct, with a flexible syntax supporting common semantic needs. Among these variants are "**all**" and "**exists**", which have increased the language's expressive power, making it possible to express first-order predicates such as

**across** list **as** x **all** x.item > 0 end

(for what standard mathematical notation would express as $\forall\, x \in \text{list} . \, x > 0$). These notations are particularly useful in contracts, to express sophisticated specification properties.

Eiffel's agents are typical of the kind of extension that brings a major advance in expressive power. It suffices to compare the burden of implementing an instance of the well-known "Observer" pattern [282], with a class structure shown below in section 8.3, with the agent-based Eiffel solution [27]: for each type of event, the event producer defines a single *instance* et of a library class EVENT_TYPE (not a new class, just a variable denoting a run-time object); the subscriber declares its interest in the event type through et.subscribe (**agent** *operation*), where operation is the operation to be applied for events of this type; and to produce an event the producer, for example a graphical library, uses et.publish (*args*) where args are the event arguments, type-matching the arguments of operation. The simplification is dramatic. This is an example of language mechanism that did increase the size of the language but that no one ever regretted or characterized as "featurism". Both the Observer pattern and the undo-redo or "command" pattern (introduced in [292]), and their simpler agent-based versions, found their way as staples of the Introduction to Programming course.

The goal of quality means neither that using Eiffel guarantees quality results nor that it precludes bad results. More modestly, it encourages the production of good software by programmers who understand the concepts of the Eiffel method, and guards them from some mistakes (including through the principle of least surprise mentioned above). Top programmers, in their top days, will produce good software in any language. As Emmanuel Stapf has noted, the idea of Eiffel is to turn every one of your days into one of your top days.

Also characteristic of the Eiffel ecosystem is the acceptance of language change. Backward compatibility is a critical concern as soon as there is an installed base of programs, but it should not be an obstacle to moving on when, a few years after an initial decision, the community realizes that there is a better way to handle a certain programming schema. Unless one actually changes the language, the result is either to remain with unsatisfactory constructs or to keep both old and new variants, piling up features that make the language too complicated, in addition to burdening compiler writers. The Eiffel process has a number of times discarded old mechanisms, making it incumbent on compiler writers to provide programmers with a reasoned transition path and, if appropriate, translation aids. This is the principle; the implementation requires careful engineering and delicate choices; there have been a few mishaps, but the general goal has been both to preserve existing program investment and to permit progress.

As a result, Eiffel today is both the same language as the original, retaining the same spirit, architecture and view of the software development process (class-based structure, straightforward keyword-based syntax, simple OO with multiple inheritance and genericity, Design by Contract, information hiding, precise language definition), and a visibly different one in its range of features. The focus on simplicity has led not only to removing some features but also to making the rest more general and consistent, removing limitations and avoiding special cases.

## 7.2 Void safety

Other than agents, the most important development of Eiffel in the past fifteen years has been void safety. Null-pointer dereferencing — the program crash or exception resulting from a call x.f where x is a null, or "void" pointer or reference — is (for Eiffel, was) the major remaining threat to the safety and security of program execution, including with object technology. The following figure from Alexander Kogtenkov's thesis [272] shows the security attacks, in the Common Vulnerabilities and Exposure database, that involved null-pointer dereferencing:

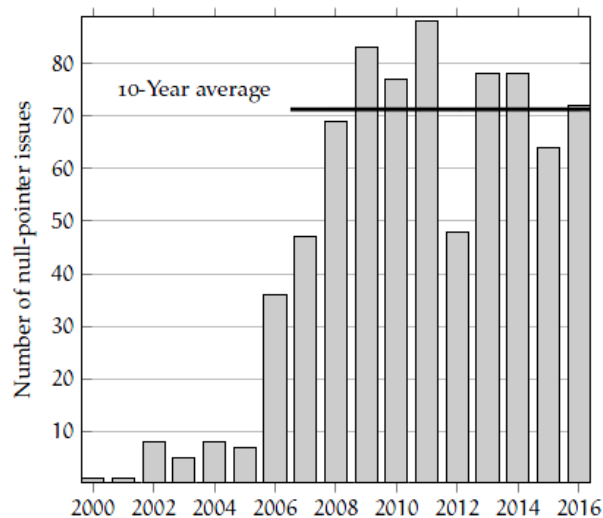

No less an authority than Tony Hoare wrote (cited in [133]):

*I call it my billion-dollar mistake. It was the invention of the null reference in 1965. […] [It] led to innumerable errors, vulnerabilities, and system crashes, which have probably caused a billion dollars of pain and damage in the last forty years.*

I had looked at the problem earlier [300], but with no actionable result. The decisive step was a late-afternoon chat with Erik Meijer during a visit at Microsoft

Research in Seattle, where he introduced me to a void-safety mechanism that had been devised for one of the research languages, Cω I think, then under study at MSR. The mechanism never found its way into C# or any of the actual production languages of Microsoft, but this short demo clicked: when you see that someone else has been able to find a solution to a problem that you had consciously assigned to the "death and taxes" category (that is, not solvable any time soon), a mental block goes away and you may be able to find your own solution. I presented ours in an ECOOP keynotein 2005 [39]; it was a paper design, which we implemented in the years afterwards, also revising it extensively as a result of both practical experience and discussions in the standards committee. Once we had an implementation at Eiffel Software, we (including the two principal contributors to that implementation and to the discussions, Stapf and Kogtenkov) wrote a second paper of the topic on the basis of that experience [133]. Entitled "Avoid a Void", suitably published in the Festschrift for Hoare's 75th birthday, it states:

> *Devising, refining and documenting the concept behind the mechanism presented here took a few weeks. The engineering took four years.*

That sentence, written in 2010, was optimistic. The design may have taken (in fact) an afternoon, but engineering took another seven years after 2010. Thanks largely to the work of Alexander Kogtenkov, we consider it final today (2017) — who knows.

The reason things took so long is not the basic idea, which is simple, combining a type system mechanism, attached types, and compiler-performed static analysis to sanction guaranteeably safe cases, called "Certified Attachment Patterns". The problems that have taken all that time are:

- On the theoretical side, to take care of initialization. If you declare a variable as attached, meaning that it should never become void, you have to make sure that it is non-void on object initialization. Work by Peter Müller identified some of the problems and helped us finalize the Eiffel mechanism.

- On the practical side, to handle existing code. Once you understand the basic idea, it is not hard to write void-safe code. The void-safety style becomes natural. But converting non-void-safe code, which by definition did not use this mindset, is painful. It took several years to convert the fundamental Eiffel libraries, and even today the core code of EiffelStudio is not void-safe. The effort will have to be performed at some point but it is large and tedious (and hard to justify since, through trial and error, the EiffelStudio code seems to be largely free of null-pointer dereferencing, although only void safety would turn this impression into a guarantee). All new code, however, is void-safe from the start.

These difficulties explain why a significant part of the Eiffel community was initially reluctant to adopt void safety. As late as 2009 a message on the user forum stated:

> *Last night I had a dream. I was programming in [pre-void-safety- Eiffel 5.7. The code was elegant. There was no need for defensive programming just by taking full advantage of design by contract. Thanks to these contracts the code was easy to reuse and to debug. I could hardly remember the last time I had a*

> *call-on-void-target. It was so pleasant to program with such a wonderful language.*
>
> *This morning when I woke up I looked at the code that had been modified to comply with void-safety. This was a rude awakening. The code which was so elegant in my dream now looked convoluted, hard to follow. It looks like assertions are losing all their power and defensive programming is inviting itself again in the code.*

This criticism, which at first struck me (with my enthusiasm for void safety) as a refusal to move on with modern technology, were entirely justified. The mechanism had not been engineered well enough in its details to address the needs of practitioners with extensive Eiffel experience, deep knowledge of Eiffel and OO principles, and a large code investment. The work of Kogtenkov, Stapf and others addressed this issues and it is fair to say that today the dissent is gone and every serious Eiffel programmer, including the former skeptics, not only understands the conceptual value of void safety but has accepted the mechanism as it is and develops void-safe code from the start.

Kogtenkov's thesis is also an example of the growing use of formal methods in our work. While separate from the bulk of the verification work in AutoProof (section 10), it demonstrates that the void-safety mechanism actually makes Eiffel void-safe, in other words guaranteeing that execution cannot produce a null-pointer dereference. This result is far from trivial since the mechanism has become intricate, after all these years of tuning the initialization semantics to ensure that programmers can still use programming schemes that sound natural to them, including many which simple rules would wrongly flag as void-unsafe. The proofs have been mechanically verified using the Isabelle-HOL system.

## 7.3 The standards process

In the early days of Eiffel I was in charge of the language definition, reflected in the book "Eiffel: The Language" [293] and the working version of its planned revision. A process of international standardization made it possible to unproprietarize the language definition several years before the EiffelStudio implementation went open-source (as was described in section 3). In 1999, before ETH, I had become involved at Eiffel Software in a cooperation with Microsoft as part of the development of .NET, initially under non-disclosure; I was associated with the formal public introduction of .NET by Bill Gates and published an early article on it just before I went to Zurich [301], as well as a video book, a kind of MOOC *avant la lettre* [2]. The immediate goal of the cooperation was to produce a version of our Eiffel compiler for the .NET platform, but we became seriously interested in .NET and participated actively in discussions with its designers, particularly about the design choices in the object model[37]. Emmanuel Stapf, head engineer at Eiffel Software and a compiler expert, was the team member most active in this cooperation. When Microsoft decided to submit key aspects of the .NET technology to international standardization, through the Ecma consortium, he was invited to become a member of the committee and played an active role in the standardization of the. (In particular, since we had with Eiffel a powerful generic mechanism, different from Java's, he made sure that the .NET generics were reasonable.)

---

[37] The discussions also involved other groups working with Microsoft, such as Jürg Gutknecht's Oberon group at ETH. These were actually my first ever interactions with ETH and played a role in my joining the institution a couple of years later.

Ecma (formerly the European Computer Manufacturers Association, but no longer expanding its name) is a consortium of IT companies that stands behind many of the main standards in information technology. It is goal-oriented and less bureaucratic than the International Standards Organization, which must accommodate all the national standards organization that it federates. But (because in the early days ISO was not prepared to deal with the huge and urgent need for IT standards, while Ecma was) it has a special agreement with ISO enabling Ecma standards to become ISO standards through a fast-track process, avoiding the typical ISO multi-year approval course. The work in the .NET committee was friendly and effective. When the issue came up of standardizing the Eiffel language, Stapf naturally proposed, on the basis of his experience, that we approach Ecma, which readily agreed to host us.

The committee (Ecma TC49-TG4), which started its work in 2003, was at its core a small group, with a list of participants given earlier in section 1. The first version of the Eiffel Standard was accepted by Ecma in 2005 [36] and revised the following year as Ecma Standard 367, shortly thereafter becoming an ISO standard [60]. A revised version is in progress and we hope to have it out in 2018.

While the Ecma process makes considerable use of electronic communication, the bulk of the work occurs in face-to-face meetings. Most of our meetings occurred at ETH, in California (Santa Barbara, site of Eiffel Software, or Orinda, site of Axa Rosenberg), or in Nancy or Villebrumier (France). The process leading to the current (2006) standard involved eighteen 3-day meetings (with several more since then). In other words, the committee members, most of whom were not married to another member, lived together for two full months of their lives. This figure is just an indicator of how intensive such a standards process is, compounded in our case by the passion and focus on exactness that characterized our particular group. It was not rare to spend an entire afternoon on a paragraph, in a succession of fiery verbal fights and long periods of silence (as happens when every one of the members realizes, however passionate the last outbursts, that the problem is actually more complicated than any of them thought), and restart on the same paragraph from scratch the next morning.

The base work is exacting (the nice way of putting it) and tedious (the more immediate feeling). For Eiffel we pursued the goal of reaching as much precision as is possible without resorting to a fully formal (mathematical) specification. The practice of formal specifications helped considerably. One of the irritants in reading specifications of other languages is their practice of "only if" rules: listing necessary conditions for a construct to be valid. For example you can assign an expression e to a variable x only if their types are compatible. Oh, and x cannot be read-only. And, by the way… I do not want to be told, piecemeal, what I may not do! I need to be told authoritatively what I *may* do. I want a rule that tells me "the assignment x := e is valid **if and only if**…", followed by an exhaustive list of conditions. This "if and only if" style has been the rule for the language specification ever since the original Eiffel book [293] and the committee stuck to it. Here for example is the validity rule for feature bodies in the standard:

---

**Validity: Feature Body rule**                             **Validity code: *VFFB***

A Feature_value is valid if and only if it satisfies one of the following conditions:

1.   It has an Explicit_value and no Attribute_or_routine.
2.   It has an Attribute_or_routine with a Feature_body of the Attribute kind.
3.   It has no Explicit_value and has an Attribute_or_routine with a Feature_body of the Effective_routine kind, itself of the Internal kind (beginning with **do** or **once**).
4.   It has no Explicit_value and has an Attribute_or_routine with neither a Local_declarations nor a Rescue part, and a Feature_body that is either Deferred or an Effective_routine of the External kind.

---

*Informative text*

The Explicit_value only makes sense for an attribute — either declared explicitly with Attribute or simply given a type and a value — so cases <u>3</u> and <u>4</u> exclude this possibility.

Such a rule represents a *contract*: it does not just tell the programmer the "only if" part, the conditions he has to satisfy for his program to be valid, but also provides a guarantee (the "if" part) that if these conditions hold then the compiler will accept the program and give it a meaningful, well-defined semantics for execution. That is why the discipline of producing such rules is so hard: you have to make sure you have envisaged all possible cases.

The description is divided into clearly distinct levels: lexical, syntactic, validity and semantics. Validity consists of rules such as the one above expressing conditions (such as typing constraints) imposed on constructs that are already correct at the syntax level. Each rule has a code, such as VFFB above, for precise reference. Semantic rules describe the execution-time effect of syntactically correct and valid constructs.

Standardization efforts primarily document, disambiguate and solidify existing technology. They usually stay away from innovation (in the same way that Wikipedia articles are not supposed to include "original research"). The period of Eiffel standardization coincided with a period of language evolution, we decided to allow ourselves more freedom and initiative than is normally the case in standards committee. Encouraging this approach was the small size of the committee, its informal nature (we were responsible to our institutions and, particularly in the case of industry representatives, to practicing Eiffel programmers with a vested interest in the stability and quality of the language, but did not have to report to national standards organizations with their specific agendas), and our sharing of Eiffel principles. As a result, many of the novel features of today's Eiffel were introduced in committee discussions. We instituted a rule, inspired by the practice of Internet standards, that we would not adopt any new mechanism until at least one compiler had implemented it successfully. In practice we did not always follow that rule (sometimes, I must admit, out of my out premature enthusiasm for a new idea). More generally, we made mistakes, and many times we had to revisit a matter that we had thought settled. While aware of the need for a revised version, however, we are proud of the achievement. The standard has stood the test of time and is the basis for all Eiffel work today. I believe that beyond Eiffel the style and rigor of this work sets an excellent reference and example for other standards efforts.

Standardization work is not a frequent activity for academics, and for institutions that count things like publications and patents does not give any academic brownie points. Certainly not a good use, for a career-conscious researcher, of eighteen full-time 3-day stints. If your goal is to increase your Scopus h-index, this is just about the last activity to consider. Yet I never for a minute regretted investing my time in this demanding but rewarding work.

### 7.3 Formal specification of Eiffel semantics

A fully satisfactory language description should have a version that is not only in the style of formal specifications but actually formal. When and if the Eiffel standards committee embarks on such a project it will benefit from earlier efforts.

Incentives for precise specifications arose from the intricacies of concurrent programming.

One of the principal contributions of Piotr Nienaltowski's thesis [76] was to replace the ad hoc and somewhat messy rules for avoiding SCOOP traitors in my earlier work (chapter 32 of [299]) by a simple and elegant type system with inference rules. (A "traitor" in SCOOP is a variable that denotes an object in another thread, hence subject to special semantics, but is not declared accordingly, so that its treatment will be wrong.)

As part of his own work on concurrency, Benjamin Morandi needed more precision than was available in the existing definitions of SCOOP semantics. He turned to José Meseguer's operational semantics and MAUDE system [289] and produced (with Sebastian Nanz) a detailed specification which includes many non-concurrent aspects as well, resolving many potential ambiguities of an informal description. He took a new look at the delicate matter of exceptions in a concurrent context [170], providing what I believe is the proper conceptual basis for addressing this issue.

Also notable is the work (particularly by Martin Nordio, initially in his thesis work on proof-preserving program transformation in the context of .NET, and Julian Tschannen, also initially involving Peter Müller and Cristiano Calcagno) to verify advanced Eiffel mechanisms such as exception handling and agents [106] [127] [177]. The work on specifying exceptions formally also involved looking into the mechanisms of other languages. Nordio gained considerable insight into the Java and C# exception schemes. A quiz he devised involving the Java/C# "finally" concept illustrates the subtleties; inevitably, when one presents the examples to expert programmers in these languages, they guess some of the program results wrong.

My work on aliasing is discussed in section 10.10.

### 7.4 Language translation

A language-related development is Marco Trudel's work, starting with his master's thesis, to provide production quality translation between programming languages. The source language is typically C and the target language Java or Eiffel. The goal of this work [169] [178] is, as expressed by the title of the second of these references, to go "*beyond the easy stuff*". It is not that hard to produce a translator from C to an OO language that processes a significant subset of C. But C as used in practice includes many tricky constructs such as function pointers, pointer arithmetic and signal handling. The tools cover all of C and do not just perform a literal translation; they analyze the source for meaningful data abstractions so as to produce not just a program expressed\ in an object-oriented language but an object-oriented program.

## 7.5 Theory of Programs and FLIP

Recent work, still in progress, addresses the general nature of programs and programming language. What I found is that it is possible to describe all programming concepts − truly all, including concurrency − on the basis of extremely simple ideas from elementary set theory; in fact, three operations suffice, defined from elementary set operators such as union and restriction.

This effort, described in [253], can be called a theory of computation; it does not compete with the traditional theoretical approaches (Turing machines and such) because it has a different focus: to model the concepts of programming and programming languages in a way that reflects their practical use.

The theory has the potential to explain all form of programming, object-oriented for example, although the existing article only covers the basics. It could also (a claim not supported by concrete evidence so far) serve as a new basis for verification. An as yet unimplemented idea is to use the theory to build a "Formal Language Innovation Platform" (FLIP, where "formal" applies to "innovation platform", not "language") to foster experimentation with language features, with support for of immediate generation of prototype compiler generation and formal semantics subject to mechanical verification of consistency, and the availability of construct libraries (assignment, for example, has a general semantics and does not need to be re-specified and re-implemented for every new language variant). Verification is essential: when you play with a new language feature, you want to be able to prove (without having to produce an entire theory of the language including all the constructs irrelevant to your new idea, hence the importance of libraries) that it satisfies certain properties. For example, Eiffel is void-safe, as demonstrated in Kogtenkov's thesis and SCOOP is free of data races. As another example, [303] shows that the revised Java memory model is sound, but the effort requires an entire book; the FLIP idea is that you start from a verified description of the rest of the language and just add the elements needed to verify a new specific property.

 Or maybe the theory has no immediate practical application, but I believe that it illuminates the nature of programming and could at least serve as the basis for teaching the discipline. And if even that does not happen, the article has helped *me* clarify, if only for my own benefit, a view of programming, until then informal, that has for a long time underpinned all my theoretical and practical work.

## 8 Software process, methodology, agile methods, requirements

Programming methodology, a thriving field in the seventies and eighties, is somewhat out of fashion; even the IFIP working group WG2.3 that bears this name, which provided us with so many insights and whose meetings were always such a fascinating experience, does not much address its own official topic but is mostly about software verification. Methodological issues, however, have constantly been at the center of our concerns.

## 8.1 Agile methods

In one of the very first lectures I gave, probably in the spring of 2002, I covered principles of object-oriented design. A third-year student came to me at the end of the lecture and asked me why I was even bothering to broach these topics. (ETH students are polite, but he clearly meant "why are you wasting our time?") No one does design these days, he said. Everyone knows (he said "everyone" but his tone clearly meant "everyone but professors of software engineering") knows how to proceed: produce the "*simplest [program] that could possibly work*", per the Extreme Programming (XP) slogan, then refactor until it is good enough.

I was stunned. I had not realized how far XP and agile ideas had trickled down, all the way to undergraduate students. The comment also highlighted a feature of today's teaching of software topics: we have to contend with other sources of information. There was no teaching of agile methods (and very little of software process and such software engineering topics) at ETH at the time; students received such nuggets of knowledge, right or wrong, through summer internships in industry.

That particular nugget was of course wrong. Design is as essential as it ever was. Refactoring is a great idea, but refactored junk is still junk. The right approach is to design a first version of as high a quality as you can manage under your constraints, then take a critical look at it and improve whatever needs improvement. The encounter with that student was helpful because it alerted me to preconceived ideas that students were getting from other sources. Understanding such preconceptions enabled me to address them and make later lectures more effective[38].

In the following years I became increasingly aware of the importance of the agile phenomenon. The 2006 change of EiffelStudio's status from proprietary to open-source, mentioned in section 3, went with a change of development model for Eiffel Software, partly as a result of suggestions from Till Bay, to a time-boxed development of two releases a year, straight out of the agile principle that time overrides function. Marco Piccioni alerted me to the importance of Scrum. I realized that in the matter of software process management the gap between academia and industry was of uncommon proportions: in most companies (outside of the CMMI zone, mostly inhabited by big US defense contractors) the buzz and much of the practice was *only* about agile methods, and most software curricula universities taught *nothing* about them. *Some* gap is normal (if universities did exactly the same as companies, we would not need universities), but a total disjunction of that kind is not alarming.

I decided to educate myself in depth about agile methods; in fact I read all the books and many articles, immersing myself into agile Web sites and forums. I decided that to get into something for good you have to sing the company song. I became a Certified Scrum Master[39]. The more I learned the more I was struck with the agile paradox: its mix of very good, uninteresting and very bad ideas. In software methodology, it is usually fairly easy to distinguish productive ideas from bad ones; but here the best and the worst are inextricably mixed.

---

[38] Educators should be more aware of this phenomenon of incorrect prior knowledge. In most sessions of "Introduction to Programming", in which I often used lecture breaks to show various videos on side topics, I played the ground-breaking and eye-opening "Private Universe" video from Harvard [274].

My 2014 book *Agile! The Good, the Hype and the Ugly* [228], and the many professional seminars I gave on the topic, are a result of this experience. (I worked again from the principle that I have often applied: that a good way of learning about a topic is to teach it, and an even better way is to write a book on it.) The software architecture/engineering and DOSE courses took in a growing amount of material on agile methods, and in the DOSE context we made a study of their effectiveness in a distributed context [171].

The book is both a tutorial and a critical analysis, with "critical" in the sense of "critique" and not of "criticism". Almost all of the agile literature is the adulatory kind, with authors exhorting their readers to get on their knees and start applying the gospel. I was neither in the preaching business nor intent on criticism for criticism's sake. I was interested in analyzing agile methods from a software engineering perspective, as another contribution to the field and not, as one would sometimes believe from reading agile authors, a revolutionary replacement for everything that came before. I was even more interested, with a practitioner's focus and the benefit of concrete experience with agile methods, to help people sort out the useful and in some cases brilliant elements of the agile advice from the inconsequential parts and the truly damaging ones.

Two examples illustrate these extremes (read the book for more examples, and more analysis of these two):

- The most catastrophic advice is the rejection of upfront work, particularly the agile scorn for upfront requirements and the rejection of upfront design, which had found its way into the mindset of that student in 2002. Like several others, this agile rule is based on a valid criticism of some exaggerated traditional practices, specifically the temptation of many traditional projects to spend too much time in foreplay and not enough on the real thing (writing programs); the phenomenon has given rise to the term "analysis paralysis". But highlighting the importance of code does not mean throwing away all the rules of good engineering. I immodestly believe that the "cluster model" of the software process which I first described in the nineties (prompted by ideas from Jean-Marc Nerson, co-founder of Eiffel Software, and Frédérique Sada, who had been using Eiffel in her project at Thomson), particularly in my book "*Object Success*" in 1995 [298] and in more detail in the second edition of OOSC [299], and which Eiffel users have widely applied, provides a more sophisticated and effective guide for how best to reconcile the need for early production of code with the indispensable role of upfront tasks.

---

[39] Revelation: of the exams I have taken in my life, this one was not intellectually the hardest. It was definitely the most expensive; you have to take a workshop, which was in fact worth the high price tag, and pay your certification renewal fees every two years. The business scheme, populating the industry with Scrum Masters who want to recoup the investment and become apologists for Scrum, bringing new recruits to the fold, is brilliant.

- Among the brilliant ideas is the "closed-window rule"; the name is mine, since the agile literature does not have one, but it does have the concept itself. It is the rule that during a project iteration, a "sprint" in Scrum, no one, regardless of rank, is permitted to add functionality. It can only work under two assumptions: the iterations have to be short (a few weeks), otherwise the wait would become intolerable; and there has to be an escape mechanism for truly urgent changes, in the form of the possibility to cancel the sprint and start afresh. The closed-window rule is a powerful stabilizing force for software systems, forcing everyone to think twice about proposing new functionality.

For me, *Agile!* also served as reassurance that I was able to produce short books.

## 8.2 Contract discovery

Among other work performed by the Chair that falls in the "methodology" category, two important early projects, performed with Karine Arnout, explored methodological hypotheses. The first was a kind of sanity check. Eiffel programmers see contracts in every problem. The rest of the world typically does not. Are the contracts just in our minds? In less pleasant words, are we like the psychiatrists in the first scene of *One Fly Over the Cuckoo's Nest*, who turn out, when the asylum's real doctors show up, to be the inmates? To answer that question we turned to the .NET collections library, which for fundamental data structures has classes in one-to-one correspondence with those of EiffelBase, --ARRAY, STACK and such. We looked at the .NET version, and, by analyzing the code, found that, sigh of relief, we are not the crazy ones: the contracts are there, non-Eiffelists just emulate them by various awkward mechanisms such as exceptions or error messages. For example if a structure is full you cannot insert an element into it. That is not a design decision but a constraint. (A "domain" rather than "machine" property in the classic Jackson-Zave requirements work, see e.g. [284].) You can address it in various strange ways, but the simple solution is to have a precondition **not** *is_full* in the insertion procedure. Anything else obfuscates the essential property.

The resulting article [22] [26] attracted some attention, but the reaction took me by surprise. While I thought the study supported the view that one should make contracts an explicit part of the software specification and design process, people took it to mean that they can be extracted from non-contracted code! Tony Hoare, for example, told me that this work was a great first step and that now we should work on a tool to analyze code automatically (our study was a human analysis, although with some tool support) to produce the contracts! The focus seemed to me, as it still does, misplaced: contracts are a methodological tool and a powerful tool in analyzing problems and building software.

Extraction tools can of course help. Partly thanks to Hoare, I became aware in Michael Ernst's Daikon contract extractor [279] and we (particularly Ilinca Ciupa and Nadia Polikarpova, who built an Eiffel front-end for Daikon, CITADEL) used it extensively in our work [108]. But contracts cannot all be inferred: were inference the only source, all the contracts would reflect is the program as it is, bugs included. The aim of contracts, and specifications in general, is to provide a description, separate from the code itself, of what that code is supposed to do. Correctness in soft-ware (as the first lecture in "Software Verification" always recalled) means adequacy to a specification; the specification comes from the code, the code is correct by definition. (All that contract extraction can yield in this case is the discovery that the inferred specification is *inconsistent*, a useful but limited result.) The code becomes defendant, executioner (AutoTest), judge and jury (AutoProof).

Contract inference is, then, inappropriate for the key specification elements of a software system: routine pre- and postconditions and (possibly) class invariants. The programmer should specify what the code is supposed to do. Inference ("what the code should do is what it does") would be useless or pernicious at that level. Where inference is useful is for contract elements that are a technical necessity of the verification process rather than fundamental elements of the specification, particularly loop invariants (the focus of Daikon), which many programmers find tedious to write.

The results of the empirical study reported in [108] confirm this analysis. The study compared contracts written by programmers (in Eiffel code) and those inferred by CITADEL/Daikon. The two categories are largely disjoint.

Loop invariants area fascinating concepts. I have long (at least since 1980 [290]) been convinced that the invariant is the key property behind any loop, the loop itself being just an operational appearance. Over the years I edited algorithm entries in Wikipedia to add the invariant. (For example the entry for Levenshtein distance [304], if only because I only understood the algorithm myself once I was able to figure out the invariant.) The 1980 article described the invariant as a weakened version of the loop's goal (postcondition), and described a few weakening heuristics. Carlo Furia and I took up this work and tried to make it more systematic [131], realizing that there are very few other useful heuristics. The resulting techniques of static invariant inference were implemented but are not yet integrated into AutoProof. Invariant inference, which may use dynamic [216] as well as static techniques, is one of the most important open issues in making full software verification practical (see section 10).

The other kind of invariant, the class invariant, is just as important. Work still in progress [265] attempts to develop a comprehensive yet simple methodology for class invariants in the presence of aliasing and other OO mechanisms.

## 8.3 Contracts do not crosscut

In early work in the Chair, Stephanie Balzer, with the help of Patrick Eugster, examined an assertion often made by the proponents of the then-fashionable methodology of Aspect-Oriented Programming (AOP): that aspects could serve to introduce contracts when the programming language does not support them. That assertion is not central to AOP; it was typically an example among several of what AOP could potentially do. It was common in the aspects literature, though, so it was fair to look into it.

What the analysis found and reported in [53] [40] is that the simulation of contracts by aspects is not practical. Specifically, useful aspects must be "crosscutting", meaning independent of each other. As the paper demonstrates, contracts do not crosscut other important aspects.

The AOP community has ignored the paper and the result.

## 8.3 Pattern componentization

The efforts just described are from recent years. Coming back to the early days, the other joint work with Karine Arnout was on the componentization of design patterns. Patterns [282] are a major advance in programming technology, but from the start I was troubled by their nature: methodological recipes, not reusable components. Much of the pitch for OO (particularly from me [291] [292] [297]) has been based on the prospect of reusing software elements, in the sense of true "as is" reuse. Compared to components, patterns seem like a return to pre-reuse days when, to sort an array you would look up a textbook and program the algorithm. Patterns are at the level of system architecture rather than individual algorithms, but the idea is the same: a pattern is a pedagogical concept (a "Truc" in the sense of section 5.1) For example, the Introduction to Programming course explains the Observer pattern, teaching students to understand and apply the following class architecture[41]:

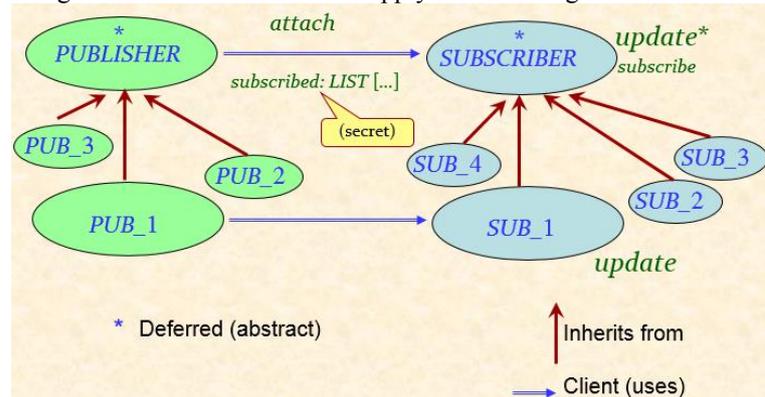

The concept's inventors are adamant that patterns are not components[42].

A component is better than a recipe. The recipe must be learned anew by each programmer and implemented anew for each new program. The component is there once and for all. Could we address the technical limitations and turn the fundamen-tal design patterns, starting from the "classical" ones in [282], into off-the-shelf library components? This was the task set to Arnout and led to the core of her PhD [34]. The answer was largely positive: most of the classical patterns can, under the right conditions, be turned into reusable components, which make up the patterns library that she produced.

---

[40] This paper was not checked carefully enough and has small problems of form; for example some of the program texts are not in correct Eiffel. The analysis itself is solid, however. It had proved its point, so there was no further development of the work.

[41] Extract from a PowerPoint slide for the English version of the course. See section for a simpler solution using Eiffel's agents.

[42] Erich Gamma is based in Zurich, a pleasant coincidence (I once bumped into him at the intermission of *Die Zauberflöte* at the opera house) which provided many opportunities for spirited discussions.

The main condition, self-serving as it sounds, was to use Eiffel constructs. In particular, Eiffel's agents were essential; they had no counterpart in Java at the time, so that our componentization could not have worked in that language. (A 1997 white paper from Sun Microsystems proudly declared that agents/closures/delegates were a terrible idea and that Java would not have them. It took fifteen years for wisdom to set in and closures to come into Java[43].) Contracts, although not formally indispensable, also helped considerably.

The first pattern composition paper, showing the way, was just by me [27] and was an in-depth study of publish-subscribe schemes including a componentization of the Observer pattern, with many other methodology topics along the way. It includes a discussion of where agents/closures fit in the object-oriented approach and complements other OO mechanisms such as polymorphism and dynamic binding. (In [24] Volkan Arslan, Piotr Nienaltowski and Karine Arnout generalized that work, which handled the sequential case of event-driven computation, to a concurrent setup.) It is a sign of the interaction between research and education in our group (and of the whole idea of a research university) that the result immediately became, in suitably adapted form for our *first-semester* Introduction to Programming course; the treatment appears in chapter 18 of the *Touch of Class* textbook. Next two joint papers based on Arnout's elegant work applied the componentization idea to two other fundamental patterns: Factory [54] and Visitor [56].

## 8.4 Multi-requirements

Beyond the simplistic and damageable "user story" techniques of the agile school, everyone in software recognizes the importance of requirements (the proper definition of a system's goals and constraints). In the software engineering community, requirements engineering forms its own community (and has its own IFIP working group). While not exactly part of it, I have always devoted much attention to requirements and have benefitted from numerous discussions with such requirements luminaries as Pamela Zave, Michael Jackson, Axel van Lamsweerde and Martin Glinz.

The object-oriented approach as I see it encompasses the full software lifecycle and is particularly productive at the requirements stage, based not on popular ideas such as use cases, but on the application of the same OO abstraction principles and constructs (classes, inheritance, Design by Contract) as steps of design and implementation. A core idea is "seamlessness", the ability to move smoothly back and forth between all those steps − especially *back*, in the case of change − without having to change concepts and notation. Eiffel intends to be that common notation, accompanying the development throughout.

---

[43] The page used to be at java.sun.com/white/index.html and has (also wisely) been removed. archive.org yields nothing, probably because of the dynamic (JSP) nature of the page. To avoid a copyright violation I am reluctant to make my own snapshot of it publicly available, but anyone interested can ask me for a private copy.

These ideas have been around for a while and are explained in pre-ETH publications. (The EiffelStudio environment provides tool support for them in the form of the Eiffel Information System, developed for a large part by Tao Feng at Eiffel Software, though which one can directly link elements of the Eiffel code and specific sections or paragraphs of a requirements document produced in Microsoft Word, PDF or other formalisms, providing full traceability from requirements to design and code. Further work is in progress through Florian Galinier, Jean-Michel Bruel and Sophie Ebersold at the University of Toulouse.) The new development at ETH is the concept of multirequirements, published in the 2013 Festschrift volume for Glinz [196]. It puts together into a comprehensive framework a number of techniques that my colleagues and I had been applying for a while. The core idea is that different people need different views of requirements, some graphical (UML or often for Eiffel users the BON notation), natural-language (as with a requirements document written in English) and formal. They should be compatible, however, to avoid divergences and inconsistencies. The formal version, expressed in Eiffel according to the seamless method, serves as the reference, since it is the only one with a sufficiently precise semantics. Others can be generated from it or (at least in the case of graphics specifications) translated into it. The presentation of requirements uses a freely interlaced narrative of the three views, a bit in the style of Knuth's Literate Programming [286] although based on a radically different view of programming.

The article has not been particularly well received so far. Part of the reason is its style of presentation: possibly a bit too clever for its own good it uses, without advance explanation, the very approach that it is promoting, using the notion of requirements as its own example "system". The result is disconcerting, if only because you may have to read the article twice to understand it. But the form of the article is not the only obstacle. The concept of seamlessness shocks most people, particularly requirements experts. The idea that a programming language, at least one designed with goals of expressiveness, is perfectly able to express requirements just as well as implementation and design, breaks conventional wisdom and is hard to swallow.

In recent work at Innopolis University, Alexander Naumchev has taken the idea further by adding verification mechanisms thanks to AutoProof (section 10), and better integrating domain constraints.

## 8.5 Other methodology work

Among many other methodology-related developments, three projects are worth evoking:

- The canard that "real programmers don't write contracts" persists, in spite of Chalin's study [276] confirming that programmers given a notation and methodology naturally include contracts in their code. Estler, Nordio, Piccioni and Furia performed an extensive empirical study [222] of millions of lines of code in Eiffel, JML (contract-equipped Java extension) and C# with Microsoft's

"Code Contracts", to analyze precisely how programmers use contracts and. since the study includes revisions, how contracts affect program evolution.

- I performed a semi-rigorous comparative analysis of functional and object-oriented programming. Having done some of my early programming in Lisp, closely followed later developments such as Scheme, Miranda, ML, OCaml and Haskell, often from talks by the designers, I was sensitive to both the attractions and the limitations of the functional approach. The recent re-emergence of functional languages is intriguing. A request to a contribution to a book on "beautiful architectures" provided the opportunity to contrast the functional and object-oriented approaches to architectural design [99]. The chapter is somewhat in the style of my later *Agile!* analysis [228]: trying to sort out "the good, the hype and the ugly". Starting from a published example in an article intended to showcase the benefits of functional programming, it analyzes and how the result lends itself to extension and reuse, in comparison to an object-oriented version. To me the OO approach is the clear winner; in addition it subsumes it thanks to functional-style mechanisms such as agents (7.1), which make it possible to treat functions of any level of abstraction as objects.
- In connection with Ivar Jacobson (the enthusiastic driver of the effort) and Richard Soley, I helped start the Semat project, an attempt to provide a general framework for software projects and their terminology. I co-wrote the original article [112] and organized the initial meetings and Web site (with Carlo Furia) but let others take over afterwards and am no longer active in Semat, although I remain interested in the goal of a comprehensive framework.

## 9 Object persistence and databases

Another line of research lies at the border between our official area of software engineering and another discipline, databases.

The work was hard to start because of fashion effects; as soon as funding agencies saw the words "database" (or "persistence") and "object-oriented" in the same document, the immediate and final reaction was "*been there, done that*": "*it failed, go away, get a life and stop wasting our time*". (Paraphrasing, from various rejection letters.) The effect was just the same on prospective PhD students; one PhD effort, then another, started on the topic, only to stop after a year or so when the students decided to leave for something sexier. Finally, with Manuel Oriol's help I found someone, Marco Piccioni, who cared about the problem, fashionable or not, understood it, and got down to work.

Along with the rest of the programming community, I had in the 90s seen the fanfare announcing the arrival of object-oriented databases, then the quick demise of the idea due (along with the natural reaction to a technology that has been hyped before being ready to deliver) to two factors: the neglect of many issues fundamental to database practitioners, such as transactions, locking, concurrent access control and, more generally, the so-called ACID properties; and the quick moves of Oracle and other relational-database companies, which convinced the market that an incremental evolution of their offerings was a safer bet. The result was the emergence of object-relational interfaces, based on a simple idea (take an object in the OO sense and store it as a row of a table in the relational sense), which turned out to work in practice. The EiffelStore library, based on an idea of Jean-Marc Nerson, was the Eiffel version of this technique, present almost from the beginning. Because OO database advocates had cast the issue as a battle with relational technology, and lost that battle, many people considered the matter solved, explaining the reluctance to fund any research on related topics.

Issues remained, though. Two were of particular relevance: seamlessness and versioning.

"Seamlessness" is a convenient term for the general goal of facilitating the interplay between the OO and relational sides. Right from the start, Eiffel has had a powerful persistent mechanism to store ("serialize") the entire object structure in one instruction (I had discovered the power of such a mechanism when using the SAIL language, which had it, at the Stanford AI lab in the seventies, and would not have wanted to work without it). That mechanism, known as "STORABLE", had several versions, the original one written in C and a more recent one all in Eiffel. There was also the separate EiffelBase object-relational handle. Too many tools. It was urgent to unify everything. My original idea was that one should just program in an OO way and forget about persistence, which would just happen automatically; all that one might want to specify is which objects, occasionally, should not persist. That view may come back at some stage but in the current state of the art it turned out to be naïve. The more immediately useful task was to unify all the models and provide a consistent, sound and practical solution. This is what Piccioni's work did. It is reflected in a publications [93] (based on a student project by Ruihua Jin) and his thesis [187] and just as importantly in library developments such as ABEL.

This part of the work actually had some trouble getting accepted by the department's doctoral committee at the stage of Piccioni's thesis plan; the comment was that it sounded too much engineering and not enough as research. What were the research questions to be addressed? Our first thought was, well, yes, we are a chair of software engineering! Indeed much of our work is applied; the typical PhD thesis from the group does not have many theorems, although some do. That was not a good answer. We soon realized that the doctoral committee (through its chair, Moira Norrie) was right. Even applied work must, if it is applied *research* work, define the research questions clearly. A good software development project, however successful and productive, does not necessarily make a good PhD thesis in applied software engineering. Piccioni's goals were indeed research goals, so we were able to define the research questions clearly and get the plan approved. I am mentioning this episode because it illustrates again where the quality of a top academic institution lies. While the basic ingredient of quality is to hire the right professors and the right students, the process also requires checks at some strategic steps. Many universities that I have seen institute too many of these checks, and, more damagingly, the wrong kind of checks: bureaucratic requirements of providing documents, certificates, notarized signatures, and of collecting signatures of people who are too overwhelmed by administrative requests to give much attention to what they sign. Such checks are proxies for the real ones; they focus on form, not substance. Simplifying the process makes it possible to insert, at one or two crucial steps, a useful check focused on content. The committee did not just check that the forms had been filled properly but read the thesis plan and came back with criticism. What matters is not just that the criticism was justified (had we disagreed, there was room for discussion) but that focus on the substance of the research. No one, by the way, was intruding into the research itself, or the critical advisor-to-PhD-student relationship. The members of the committee, while all professors in the computer science department, were not trying to influence the topic or second-guess the thesis advisor; their task was simply to enforce standards of scientific quality, applicable across all areas of the discipline.

The other persistence-related topic, which did not raise such problems, was versioning. There is little discussion of this topic in the literature, aside from my own analysis in Object-Oriented Software Construction (chapter 31 of [299]), which in my opinion (in line with the general absence of any pretense at modesty in this article) still provides the appropriate conceptual framework. Popular or not in the literature, the problem is fundamental in the practice of software development. Assume you are a major bank and have 100 million "bank account" objects, instances of the class ACCOUNT, stored in your database. A programmer adds or removes an attribute (field) in the class. What happens to the stored objects?

My guess is that most people use some ad hoc solution, for example adding a field with a default value. But if you have any concern about correctness, that is dangerous. If the change was (as a simplistic example) to add a "balance" attribute, whereas account objects until now only had lists of deposits and withdrawals (from which the balance could be computed, but now we want to store it at all times in the object), then 0 is not a proper value for the fields added to the existing objects. It is in fact very wrong. The right value is the sum of deposits minus withdrawals.

The versioning work produced two papers, [114] (incorporating the results of a student project by Teseo Schneider) and [188], as well as an IDE extension, ESCHER, to visualize and control what happens to objects over successive evolutions of the software.

## 10 Verification, static and dynamic

Verification may in this article have waited until section 10, but it was always (and for me, long before ETH) at the center of our attention.

Various subdisciplines have different ways of understanding the term "verification". We will use it in the most general sense, covering both parts of what in the software engineering community is often called Verification and Validation or V&V, with the first component assessing the "how" and the second the "what". In the academic programming community verification has come to mean proofs, while for people in industry it basically evokes tests; we are interested in both.

### 10.1 The vision: Verification As a Matter Of Course

Many research teams interested in verification focus on one approach, such as model checking, or testing, or Hoare-style proofs. Our view is that the problem is hard enough to justify summoning any idea that helps. The broad spectrum of our effort lead to the slogan "Verification As a Matter Of Course" or VAMOC, with two key ideas:

- Unlike some heavy-duty approaches based on formal specifications and proofs, do not try to impose a radically new software process and tools, but let developers work in the context of modern software development, with a good IDE and a language offering all that makes it worthwhile to get out of bed in the morning and go to work: classes, single and multiple inheritance, dynamic binding and the like. Hence "As a Matter Of Course": the process should be to develop software normally, adding a verification component.
- The verification machinery should support verification in a discreet way, applying different tools, some dynamic (tests), some static (static analysis, proofs) to comb the code for bugs, filtering them to avoid overwhelming programmers with information, and reporting them.

This VAMOC vision could also be called "verification for the people".

It is close to Tony Hoare's "Verifying Compiler Grand Challenge" [283] and was influenced by it. In the early 2000s, Hoare was agitating for launching a large international cooperative project, in the style of human genome sequencing, to produce a compiler that would also verify the code it compiles and, if successful, guarantee the correctness of the result. No funding agency launched such a grand project. Hoare's talks encouraged verification research, however, and our effort is a direct attempt to implement the Grand Challenge.

We often used, to explain the VAMOC idea of complementary techniques, a metaphor apposite to Switzerland. The Gotthard tunnel connects the (sometimes rainy) German-speaking North to the (sometimes sunny) Italian-speaking South. At the time of construction in 1882, the two drilling teams met and held a big celebration. In just the same way, the EVE environment presented next connects dynamic and static techniques of verification. It is not just that we use both, but that we use them together; some of the ideas we developed, such as model-based specifications (section 10.9) are just as useful for one as for the other; [153] describes some of that interaction.

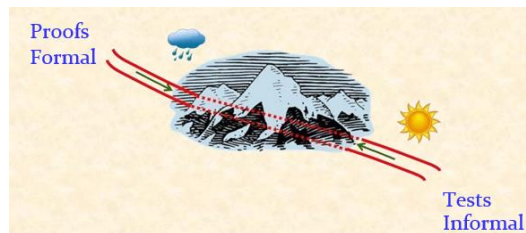

The following picture illustrates the VAMOC and EVE scheme.

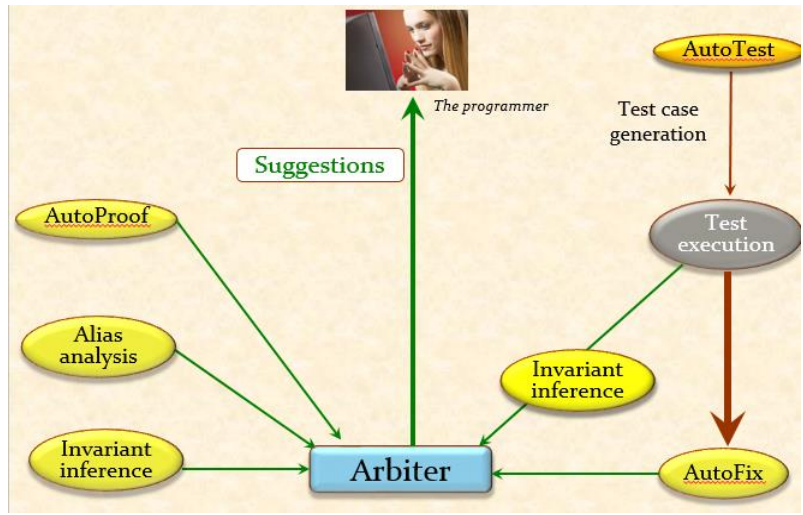

We applied both the static techniques on the left and the dynamic ones on the right. As explained in [256], the idea of the "arbiter" is to assess how significant a verification result. There is a full scale from -1 (demonstrably incorrect) to +1 (demonstrably correct). A successful proof yields +1 if the prover is sound (AutoProof comes close), a failing test -1. With a complete prover, a failed proof would yield -1 (no sound prover can be complete because of undecidability results). A passing test slightly increases trust.

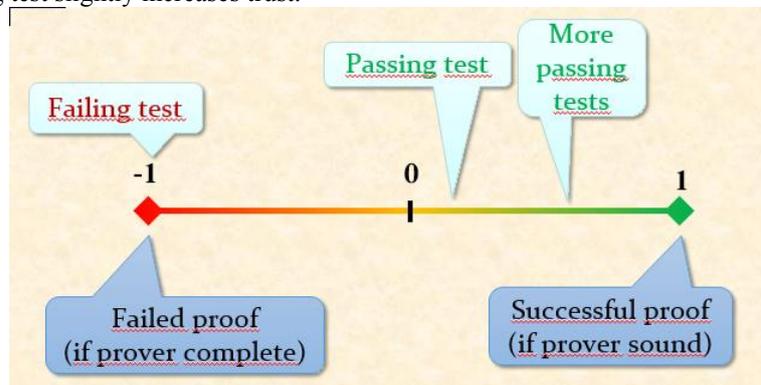

The arbiter idea is the future of verification in the VAMOC style. It was implemented, although not yet included in any released version of the environment.

## 10.2 EVE

What does "released version" mean? For a Chair of software engineering the results include not only papers, the traditional academic artifacts, but programs. Where do these programs go? Most teams produce research prototypes; that is what we did too, until I realized that the approach was self-defeating. Typically, the prototype would work well enough for a demo in the master's project presentation or the PhD defense, and the day after, or when the student left, there was no usable result. The code was in some repository, but it no longer compiled with the current version of the compiler, or it had a dependency, documented or not, on some external software product that was no longer available, or some piece was missing. It was hard or impossible to help someone who wanted to use the tool after seeing an impressive demo or reading the published paper; another student asked to take up the work would waste time getting it back into shape.

I made the decision to forbid research prototypes. No exceptions: all software had to go into a single base product. Having become open-source (section 3), EiffelStudio could have been that product, but that approach was not realistic: EiffelStudio, whether closed- or open-source, is a production tool used by companies for mission-critical projects, and undergoes a carefully managed release process. It would not be acceptable to endanger it with the latest results of research, where creativity is encouraged but can lead to mistakes.

The solution was to create a research branch of EiffelStudio, called EVE, the Eiffel Verification Environment. In general, branching is a terrible idea in software engineering [228], leading to the merge nightmare evoked in section 6, so we had to devise the process carefully. Our process would not affect EiffelStudio: EVE would be free to include all the additions and adaptations we wanted, not interfering with the production branch, but it had to compile and run under its current version: they did not have to care about our work, but we had to be adapt it to theirs.

While the idea sounds simple, there had been so much divergence in the Chair's early developments that it took a year and a half for Yi (Jason) Wei to implement the EVE process and integrate the major software results produced thus far, in particular the early versions of AutoTest. Once all was in place, however, keeping EVE in sync became a well-oiled process, successively handled by Wei, Tschannen and Schill, of reconciling the EVE branch with the EiffelStudio "trunk" every Friday afternoon before leaving for the week-end.

Some of the EVE tools, notably AutoTest, migrated to EiffelStudio, after undergoing productization (what is good for research and student use is usually not good enough yet for production use). Other tools, notably AutoProof, have not crossed over yet.

## 10.3 AutoTest

Eiffel has always used contract elements (preconditions, postconditions, class invariants and others) as tools for testing. By enabling run-time contract monitoring, programmers find bugs much faster, and much closer to the source, than with the usual methods of programming. It has long been for me a source of bewilderment, and continues to be, why the whole world has not switched to this mode of developing software[44].

---

[44] It is also fair (even if grossly vain) to mention that I can hardly give a conference talk in a new place, however exotic, without seeing some locals come to me at the end to tell me that discovering these ideas from my books and articles changed their lives.

The ideas (the root of which comes from verification work in the seventies — Wirth and Hoare's Algol W already had an ASSERT instruction) have spread in part to other approaches, from testing-oriented assertions in JUnit to the Java Modeling Language and Microsoft's Code Contracts, exposing a wide audience to some of the benefits. The full idea of Design by Contract, however, treats the corresponding constructs as a built-in part of the programming language, method, requirements process (see section 8.5) , implementation process, proof process when available (as discussed below), and, most relevant for this discussion, testing process.

These elements are not new. The new question was: how much can we automate the test process further? "Test automation" came into vogue with JUnit and its predecessor "XUnit" tools. But they only automate one part of the testing process: running the tests. This step was a significant advance, but it left to manual work two of the most labor-intensive and tedious parts of testing:

A. Generating test cases.

B. Generating test oracles, the test success criteria. You can run a million tests and learn nothing if you do not determine which ones succeeded and which failed.

The impetus to innovate in this area came from a suggestion by Xavier Rousselot, during a discussion with Emmanuel Stapf and me when he was finishing his student internship at Eiffel Software in Santa Barbara in June of 2003. His idea, which we found to be more suited to an academic research project at ETH than for a commercial development, was simple: use classes and their routines for A, and contracts for B. Then we can have fully automatic testing.

After a proof of concept by Karine Arnout under the name "Test Wizard", the full-throttle AutoTest project started for good, involving in particular the PhD theses of Ilinca Ciupa, Andreas Leitner and Yi (Jason Wei), with key supervision by Manuel Oriol and help from Stapf and others at Eiffel Software[45].

A good summary of AutoTest appears in an IEEE Computer article, "Programs that Test Themselves" [113]. Testing-related references in the bibliography of this article include [55] [63] [64] [65] [67] [71] [81] [82] [83] [92] [97] [98] [108] [113] [123] [149] [156] [164] [186] [189] [193], plus the bug-fixing references cited below in section 10.6.

The core ideas behind AutoTest are:

- Add automatic testing (in the sense outlined above) to a basic XUnit-style framework. In other words, manual tests remain; automatic tests are a complement, not a replacement. Our empirical work [64] [65] shows that they are complementary in uncovering bugs.

- Generate objects by using the constructors of the classes under test.

- Generate tests by calling routines on objects.

- To select objects on which to call routines, alternate between newly created objects and objects from a pool resulting from previous creations and routine calls. The pool is necessary to avoid working only with young objects. For example if you are testing operations on a list class, you want to make sure that some of the tests will use big lists, which can only result from repeatedly calling insertion routines on previously created lists.

---

[45] Microsoft Research's PEX, by Bjorner and de Halleux, is an elegant effort starting from some of the same ideas and combining them with symbolic execution. It is now part of Visual Studio under the name IntelliTest.

- To select arguments for routines, use various heuristics for basic types (for example, in the case of integer arguments, 0, +1, -1, MAXINT, MININT and so on), and for object (reference) arguments use the pool, with heuristics such as object distance discussed below.

- If a precondition is violated in a call by AutoTest, skip the call: this is just AutoTest trying to call a routine in a case that the routine's specification has explicitly ruled out, so it is just wasting CPU cycles.

- Minimize such waste. This is the whole matter of optimizing object selection for precondition satisfaction, which Wei in particular studied in depth with Manuel Oriol [123] and implemented in AutoTest.

- Shoot for postcondition violations. They are the real prize! If AutoTest calls a routine with its precondition satisfied and, on exit, the postcondition fails, we have hit Bingo. Also precondition violations, in routines called by other routines (not directly by AutoTest), and invariant violations.

The implementation of these ideas took several years and PhD theses, supported by many auxiliary efforts. A sign of success is that today the essential elements of AutoTest found their way into the standard delivery of open-source EiffelStudio. An even more concrete sign is the number of bugs that AutoTest uncovered in an entirely automatic fashion. Some of these bugs were in EiffelBase and other widely used libraries where they had laid dormant for many years[46].

EiffelBase was indeed our major practicing ground for AutoTest, and for AutoFix as discussed below. Possibly too much; I yearned for more testbeds, but one of the limitations of AutoTest in its current form is that it works best with mostly self-contained software. If you have too many dependencies on external stuff, as in an application library, you need some way to give AutoTest information about it. We had ideas on how to address this issue, but ran out of time before we could try them.

The other main issue with current AutoTest is performance. For reliability and resilience (the testing tool should not fail when a test fails, since failing tests is indeed the whole purpose of the game!) the tool uses a multi-process architecture, which penalizes speed. In current uses of AutoTest, which we call "test while you lunch", this is not a major limitation: periodically, start an AutoTest session on a number of classes, go to lunch, and when you come back the tool has found the bugs for you. In the VAMOC spirit of a guardian-angel kind of tool that is constantly at work, behind the scenes, to dissect your code and alert you to potential problem, the speed is typically not sufficient for such interactive feedback. This is largely an engineering problem, but we have not yet been able to get to it.

---

[46] About one of our testing papers relying on EiffelBase, an anonymous referee commented that he did not believe the results because he had never seen production code with so many bugs. Perhaps EiffelBase was indeed substandard. Or perhaps not (as I suspect). Without a systematic application of such techniques to other software, which also assumes that it is *specified* by extensive contracts such as those of EiffelBase. So we do not know for sure. The paper was accepted.

## 10.4 CDD

Complementing the basic AutoTest ideas discussed above is the concept that Andreas Leitner proposed under the name "Contract-Driven Development" or CDD [71], a response to the Test-Driven Development (TDD) of Extreme Programming. In TDD writing code follows from writing individual tests; the idea of CDD is to go one level of abstraction higher by using specifications (contracts) instead of tests. A central insight was the following observation on the debugging process. Typically, you are trying out your program interactively; you hit a bug:

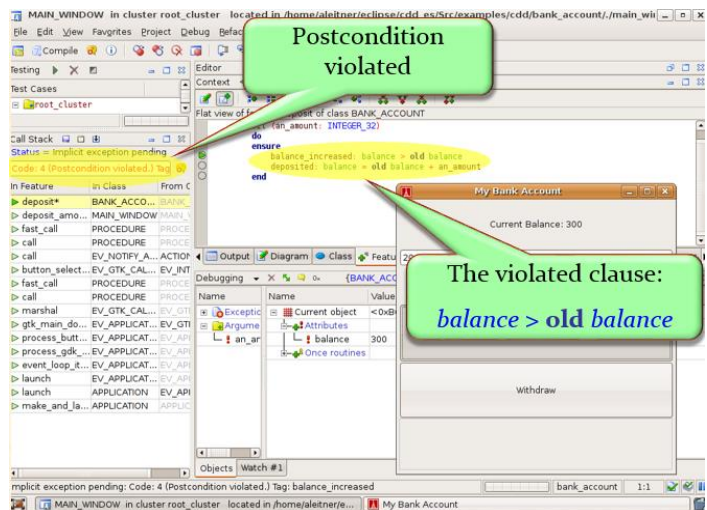

What happens next? You find out what was wrong, you fix the bug, and move on with your program and your life. But that is a pity: a piece of wisdom that the software engineering community has learned over the years is that bugs should remain part of your life, like the misdeeds of your youth of which your great-aunt never fails to remind everyone in every family reunion. The software engineering notion is "regression suite": the collection of useful tests, to be run for each new iteration of the code. It is particularly important for the regression suite to include every test that has failed at any time in the history of the project, because of the particular software problem of "regression": old bugs coming back. But in the informal interactive approach to debugging suggested above, the failure occurred during an interactive session, not as part of a formally recorded test.

The "Test Extraction" mechanism of AutoTest, following from the CDD analysis, automatically generates a test from a failure in such a case. Going from the idea to its realization required addressing a number of delicate problems: every execution is different, and the program evolves, so we have to do our best to recreate the original conditions when possible. Test Extraction is part of the research version of AutoTest but not yet of the production version in EiffelStudio (largely because of these difficulties).

## 10.5 General testing results

The work on AutoTest, and its systematic validation on numerous libraries and other programs, moved our group's work firmly into empirical software engineering, and enabled us to address many general issues of testing, far beyond the application to AutoTest and its target language. Here is a small selection of the insights we gained.

First, I should mention a principle that we always applied to our testing research: we never once seeded a single bug. Most empirical testing research works on software into which faults have artificially been inserted ("seeded", or "injected"). Although research by Lionel Briand provides some empirical justification for fault injection, this approach always felt wrong to me, since it is difficult to guarantee that the injected bugs are representative of real ones. We did not need to inject anything since we were working with real software (mostly, Eiffel libraries) with a 25-year record of its evolution including all actual bug reports (in the EiffelStudio repository) and which in some cases still had some bugs. It is always better to try your ideas on the real thing rather than a mockup.

Our empirical studies derived several non-panacea (or "no silver bullet") results. For example [92] compared "Finding Faults: Manual Testing vs. Random+ Testing vs. User Reports" (the title), respectively using manual test results from the repository of EiffelStudio evolution (a rich source of empirical software engineering data, going back to the early nineties), and early version of AutoTest, and bug reports from user in the fields. It found that they uncover largely complementary kinds of faults (bugs). For example AutoTest systematically tries extreme values such as MAXINT but human testers almost never do.

Another seemingly disappointing result comes from early work, performed in part by Raluca Borca-Muresan, a student intern from Romania [65]. The idea was to look into failures found by AutoTest and decide, on the basis of human analysis, what the source of the fault is in each case. The study found that a substantial proportion of contract violations reflect bugs in the contracts rather than the code. On the surface, this results seems to provide an argument against Design by Contract: why bother adding assertions if some of them are going to pollute code that was OK to start with? The argument does not hold, however: "OK" is a subjective assessment, which can only be based on some higher-level human view of what goal the code was truly trying to achieve. If you cannot specify that goal precisely, there is something wrong, and you will not be able to verify the code, since verification means comparing implementation to a specification. The more directly actionable result of the study is that (as anyone who has tried his hand at formal specification knows) writing contracts is hard too, even if not of exactly the same kind of difficulty as writing code. [65] is a short paper and the study was limited; it would deserve to be run again, with the benefit of recent empirical work on the practical use of contracts such as [222].

Yet another comparative study of different kinds of contract leading to a complementarity, no-panacea result is [108]. In this case, the techniques found to be complementary are contracts written manually and those that, in the light of our experience with DAIKON, can be inferred automatically using dynamic techniques.

The search for effective testing techniques in AutoTest led us to improve existing techniques of "Adaptive Random Testing". ART [277] (also the subject of scrutiny by Briand, with Arcuri) uses values selected largely at random but with some bias so as to provide reasonable coverage of the input domain (for example, in the case of two integers, the four positive/negative combinations). Ilinca Ciupa came up with the idea of "object distance" to make ART possible not just for basic values but for objects as well. The idea, described under the acronym ARTOO in an ICSE 2008 paper written with Andreas Leitner and Manuel Oriol [83][47], is to spread objects too, the way ART spreads for example integers, by maximizing their distance, defined through a number of parameters of both the corresponding classes (for example how far apart they are in the inheritance structure) and the field values in the objects themselves. Our studies did evidence the benefits of applying ARTOO to vary the objects in the pool.

Another insight that I learned largely from Ciupa is about the proper focus in testing research. Many articles we saw proclaimed victory when they could evidence a shorter "time to first bug" than the previous method. This criteria is not the right one for two reasons:

- We need to find all bugs, not just the first. Maybe the first bug comes very quickly and the second bug will take two hours.
- Many of the strategies that are so good at finding the first bug get there thanks to a complex setup. But the time for that setup may eclipse the actual testing time! Any objective results must take everything into account, setup time and testing time.

These observations led us to a strict discipline in our testing work. We learned not to fall in love with our own brilliant ideas of testing strategies. What counts is not how convincing the idea sounds on paper, but how many bugs it will find in how long. We decided that this criterion − B (t), the number of bugs (real ones, not seeded faults) found after t seconds of test execution, setup and teardown time included −would be the only one we applied to our empirical testing work.

Another lesson we learned, with initial insight coming from Manuel Oriol, was had to do with testing coverage. Industry mostly uses branch coverage as its estimate of how extensively a program has been tested. A common rule of thumb is "release at 80% coverage". Common in theory at least; in projects that I have seen the threshold can be as low as 30% or even 20%. But even a higher threshold is, in light of our experiments, not that promising. We had the software to test (Eiffel libraries); we had the bugs (we often used older versions from the repository, with many faults corrected later); we had the completely automatic technology (AutoTest); and we managed to acquire the computing power: a good dozen servers that, for a while, did nothing else than run AutoTest of code, day and night. This infrastructure was necessary in part because AutoTest relies on a number of heuristics, such as how many values to use in ART and what proportion of objects to create versus reusing objects from the pool, so we also used the process to optimize these parameters experimentally.

---

[47] I told Ciupa that the paper was not ready for submission. She sent it to ICSE anyway. At the time of writing it has 177 citations on Google Scholar.

The following experiment results by Oriol and Yi Wei, published in "Is Coverage a Good Measure of Testing Effectiveness?"[164], show coverage as the gray line and cumulative bugs found — the "B (t)" mentioned earlier — as the thick line. The various figures correspond to different classes in EiffelBase.

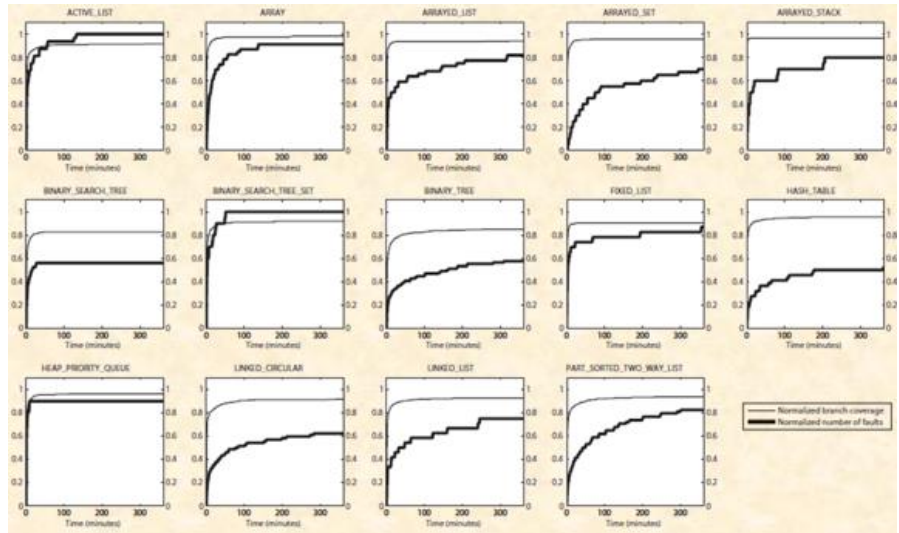

The picture is striking: in all cases the tests, all generated automatically by AutoTest (there were no manual tests) reach high coverage quickly, sometimes 100% coverage very quickly. But in most cases the process continues to find new bugs!

With all the due reservations on the specificities of the study, it seems to provides strong evidence that coverage is not good enough.

If coverage does not provide the answer, what information can industry projects use to address the crucial practical problem of when to stop testing? This study and my experience suggest that the best predictor of remaining bugs is past bugs. Looking at the charts above, I think that if I were following the results of AutoTest for any of the given classes I would have a good intuition for when the likelihood of remaining bugs has reached a low enough level.

A precise, actionable answer to this question would be of extraordinary interest to software practitioners, particularly managers, for whom the question of when it is safe to stop testing is paramount. The same team of Wei, Oriol and me, with Carlo Furia as well as Andrei Tikhomirov from ITMO took the work further, generalizing the empirical basis by including results from Yeti, an AutoTest-style tool devised by Oriol at ABB after he left our Chair [189]. It got it closer to the magical formula, but work remains.

Testing was largely a new research area for me when I came to ETH. As noted, it played an essential role in introducing our group to empirical research. My "Seven Principles of Software Testing" in IEEE Computer [89] (see also a follow-up [109] in response to reader comments) distills some of the theoretical and practical wisdom that our work (up until 2008) on this key area of software engineering,

## 10.6 Automatic bug fixing

Towards the end of his thesis work on AutoTest, Andreas Leitner came up with an intriguing suggestion: move on to not only detecting bugs but proposing fixes, again building on the Eiffel advantage of contracts, which give a specification of the intended semantics. He also discussed the concepts with Andreas Zeller in Saarbrucken, who was keenly interested in the topic. We soon started parallel connected projects (using provided by the German and Swiss national research agencies for connected projects in both countries). Yi Wei devoted a substantial part of his thesis work to the tool naturally called AutoFix, and Pei (Max) Yu, who joined us shortly after, almost all of his. Martin Nordio and Carlo Furia also played important parts, and we worked in close collaboration with Zeller and his student Valentin Dallmeier. AutoFix publications include [115] [128] [154] [155] [215][237][246].

AutoFix, finding a bug through AutoTest, suggests fixes based on a comparison of the passing and failing states. The aim is not just to find a fix, which by itself is trivial (just add an "if" instruction to yield the expected answer in the so far erroneous case) but a high-quality answer. The ultimate criterion of quality is, in conformance to our cardinal rule of working with real bugs on real software, whether the fix suggested first by AutoFix will match the fix that the human programmer actually applied. In tests with earlier versions of EiffelBase from repository, AutoFix met this goal in an impressive 42% of cases.

We were not the only ones to get into automatic bug repair; this new research field was born and quickly developed just as we got our first results. We wanted to ascertain these results and did not rush to publication; others who did and received visibility had to retract some of their claims later on.

AutoFix suffers from the same engineering need as AutoTest, mostly in fact because of AutoTest and not of the AutoFix process. This issue must be addressed before AutoFix can fit the VAMOC vision of identifying problems in real time while a programmer is developing code, and providing immediate feedback, in this case a suggested fix. But even without this interactive capability AutoFix is a significant advance, built on solid ideas and validated through meaningful examples, and a potential basis for significantly improving the programmer experience.

## 10.7 Proofs: the road to AutoProof

Using contracts as dynamic test oracles is great, but the idea and ideal had always been that they should lead to correctness proofs. With the progress of proof technology, this ideal is within reach. Our main tool towards it is AutoProof [127] [132] [151] [153] [177] [186] [193] [194] [221] [230] [236] [239] [256] [257] [259] (autoproof.org [273]) is the regularly updated Web site).

We never attempted to develop a proof engine. The reasoning was that provers are a specialized technology and that others with expertise in the field were already working hard on proof tools such as Isabelle, PVS and Coq. Our expertise was on the programming side; we should rely on their efforts.

The first attempt at using an existing prover failed. The idea behind bringing the Event-B project to ETH was to combine B and particularly its powerful prover with object-oriented development. It takes two to tango; the Rodin team had its own

| Feature | Line | Result |
|---|---|---|
| ACCOUNT (invariant admissibility) | | Successfully verified. |
| ANY.default_create (creator, inherited by ACCOUNT) | | Successfully verified. |
| ACCOUNT.deposit | | Successfully verified. |
| ACCOUNT.withdraw | | Successfully verified. |
| ACCOUNT.transfer | 49 | Postcondition balance_decreased may be violated. |

interests and the B prover remained bound to its original goal, supporting the refinement-based B process. Object-oriented development is more geared to a bottom-up process which favors reuse and change [291] [292] [299]. I believed that it would be possible to overcome this obstacle[48] and bring the two approaches together but that did not happen.

The search for a prover continued; one of the benefits of hosting so many seminars, conferences and summer schools is that one gets exposed to many good ideas and developments. One of them was the work on verification done at Microsoft Research, by Rustan Leino together with Mike Barnett, Wolfram Schulte and others on the Spec# language. Spec# was an extension of C# with Eiffel-like contracts, in the line of earlier work on ESC/Java at Digital's research center in Palo Alto and, like ESC/Java, incorporating a proof tool. Spec# as a language did not have a long life because it was geared to a particular version of the C# language and did not follow later evolutions of C#. But the prover, Boogie, turned out to be remarkably successful thanks to an API design that did not tie it to the source language, defining instead an Intermediate Verification Language (IVL) which verification systems for languages other than C# can target. Bernd Schoeller, supported by Peter Müller who knew Boogie well, was the first in our group to point out the suitability of Boogie for verifying Eiffel. Boogie is the prover at the heart of AutoProof, as built in the following years by Julian Tschannen, Nadia Polikarpova with the help of Carlo Furia and Martin Nordio.

## 10.8 AutoProof

The best way to get familiar with AutoProof is to go to the autoproof.org site and try the tool online. Thanks to the online programming tools presented in section 5.3, AutoProof is available there, serving as a tutorial but also simply as an online version of the tool. The tutorial includes examples of increasing difficulty. One of the first is a bank account class which I have often used in talks. In trying to verify the original text, AutoProof proves the correctness of the routines *deposit* and *withdraw* but fails on *transfer*:

---

[48] I owe to Michael Butler the observation that in a B-like refinement process the *invariants* can serve as the basic guiding tool for handling change.

If I ask the audience what is wrong, experts in verification will usually spot the problem quickly, but many times an audience of even experienced OO programmers goes blank. The routine is written as follows[49]:

```
transfer (sum: INTEGER; other: ACCOUNT)
            -- Move `sum' from Current account to `other'
    require
            sum <= balance
            other /= Void                    -- The other account exists
    do
            withdraw (amount)
            other.deposit (amount)
    ensure
            balance_decreased: balance = old balance - amount
            other_ increased: other.balance = old other.balance + amount

    end
```

The implementation and specification seem right at first (you are invited to ponder them before reading on). The reason AutoProof cannot prove its correctness is that in fact it is not correct: if the `other' account is the same as the current object (Current in Eiffel, "this" in Java etc.), the first postcondition clause will not hold. Once spotted, the problem is easy to fix: either add the precondition clause other /= Current or adapt the implementation so that it does nothing when other = Current. This example is illuminating since it illustrates a type of bug that is both:

- Very possible to produce in the practice of program writing: when writing the transfer operation, a programmer typically thinks by default about the standard case of transferring to a different account (why would one transfer money to the same account?). The experience of AutoProof talks — when no one in an audience of experienced programmers sees the problem after looking at the code for a few minutes — supports this conjecture.
- Very possible to arise in the practice of program execution: sooner or later, the same-account case may occur.
- Often hard to debug: such wrong run-time effects typically manifest themselves rarely (like concurrency bugs) and may remain undetected for a long time.

AutoProof finds it right away in its attempt to prove the correctness of the class.

One of the characteristics of AutoProof is that it covers the full Eiffel language. Other successful verification tools, such as Spark Ada, limit the programmer to a restricted programming language, or, as in the case of Event-B, a specific mathematical notation, not benefitting from advances of modern programming languages, and translated to an actual programming language only at the very end of the process. The VAMOC idea is to let the programmer benefit from the full power of a powerful OO language, but then it means the tools must support all of it. In such efforts the last mile − the last constructs to be equipped with a formal semantics and the corresponding verification rules − are among the hardest. Work on agents [127] and exceptions is typical; see also [177] and [239]. Because a programming language is an evolving target it is hard to guarantee 100% coverage, but AutoProof handles all the constructs that Eiffel programmers use in practice.

---

[49] With void safety (section 10.2) this second precondition clause is not necessary. AutoProof does not yet take advantage of the void-safe mechanism.

AutoProof handles full correctness: proving that a program meets a certain specification. Other forms of verification, such as model checking and abstract interpretation, do not require writing such a specification, but only provide specific results, in the form of a guarantee that the program will not cause certain undesired event. Full correctness means that we have to provide more annotations, in the form of contracts, but then the prover guarantees what the program actually does.

In the VAMOC spirit of integrating every bit that helps we also used model checking, particularly for concurrency under the influence of Chris Poskitt [258] [266]. Another technique complementary to the Hoare-style framework, not used in AutoProof, was separation logic in Stephan van Staden's and Cristiano Calcagno's work [126] [214].

Object-oriented programming, with its rich run-time object structures, raises delicate verification problems, particularly in connection with aliasing and the semantics of invariants. A considerable literature exists on these issues, proposing many solutions such as "ownership types", but not converging on any single one that has been widely accepted. Started by Nadia Polikarpova, the "semantic collaboration" approach [221] [236] attempts to unify and simplify the best of these ideas and is the basis for the handling of invariants in today's AutoProof. I find the concepts still too difficult for daily application by ordinary programmers, in the VAMOC style, and am working on a simpler solution [265], but it is for future versions of the tool.

The most visible achievement of AutoProof is Polikarpova's complete correctness proof of the EiffelBase 2 library [257]. EiffelBase is the standard Eiffel collections library (arrays, stacks, queues and other "Knuthware"). It soon turned out that taking the original EiffelBase and verifying it would not work. EiffelBase 2 is a rewrite, largely but not fully API-compatible with it and using a simpler inheritance structure. It covers the same ground, and is used instead of the original EiffelBase, for example, in the Traffic library (section 4.5). It may one day completely replace EiffelBase.

The verification of EiffelBase 2 shows the power of the AutoProof approach. The library is sophisticated software, using the most advanced language mechanisms and dealing with complex data structures. The result also shows the limits of the approach: it required many months of work by experts in the field (mainly Polikarpova), in fact by the very builders of the technology. Some of their experience can be taught to others, but generalization to the programmer community at large, in the VAMOC spirit, requires further simplification. That is the aim of my current work mentioned above.

These observations detract in no way from my view that the mechanized proof of EiffelBase 2 is a milestone in the history of verification and more generally of the quest for quality software.

### 10.9 Model-based specifications

Proofs of full correctness require a specification of the full behavior. The basis for our specifications is the notion of contract as expressed by the preconditions, postconditions, class invariants, loop invariants and "check" instructions that Eiffel programmers include in their code. Are contracts enough?

Potentially yes, but not in the way they were traditionally used. A typical postcondition for a "push (x)" operation in a STACK class states

       top = x                     -- The top element is now x
                                   -- (Last-In-First-Out properties of stacks)
       count = **old** count + 1       -- One more element than before

which is sound but leaves out the requirement that the push operation should not mess up with any of the existing stack elements. A devious implementer working from this specification could change any of those elements. A "full specification" is one that, in such cases, expresses all relevant properties, not just the most interesting ones.

The solution we devised, first published in [49] and further developed in [132] and [193], uses "model queries". Its idea is to associate with the abstraction defined by a class a mathematical interpretation, or "model", and express specifications, when needed, in terms of that model. The model can be a set of features already present in the class; for example, we may select the attributes "owner" and "balance" (and possibly others) to form the model of a class BANK_ACCOUNT, stating that they capture the essential semantics of the concept. In other cases, the model queries have to be added to the class. For STACK we use as a model the sequence of values in the stack, from the top down; we introduce in the class a new attribute

       model: MML_SEQUENCE [T]

where T is the type of the stack elements. Such model attributes or functions do not need to penalize the implementation in space or time: they can be declared as "ghost" features relevant only to the verification. (Model features are simply features marked "model" in a "note" clause of the class.) The postcondition of push (x) becomes simply

       model = <x> + **old** model

in other words, the new sequence after a push is the concatenation ("+") of the single-element sequence <x> and the previous sequence. The properties of the traditional specification, expressed in terms of "top" and "count", become consequences of this more general one (theorems), deduced from the properties of sequences and the model-based specification of "top" as the first element of "model" and "count" as the size of "model".

Model queries solve the problem of full specification and enable AutoProof to fulfill its goal of full correctness proofs. They also had a substantial effect on the dynamic (AutoTest) side of our verification effort, enabling AutoTest, as reported in [193], to uncover significantly more bugs.

### 10.10 Automatic alias analysis

Some of the practical difficulties of verification mentioned above arise from aliasing: the possibility for two reference variables (pointers in some languages) to become attached to the same object at run time. The case other = Current in the `transfer' example was an example. In that case the problem has an easy solution: disallow aliasing through the precondition, or handle that special case in the implementation. But often, with complex data structures, aliasing is inevitable. For example in a circular linked list the first element is aliased by construction, pointed to both by the list header and by the last element.

Semantic collaboration and ownership, used in AutoProof, address the problem, but at the price of added complexity. Another technique that has attracted attention is separation logic, to which we were exposed in depth at the 2008 LASER summer school, catching Stephan van Staden's attention. The resulting work [126] [214] does an excellent job of applying separation logic to object-oriented programming. It was important to perform that job, but my personal conclusion from it is that separation logic is not sustainable in a practical approach to verification, particularly in the VAMOC spirit of verification for the people. The extra annotation effort (specifying that certain parts of the object structure are disjoint from each other) is just too formidable. It is also, in my opinion, too low-level, forcing programmers to discover and express properties of the run-time structure (often called "the heap" in separation logic, as if to reinforce the implementation-oriented nature of the approach) from which object-oriented programming, in its effort at abstraction, normally shields them.

Uncovering these properties should, in my view, be the business of the implementation rather than the programmer. Based on this idea, I devised an *alias calculus* [122] [147] [241] complemented by a *change calculus* (see the last of these reference). The alias calculus is a set of rules defining a transfer function on programs, though which it is possible to compute the "alias relation" at any program point. The alias relation is the set of pairs of expressions (of reference types) that *could* be aliased to each other (have values that point to the same object) when an execution of the program reaches that program point. For example, after the assignment ref1 := ref2, the pair [ref1 , ref1] should be in the alias relation. The calculus should yield an alias relation that is:

- **Sound**: if there is any execution for which e and f could be aliased at the given program point, the relation should include [e, f].
- **Precise** to the extent possible: avoid including in the relation any pair [e, f] which no execution can alias. Since it is impossible for a general-purpose program analysis to satisfy both soundness and perfect precision (the problem is undecidable), precision is an optimization criterion: we try to get the best precision we can. In other words it is a criterion of realism: an analysis that puts all expression pairs in the alias relations is sound, but useless.

The rules of the alias calculus yield, for every construct of the language, the induced transformation » of the alias relation. For example the rule for sequential composition, as expressed by the language operator ";", is

a » (p ; q) = (a » p) » q

where the right side expresses the alias relation that will hold if the program executes `p ; q`, that is to say p then q, from a state in which the alias relation is `a`.

The alias calculus complements standard Hoare-style reasoning with assertions: it can inform it, by deducing properties of the form e /= f, and can be informed by it. Aliasing is at the center of many problems in verification, in particular:

- Providing a simple verification technique for class invariants.
- Deadlock analysis in concurrent programming, particularly in the SCOOP context as described in [244].
- Frame analysis, as discussed below.

I wrote a first implementation of the alias calculus to test the ideas. A more serious implementation in EVE was then produced by Alexander Kogtenkov. A conceptual problem delayed the progress of the implementation. My original papers [122] [147] described both alias relations and "alias diagrams", intended as a graphical illustration. Sergey Velder from ITMO, who was collaborating on the topic (and Alexander Gerasimov also then from our ITMO lab) argued that it should be used as the mathematical model as well. I resisted the idea because of efficiency concerns: with diagrams, a conditional instruction, and any loop iteration, seems to require duplicating the entire diagram. But the alias relation approach has problems too. In particular, the relation can be infinite: with a loop scheme such as

    **from** a := first **loop** … ; a := a.next **end**

(common on linked lists), a can become aliased to first, first.next, first.next.next and so on. The papers handled this issue through the notion of "dot completeness", but it complicates the scheme.

I had to remove my mental block about the diagram approach. Once I started looking seriously at the approach, I realized that there was a way around the duplication issue, making alias diagrams superior to alias relations. These ideas have led both to new theoretical work, in progress, and to an implementation, developed by Victor Rivera at Innopolis (based on some initial work by Marco Trudel and with support from Kogtenkov). At the time of writing the results are not released yet in EVE but promising.

### 10.11 Automatic change analysis

One of the applications of alias analysis is to address the "frame problem". This term denotes the issue of specifying and verifying what an operation does *not* change. When working on the `deposit (sum)` routine of the bank account class, it is natural to specify that `balance = **old** balance + sum` in the postcondition. But for verification we also need (as pointed out by McCarthy as early as 1969) to know that `deposit` does not change the account number, the account owner and most other properties of the account. Writing all the corresponding `property = **old** property` postcondition clauses is not practical. We need a more implicit way.

In current verification schemes including AutoProof, the programmer specifies a frame clause, also called a "modify" or "modifies" or "only" clause, which restricts an operation's modification rights to the properties listed. For example `deposit` would specify `**only** balance`. Frame clauses can be tedious to write and (often the key argument for such cases) to adapt when the software evolves. The rest of the AutoProof team did not consider that there was much of a problem there, but to me it is essential to address it for verification techniques to spread widely. Frame clauses, like separation properties for aliasing, should not be written by programmers but inferred by tools.

The change calculus [241] is a basis for inferring frame properties. It is based on the alias calculus and part of the same implementation in progress.

The first implementation of the change calculus, written by Kogtenkov and described in [241], provided strong arguments for the approach. He ran it on EiffelBase+, an intermediate effort towards EiffelBase 2, which included manually specified frame clauses. Although the results caused some controversy within the team and the final numbers are small, the change analysis clearly evidenced some errors: both false alarms (frame clauses listing properties that do not actually change) and actual unsoundness (a few missed changes). The tool was imperfect and the chain not complete, in the sense that some steps had to be performed manually (see details in the paper), but these results are conclusive enough to call for full-fledged change analysis as a necessary part of verification. The aim of the current effort by Rivera at Innopolis is to make it completely automatic, filling the gaps in the original process.

## 10.12 Invariant inference

The other principal remaining obstacle in the road to Verification As a Matter Of Course is the inference of loop invariants. As mentioned in 8.2, many programmers find loop invariants hard to discover. This is not my view as a programmer (if anything, in the case of a non-trivial loop I tend to write the invariant first), but the tools should adapt to their audience.

There is considerable work on loop invariant inference, but almost always devoted to deducing (or guessing) algebraic relationships from the code, or in the case of dynamic approaches such as DAIKON from executions of the code. Apart from the already noted risk of documenting the bugs instead of avoiding them, this approach does not go well with the Design by Contract idea that you should specify your code. My work with Carlo Furia [131] (also the survey paper with Furia and Velder [218]) starts instead from postconditions; the idea is that the programmers has to specify the postcondition, since it expresses the purpose of the code, and the tool should compute the invariant, since it governs the verification of the implementation. This work is still not ready for integration into AutoProof. Of course nothing prevents it from integrating in the future the techniques, static and dynamic, used in the rest of the literature and our own DAIKON-style work.

## 11 Concurrent programming

Just as obsessive for me as the goal of Verification As a Matter Of Course, and for a much longer time, is the goal of simple concurrency. Concurrent and parallel architectures are ever more critical in today's IT scene, where the end of Moore's law as we knew it led to a dilemma that even made it to the front page of the New York Times: "*Newer chips with multiple processors with dauntingly complex software that breaks up computing into* [concurrently executed] *chunks*". The author, John Markoff, went on to cite computing luminaries, from David Patterson (Berkeley) to Bill Gates, to the effect that we have no clue on how to program these beasts. In the words of a 2011 report of the US National Academy of Science:

> *Heroic programmers can exploit vast amounts of parallelism* [...] *However, none of those developments comes close to the ubiquitous support for programming parallel hardware that is required to ensure that IT's effect on society over the next two decades will be as stunning as it has been over the last half-century.*

Note the reference to heroes. As with verification, top programmers can somehow get things right, at least on their top days. The issue is how to take care of all the remaining cases.

Massively, concurrent programming relies today on threading libraries as available for all major programming languages. The level of abstraction is very low, similar to pre-structured-programming with gotos, except that the risk of error is much higher. To understand a program written with semaphores and such, and to get it right, you must turn yourself into a computer; worse, a parallel computer. You must somehow manage to predict the results of all possible interleavings of possibly computations proceeding all at once. No one can do this; *sequential* operational reasoning is already hard enough. As a result, concurrent programs are plagued by the traditional errors described in the textbooks: data races (two threads reading and writing the same data in the wrong order, as if two customers of expedia.com both found the same available flight seat and because of bad synchronization both reserved it), deadlock (execution coming to a halt because two threads are each waiting on the other to free a resource), priority inversion.

## 11.1 Invariant inference

My search for a concurrent programming model applying the Eiffel principles, simplicity of programming and a promise of reliability, began early. I made a first proposal, already using the name SCOOP, with the S standing for "Sequential and", at TOOLS Europe in 1990 [294]. The root of the modern SCOOP, with S standing for Systematic, was a 1993 Communications of the ACM article [296]. The basis for today's SCOOP, "*Simple* Concurrent Object-Oriented Programming", is chapter 32 of the second edition of the 1997 book "Object-Oriented Software Construction" [299], although Piotr Nienaltowski and his successors in the SCOOP projects at ETH corrected, refined, specified, extended the model considerably and built new implementations, which live on in EiffelStudio, since SCOOP is now a standard part of Eiffel, supported by the associated tools.

## 11.2 SCOOP at ETH and the CME project

The development of SCOOP at ETH had two generations. The first, with Nienaltowski and Volkan Arslan as the prime movers, happened in the early days of the chair. The role of Nienaltowski's type system in specifying the critical notion of "traitor" has already been mentioned. The second generation came with the awarding of the CME (Concurrency Made Easy) grant in 2011.

For readers not familiar with European Union funding mechanisms, an explanation of ERC (European Research Council) grants is in order. EU funding plays a fundamental role for researchers in many European institutions; in fact they, rather than paltry and restrictive national schemes, are often their main source of research funding. But they are very heavy, typically involving research consortia with up to ten or fifteen members representing different countries and different kinds of institutions (companies, universities, laboratories). The bureaucratic complications are easy to guess. Groups from Switzerland, which is not an EU member but has negotiated an equivalent status for its institutions, actively participate. With the resources of ETH and other Swiss funding schemes, such participation is not obligatory and some ETH teams prefer to avoid the hassle. After an experience in the early nineties I had stayed away from EU funding. But ERC grants are another matter. Started in 2008 as an effort to reward excellence with a lesser bureaucratic burden, they are given to one person, not an institution, on the basis of a project description and the candidate's background. They exist at different levels, "starting", "consolidating" and, for senior researchers, "advanced". Typically, a grant is 2.5 million euros (a little more in dollars with the exchange rates of recent years[50]) over five years.

It is not hard to guess that all over Europe the competition for these grants is ferocious. The success rate for advanced grants is about 12%, from a pool of senior-level academic applications already selected by their institutions. (Roughly, the first selection, based on applicants' background, eliminates about two-thirds, and the second step, based on the project, eliminates two thirds of those remaining.) In addition to the funding, the grants are often treated as awards. It is fair to say that ERC grants have had a major beneficial effect on the effectiveness of European research.

I applied to the first call in 2008 and was rejected. The regulations then in place prevented me from applying again the next year. I applied in 2010 and was rejected. In both cases I passed the "background" test but my project came just below the acceptance threshold. The third time, in 2001, made right.

It is interesting to reveal as an aside (please do not tell the ERC) that technically the project was on substance the same each time with different names[51]. But, in addition to the luck factor (when you are that close to the threshold, a critical factor is who else is applying that year), I learned how to present it properly. The second proposal was much better than the first, in particular regarding a review of the state of the art (I had somehow arrogantly assumed that my project was so good that it did not need any of that, and was deeply wrong); and the third one was again far better in many respects. For these redesigns of the proposal I am deeply indebted to Sebastian Nanz, who with the help of Carlo Furia played a key role in bringing it to the level of professionalism required for success. What I had not sufficiently gauged was the intensity of the scrutiny to which, given the stakes of the competition, the proposals would be subjected. The right mindset was to think of what it takes to prepare an article for submission to one of the most selective conferences or journal in our field, such as ICSE, TSE or POPL, and scale it up by a factor of five. Every word mattered.

---

[50] Over the course of the CME project, the Swiss franc appreciated considerably against the euro, first slowly, then abruptly when the Swiss National Bank broke its policy of keeping the euro rate to 1.20, bringing it down overnight to around 1. Exacerbating the already important effect of high Swiss costs, this evolution complicated the finances of the project.

[51] I considered switching to the VAMOC theme and am grateful to Sebastian Nanz and Carlo Furia for convincing me to keep the existing topic and improve the proposal. It would have been a major mistake to restart in a completely different direction.

CME funding enabled us to take SCOOP to its next level by hiring several excellent PhD students, new postdocs and a research engineer, who have contributed considerably to SCOOP.

To provide an independent source of review, we established for CME a Scientific Advisory Board, selecting members from the most prestigious researchers in concurrency, including the authors of the reference textbooks; the list appeared in section 2.4. The board's two meetings, in Zurich in 2014 and at LASER in 2015, provided the opportunity for spirited discussions and invaluable feedback.

### 11.3 SCOOP essentials

SCOOP is described extensively in various references, most of them accessible from the CME project's page at <u>cme.ethz.ch</u> [184]; on that site you will also find the public elements of the CME project plan as accepted by the ERC. Practical usage information is on the <u>eiffel.org</u> site [280]. In addition to pre-ETH references, publications on SCOOP (largely coming out of the CME project) include [21] [24] [25] [76] [111] [129] [134] [135] [146] [152] [165] [170] [174] [175] [192] [197] [199] [200] [205] [223] [232] [235] [240] [243] [251] [258] [260] [269] and [271], plus the robotics software references of section 12 below.

A SCOOP application differs from the execution of a sequential program by working over a set of "regions". Each object belongs to a region. Operations on the objects of a region are the responsibility of the "processor" associated with the region. A processor is a thread of control; the concept of processor can have several implementations, typically by threads. A processor is sequential; concurrency comes from having several processors. Unlike in usual threading models, SCOOP associates the object structure, partitioned into regions, with the computational structure, based on processors.

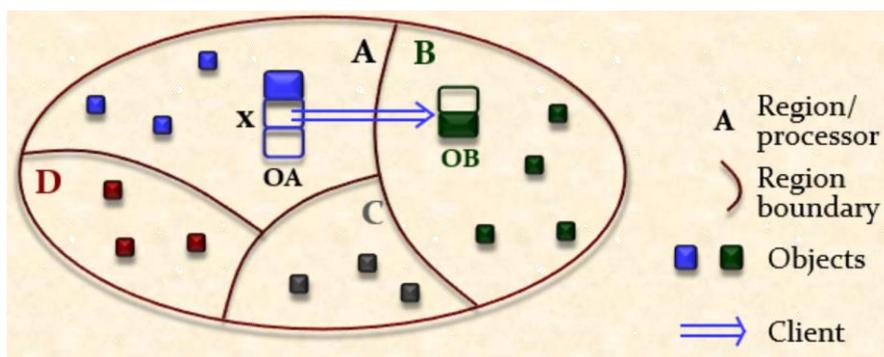

When a field of an object, such as x in the figure for the object OA, may become attached to an object in a different region, called a "separate" object, the semantics of calls x.r is different from the usual OO semantics of sequential calls, which would apply if x were guaranteed to denote objects of the same region as OA:

A. The call will be (for a command) asynchronous: the caller does not have to wait (otherwise there would be no benefit to concurrency). In terminology introduced by Benjamin Morandi, there is now a difference between the *call*, which just logs the request and enables the caller to move on, and the call's *application*, the actual execution of r, which will happen sometime later.

B. As a consequence, the semantics of preconditions changes. A precondition such as **not** x.is_full, involving a call to a separate target, becomes a wait condition.

C. If the call has separate arguments, the application will proceed only when all the corresponding processors are available. This powerful mechanism helps avoid deadlock and makes it possible to synchronize on any number of resources at once.

D. Since processors are sequential, no two calls can be in progress on the same object at any given time. As a consequence, **data races cannot happen in SCOOP**, by construction.

E. If r is a query (an operation returning a result, rather than a command), the caller cannot proceed until the application is complete. In other words, where commands on separate targets are asynchronous, queries are synchronous. This idea, which comes from Denis Caromel's "wait by necessity" [52], provides a simple communication and synchronization scheme: start any number of commands on remote targets (for example, trades in the Paris, New York and Tokyo exchanges), not bothering to wait until you need to; when you do have to re-synchronize, this will be because you need some information from one of those targets (for example the state of your portfolio), which you get from a query.

F. Since the "separate" nature of a target object strongly affects the semantics as just explained, it is essential to ensure that the program text reflects the difference between separate variables, which may denote separate objects at run time, and non-separate ones, which may only denote objects in the same region. For a separate variable, the type must be declared not as just T (denoting some Eiffel type) but **separate** T. Nienaltowski's type system ensures that at run time there will be no traitors, that is to say, a variable not declared separate will only become attached to objects in the same region.

The many examples in the SCOOP documentation illustrate the programming style that results from this model. Two examples suffice here. (See also the hexapod code in section 12.) Dijkstra's famous Dining Philosophers problem, illustrating resource contention, is programmed simply by having each philosopher execute

    eat (left, right)

where `eat` takes two separate arguments representing the left and right forks. The multiple simultaneous argument reservation mechanism (property C above) does the rest. The semantics guarantees fairness. That is all. It is worthwhile to compare this solution to the often elaborate Dining Philosophers programs, using various multi-step synchronization techniques (with semaphores etc.), in the literature.

Note that the keyword **separate** is the only one added to Eiffel in SCOOP.

---

[52] Caromel conceived wait by necessity in the early nineties when working for his PhD at Eiffel Software in Santa Barbara, where he took part in some of the early research leading to SCOOP, although he went on to design his own concurrency model.

A producer-consumer implementation will use a scheme such as, for an insertion operation into a buffer b, the precondition

**require**

    **not** b.is_full

and, for the removal operation, **require not** b.is_empty. Since b is separate, the preconditions are wait conditions (B above), yielding the desired semantics.

### 11.4 CME developments

A list of what we did for SCOOP in CME (so far − the project goes on in another setting, see section 13) would add another ten pages to this article. I will just mention a few salient points. To the names of project members mentioned, one should systematically add Sebastian Nanz, since he co-managed the project and took part in all developments, as reflected by his co-authorship of most of the publications cited.

- (Morandi) Adaptation of the SCOOP model to include "passive regions" [223], which do not need their own processors, with a considerable performance improvement as a result.
- (Morandi) Systematic treatment of exceptions in a concurrent context. Asynchronous calls (property A) raise a delicate problem in connection with exceptions: what happens if an exception arises when the original caller is already off, asynchronously, to new ventures? [170] provides a carefully reasoned answer, which has been adopted in the implementation. Further developments are due to Kolesnichenko [199]
- (Morandi) Detailed operational semantics of SCOOP using Maude [165], providing a framework for "testing" the model [197] and fine-tuning it.
- (Morandi, West, Schmocker) Detailed performance analysis of original SCOOP implementations, leading to the discovery of many inefficiencies; see in particular [174].
- (West) Complementing Morandi's improvements, major redesign of the basic synchronization semantics and implementation, using the technique of "queues of queues" and to spectacular performance improvements putting the SCOOP implementation in the lead group of data-race-free concurrency solutions [251].
- (West) "Demonic testing" for SCOOP programs [175], which goes against the conventional wisdom that "you cannot test concurrent software" by extending AutoTest ideas to include generation of possible processor interleavings.
- (Schmocker) Extensive reliability and performance improvements.
- (Schmocker) Collection of concurrency patterns for SCOOP [232].
- (Schmocker) SCOOP example programs.
- (Kolesnichenko) GPU programming in SCOOP [260] [269].
- (Schill, Poskitt) Extensions of SCOOP towards distributed processing [261].
- (Schill) support for efficient matrix programming.
- (Kogtenkov) Work on deadlock avoidance based on the alias calculus.
- (Poskitt, Corrodi) Work on deadlock avoidance and other semantic properties using graph-based semantics and model checking [266].
- (Caltais, West) Deadlock work [135] [264].

Among the remaining SCOOP tasks are the writing of a book on the approach and the development of an axiomatic-style semantics ready for inclusion of SCOOP mechanisms into AutoProof, which does not support it so far, and AutoTest.

## 12 Software for robotics

An outgrowth of the work on concurrency, obvious in retrospect but not initially envisioned, was a foray into robotics, which became increasingly important.

The connection between robotics and concurrency is paradoxical. Robotics is in principle an ideal application area for concurrent programming, since concurrency naturally exists in the application domain: robots can do many things in parallel. The first presentation of modern SCOOP in *Object-Oriented Software Construction* (chapter 32 of [299]) used an elevator system as one of the most developed examples, particularly attractive because of its extreme use of both concurrency and object-oriented concepts: every elevator cabin, the attached motors, every button on every floor, and even every button in every cabin, was described by a separate object with its own thread of contral. But in practice, as confirmed by the user survey in Andrey Rusakov's thesis [268], many people in the robotics field actually view concurrency with suspicion, because by default concurrency means reliance on a multithreading framework with all its risks. No one can be happy at the prospect of a robot arm's control running into a data race or a deadlock. As a result, authors of robotics applications tend to limit concurrency to the strict minimum.

The idea of SCOOP is to provide a safe environment for concurrent programming, in which these problems do not arise. Robotics is the perfect target domain.

I was flabbergasted when Volkan Arslan and students including Matthias Humbert implemented the exact program as written in the book, as an elevator simulation, and it worked right off the box, giving users the ability to push buttons on floor buttons and in-cabin button to their hearts' contents and see the results in real time[53]. (In fact there had been an earlier implementation, in a prototype of SCOOP written in 1998 at Eiffel Software and using processes instead of threads. It enabled me in a few conferences to give a demo together with an engineer back in the Santa Barbara office and one in yet another location, all of us competing to get the elevator cabins. This early example of concurrent programming over the Internet is in retrospect amazing. I have the impression that it was so futuristic that it must have flown over the heads of audiences, who had no idea of the power of the technology required to achieve it, and what it could do for them.)

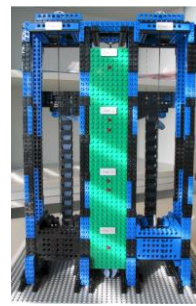

The next step was to implement the elevator in real hardware, using a Lego Mindstorm construction that for several years served as an excellent conversation starter for visitors to our offices.

---

[53] It is a telling reaction, for someone who has a long experience of both the theory of practice of software engineering, and is not Dijkstra, to be incredulous at the news that a verbatim implementation of his paper design *does* immediately work.

These experiments might have remained mere examples of illustrating the power of SCOOP had not Ganesh Ramanathan, a robotics engineer at Siemens, happened to take my concurrency course as part of the ETH extension program[54] and become excited about the robotics potential of SCOOP.

He came up with a remarkable project idea: a hexapod. He had found an article [278] in a journal that we would hardly, as computer scientists, ever have come across: *Arthropod Structure and Development*! For an arthropod with six legs, grouped in two "partner" pairs of three, the article had conditions such as those, "*derived from observations on insects and scorpions and experimentally characterized on grasshoppers and stick insects*":

- A protraction can start only if the partner group on ground, specifically:
    - Protraction starts on completion of retraction
    - Retraction starts on completion of protraction
- A retraction can start only when the partner group raised
- A protraction can end only when partner group retracted

To the SCOOP-educated mind, such a description raises a bell: it is a SCOOP precondition on separate targets, a wait condition. The SCOOP code, exactly as programmed in the routine `begin_protraction`, is

**require**

    me.legs_retracted
    partner.legs_down
    **not** partner.protraction_pending

The comparison with the complexity of any solution using traditional threading mechanisms is striking.

Ramanathan built a hexapod robot and programmed it in SCOOP, through a simple scheme based on these ideas. A YouTube demo of him showing the mechanical hexapod and driving it from a laptop keyboard through the SCOOP program is available on the Roboscoop site [185].

Our robotics software effort took a new dimension with the arrival of Jiwon Shin, a robotics expert coming over to us from down the street, the Autonomous Systems Lab, part of the Mechanical Engineering department, although she is a computer scientist by training. At about the same time, Andrey Rusakov started his thesis. That was also when the CME project was taking off, and we naturally focused on bringing concurrency to robotics software through the Roboscoop project [226]. David Itten and Iwo Steinmann, in their master's projects, also made significant contributions.

The Robotics Programming Laboratory course, presented in section 3.9, demanded considerable work from the robotics team, but provided a direct testbed for the development of Roboscoop.

To showcase our work, we developed an application intended both to test the Roboscoop concepts and for its own sake. As my mother was getting frail, I was shocked to see the low technological level of the walker that she was using, like other people with limited mobility. Surely it must be possible to do better! Thus was born the SmartWalker idea [250] [262].

---

[54] He later completed a regular ETH master's.

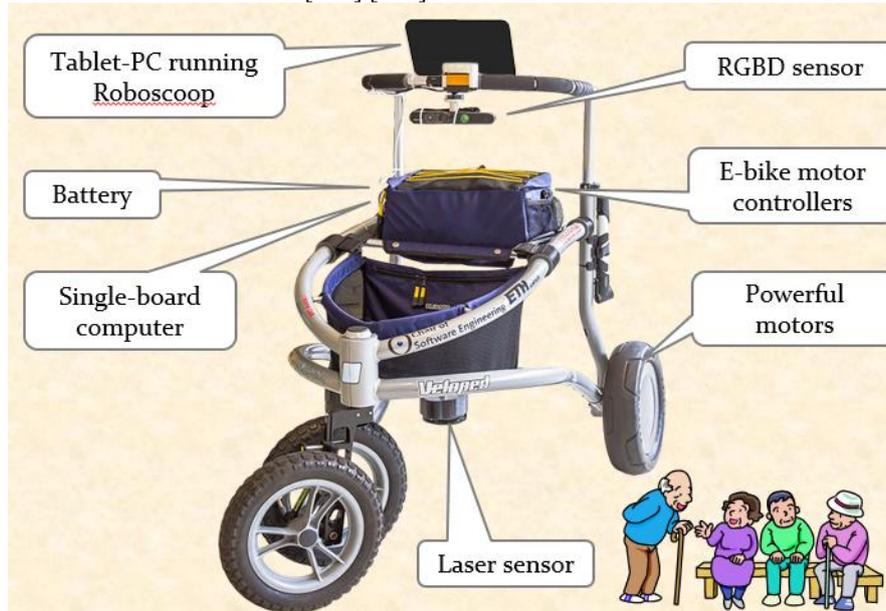

To obtain the hardware expertise that we lacked, we teamed up, in asking support from the Hasler Foundation, with the iHome lab at the University of Luzern.

The design of the SmartWalker hit many challenges, from basic issues of battery weight and capacity to software and user interface matters. At an intermediate progress presentation in a workshop of Smart World, the Hasler project that funded us, some of the other teams criticized us[55] for taking too much of a nerd's approach and not paying enough attention to our end "customers", elderly people and their helpers. We took the comment to heart, and in subsequent months Shin, Rusakov and other project members sometimes including me went around retirement homes in the Zurich area to show our toy and let them try it.

---

[55] Not in the question-and-answer time after my talk in the formal session, but during a lunch discussion. This kind of workshop, favoring researcher-to-researcher interaction, is so much more effective than the typical bureaucratic reporting process of almost all granting agencies, which require formal reports that are generally evaluated not for substance but on whether all the boxes have been checked.

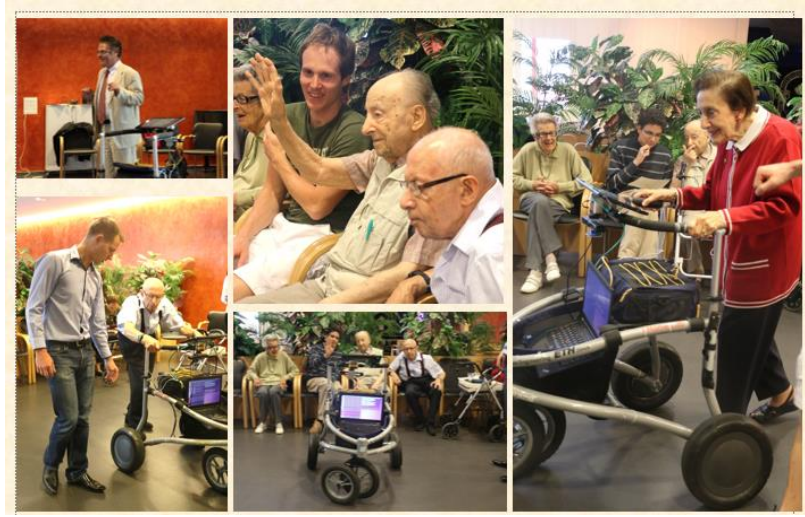

We learned a lot in these encounters. Technology proved an obstacle for some of our audiences, although not all of them, and after all many elderly people use cell phones expertly. A favorite feature was the ability to clap your hands to get the walker to come to you by itself. The personnel in one retirement home explained to us that one of their daily challenges was to collect everyone's (non-smart) walker after each meal, and that automation of that task would be a great relief.

The underlying software, Roboscoop, built by Rusakov, is a complete framework for robotics programming, relying on the ROS infrastructure (Robot Operating System) and on SCOOP for the concurrency part.

Roboscoop is one of the Chair's half-fulfilled promises, a tantalizing bit of a much larger potential contributions. Yes, we are now on to the assessment section.

## 13 Assessment

It is part of the definition of being a scientist that you do not lie: you do not lie to others, and you do not lie to yourself. A certain dose of bluster is acceptable and perhaps essential; given the day-to-day difficulty of research, unless you emphasize the bright side of things you will quickly give up in despair. But drawing attention on the positive does not mean lying. The rule also applies to assessing one's own work. The preceding sections have presented successful results, sometimes boasting; but the successes do not remove the need for a merciless overall assessment.

Our work had an ambitious goal: achieving several breakthroughs in software engineering, including "Verification As a Matter Of Course" (allowing a software process of which formal verification is a routine part) and Concurrency Made Easy. We made strides towards this goal but remain in my estimate about three years away from turning them into results readily available to the developer community for the benefit of society.

Part of the charm of Europe is its insistence on a fixed retirement age for public servants, including academics. Americans have a different practice: age discrimination is illegal; you can encourage aging professors to retire, for example by reducing their resources if they are not performing well, but you cannot kick them out. This flexibility gives US universities an edge over the one-size-fits-all strictures of Europe. To take just one example from software verification, John Reynolds from Carnegie-Mellon proposed the seminal idea of separation logic at the age of 67 and developed it in the next eleven years. A European university would have told him to go tend his garden.

I was going to hit 65 by the end of 2015 but assumed the 5-year CME European Research Council project, due to end two years later, would provide a framework for continuation; it is so far the only Advanced ERC grant in computer science at ETH. Another source of my false sense of safety was the reassuring response received verbally when I preemptively raised the issue in my entry interview 15 years earlier. The current management was, however, intent on applying the 65-year rule. Pointing out that it had its limits, since the preceding president of the university had just lifted it for himself[56], made things worse. Attempts to reach a solution failed.

ERC grants are "portable": attached not to an institution but to an individual, who can take it to another place. The process is heavy: the formal transfer of the CME project to Politecnico di Milano has taken one year and a half; at the time of writing the administrative part has just been finalized and the search for actual project members is starting. The environment in Milan, in the very department where Carlo Ghezzi and others patiently built over three decades one of the best software engineering groups in the world, is exceptionally conducive. I have also had, since late 2014, the privilege of building a Software Engineering Laboratory at Innopolis University in Kazan, with the help of my colleague Manuel Mazzara. The loss of the ETH group and environment is, however, a major blow. As noted in section 1, the group's culture was built over many years with the participation of many talented people. It is not possible to recreate it quickly.

I bear my share of responsibility. I should have obsessed more, right from the minute my flight landed at Zurich's airport on 1 October 2001, about the final deadline, however distant it seemed to lie then. In the first few years I lost some time while learning on the spot the job of a professor. I made no effort to limit my teaching load, which in fact was for many years, including when I was department head, the highest in the department; I love teaching, but perhaps I should have ruthlessly curtailed it. Perhaps I traveled too much and should have declined more keynote invitations (although experience shows that the most fruitful conferences and meetings, those where you discover a fundamental new idea or tool, are not always the ones you thought would be worth the trip). I could have focused on just one or two research topics, instead of the wide spectrum described in previous sections (although here too it is hard to predict which efforts will succeed, and in any case a common thread ran through all these projects). Early on, I did not pay enough attention to a couple of postdoc candidates who went on to pursue successful careers

---

elsewhere and could have boosted our work. More generally, I cannot complain, like so many other academics, particularly in Europe, about insufficient resources or unbearable bureaucratic obstacles. ETH is as conducive an environment as exists for those who want to do good research and teaching.

Still, one should not lie. The outcomes that matter in research are not numerous publications, best-paper awards, completed PhD theses, keynote invitations, software tools, citations and other measurable signs of progress. I was after real success, in the sense of changing the way the IT industry develops software. That was the only justification for putting in parentheses my career as a technology entrepreneur. When you have an ambitious goal, you should be judged by that goal. Such absurd pieties as "it's the journey that matters" or "what's important is to participate" (by, of all people, Coubertin, founder of modern Olympics) are even more wrong in research than anywhere else. Only the result counts. By that standard, the story told in this article is one of glaring, unremitted and probably definitive failure.

## References

The following list attempts to cover all publications by members of the Chair of Software Engineering during their time at the Chair, with the exception of PhD dissertations and of a few ETH technical reports. The information comes from the chair's collated list at se.ethz.ch/publications/ and from my personal publication list at se.ethz.ch/~meyer/publications[57]. The latter is an annotated list where you will find, for each publication, an overview of the content and a short assessment of its relevance.

Almost all publications listed are available online. The lists at the two URLs above have the links.

The most cited publications (at least 20 citations on Google Scholar, Jul 2017) appear with a title in **_bold italics_**.

A second list, beginning with reference [285], includes publication cited in the text but not part of the Chair's record.

---

[57] Although I checked as much as possible, some multi-author articles from that list may not list me in the correct author position. The published articles are the reference.

### Other publications cited in this article